\documentclass[a4paper,11pt]{article}
\usepackage[left=2.4cm,right=2.4cm,top=2.4cm,bottom=2.4cm]{geometry}
\usepackage[T1]{fontenc}
\usepackage[utf8]{inputenc}
\usepackage{tgtermes}
\usepackage{tgheros}
\usepackage{graphicx}
\usepackage{booktabs,tabularx,array,makecell}
\usepackage{amsmath,amssymb,mathtools}
\usepackage{dsfont}
\usepackage{xcolor}
\usepackage{caption}
\usepackage{enumitem}
\usepackage{microtype}
\usepackage{xurl}
\usepackage{setspace}
\usepackage{placeins}
\usepackage{flafter}
\usepackage{etoolbox}
\usepackage[colorlinks=true,linkcolor=black,filecolor=black,citecolor=blue!55!black,urlcolor=blue!55!black]{hyperref}
\usepackage[super,comma,sort&compress]{natbib}

\DeclareCaptionLabelSeparator{pipe}{\enspace|\enspace}
\renewcommand{\arraystretch}{1.08}
\setlist{nosep,leftmargin=*}
\newif\ifskipsectionbarrier
\newcommand{\skipsectionbarrier}{\global\skipsectionbarriertrue}
\pretocmd{\section}{\ifskipsectionbarrier\global\skipsectionbarrierfalse\else\FloatBarrier\fi}{}{}

\AtBeginDocument{\lefthyphenmin=2 \righthyphenmin=3 }
\RequirePackage{xspace}

\newcommand{\ModelH}{UniTB\xspace}
\newcommand{\ModelGF}{UniTB-Flash\xspace}

\newcommand{\Pp}{\mathcal{P}}

\providecommand{\tminus}{\textminus}

\newcommand{\one}{\mathbf{1}}
\newcommand{\dd}{\mathrm{d}}
\DeclareMathOperator{\diag}{diag}
\DeclareMathOperator{\clip}{clip}
\DeclareMathOperator{\softplus}{softplus}

\DeclareMathOperator{\TopK}{TopK}
\DeclareMathOperator*{\argmin}{arg\,min}
\newcommand{\authorblock}{%
\begin{minipage}{0.97\textwidth}
\centering
\normalsize
Mingkang Liu\textsuperscript{1,2},
Zhanghao Zhouyin\textsuperscript{3},
Guanghui Wang\textsuperscript{1,4},
Daye Zheng\textsuperscript{5},
Lei Shen\textsuperscript{2,6,7,*}
and Qiangqiang Gu\textsuperscript{1,4,*}\\[0.7em]
\footnotesize\itshape
\textsuperscript{1}Suzhou Institute for Advanced Research, University of Science and Technology of China, Suzhou 215123, China\\
\textsuperscript{2}Department of Mechanical Engineering, National University of Singapore, 9 Engineering Drive 1, \\ Singapore 117575, Singapore\\
\textsuperscript{3}Department of Physics, McGill University, Montreal, Quebec H3A 2T8, Canada\\
\textsuperscript{4}School of Artificial Intelligence and Data Science, University of Science and Technology of China, Hefei 230026, China\\
\textsuperscript{5}AI for Science Institute, Beijing 100080, China\\
\textsuperscript{6}National University of Singapore (Chongqing) Research Institute, Chongqing 401120, China\\
\textsuperscript{7}National University of Singapore Guangzhou Research Translation and Innovation Institute, Guangzhou,\\ Guangdong 510700, China\\[0.55em]
\upshape
\textsuperscript{*}Corresponding authors. E-mail: shenlei@nus.edu.sg (L.S.); guqq@ustc.edu.cn (Q.G.)
\end{minipage}%
}
\newcounter{snote}
\newcommand{\snote}[2]{\clearpage\refstepcounter{snote}\edef\snnum{\arabic{snote}}\expandafter\snoteii\expandafter{\snnum}{#1}\label{S-#2}}
\newcommand{\snoteii}[2]{\section*{Supplementary Note #1: #2}\addcontentsline{toc}{section}{Supplementary Note #1: #2}}
\hypersetup{
  pdfauthor={Mingkang Liu, Zhanghao Zhouyin, Guanghui Wang, Daye Zheng, Lei Shen, Qiangqiang Gu},
  pdftitle={Physical structure as an organizing principle for universal Hamiltonian learning}
}

\title{\bfseries Physical structure as an organizing principle for universal Hamiltonian learning}

\author{\authorblock}
\date{}

\begin{document}
\pdfbookmark[0]{Main text}{main-part}
\let\savedaddcontentsline\addcontentsline
\renewcommand{\addcontentsline}[3]{}
\begingroup
\setstretch{1.12}
\maketitle

\begin{abstract}
\noindent
Density functional theory (DFT) gives the electronic structure of materials, but its cost limits large-supercell calculations and materials screening. Machine-learned Hamiltonians bypass the self-consistent cycle but are usually system-specific; universal ones could cover new compositions and disordered solids. Such models must generalize across chemistry and keep downstream properties physical, while staying inexpensive for large cells. 
Here we introduce UniTB, a universal E(3) equivariant Hamiltonian representation framework that integrates physical priors rooted in density-functional tight-binding theory with deep learning.
UniTB's network, which corrects a prior Hamiltonian, generalizes across chemistry through a mixture of experts routed by that prior and a charge-equilibration head as in self-consistent-charge tight binding. A projection bounds band shifts from the prior, suppressing spurious states and keeping downstream properties physical. We provide the fastest SO(2) tensor-product GPU kernels for the network. The variant UniTB-Flash assembles its prior from two-center tables at $O(N)$ cost for $N$ atoms, against $O(N\log N)$ on a grid. Trained on 29,242 crystals, UniTB reaches a 0.48~meV matrix-element error on nonmagnetic test crystals spanning 68 elements. Without amorphous training data, both models agree with DFT optics and conductivity of amorphous Ge--Sb--Te. UniTB-Flash ranks candidates with no significant difference from UniTB, whose property values are more accurate; together they screen 16,195 crystals for five application domains of industrial relevance.
\end{abstract}

\vspace{0.6em}

\section{Introduction}
\label{sec:intro}

The electronic structure of a material determines its optical and transport properties. Density functional theory (DFT) obtains it by iterating the electron density to self-consistency, at a cost that grows as the cube of the number of atoms with conventional diagonalization\cite{goedecker1999linear}. This cost limits calculations on amorphous semiconductors and phase-change memory alloys, which need supercells of hundreds of atoms or more\cite{rosset2025asi24,zhou2025gstace24,zhou2023gstgap22}, and it limits screens over many candidate materials\cite{liu2021highthroughput,vansetten2022screening}. Machine-learned Hamiltonians bypass the self-consistent cycle: in a basis of localized atomic orbitals, equivariant neural networks predict the onsite and hopping blocks of the Kohn--Sham Hamiltonian $H$ directly from the atomic structure\cite{li2022deeph,gong2023deephe3,zhong2023hamgnn,schuett2019schnorb,unke2021phisnet,gu2024deeptb,zhouyin2025slem}. Universal DeepH demonstrated direct Hamiltonian learning on a crystals dataset spanning 24 elements in the first four periods, without spin--orbit coupling\cite{wang2024universaldeeph}. Universal HamGNN broadened chemical coverage across much of the periodic table using the residual $H-H^{(0)}$ relative to a non-self-consistent reference Hamiltonian $H^{(0)}$ constructed from neutral-atom densities\cite{zhong2024universalhamgnn}. The later Uni-HamGNN incorporates spin--orbit coupling and supports crystal screening\cite{zhong2026unihamgnn}. NextHAM also learns $H-H^{(0)}$ and additionally uses its zeroth-step reference $H^{(0)}$ to initialize the node and edge features of the network\cite{yin2025nextham}. Trained once on crystals of up to a few tens of atoms, a universal model could then be used for new compositions, disordered solids and materials screening\cite{li2025mlreview}.

These uses require one model that transfers across chemistry, keeps its spectra free of spurious states and stays inexpensive for large cells. Transfer across chemistry depends first on how parameters are shared. A universal model uses one network for thousands of element pairs, distinguished by element embeddings or assigned to sub-networks or experts by bond length\cite{wang2024universaldeeph,zhong2026unihamgnn,yin2025nextham,li2025wanet}; models trained for a single material set the accuracy that a universal model aims to reach\cite{wang2024universaldeeph,zhong2026unihamgnn,yin2025nextham,wang2024deeph2,zhouyin2025slem}. A new composition also changes the charge transferred between atoms, and the neutral-atom prior $H^{(0)}$ contains no charge transfer. The transferred charge depends on the structure beyond the cutoff of a local network, and its potential acts at long range, so a local correction captures only part of it. Interatomic potentials include charge transfer through explicit charge-equilibration (QEq) terms\cite{rappe1991qeq,ghasemi2015cent,ko2021fourthgeneration}, and a long-range Coulomb correction has been added to Hamiltonian models of single polar systems\cite{zhong2026hamgnnlr}; the same argument calls for an explicit term in a universal model. Disorder is a further test of transfer: Hamiltonian models of amorphous materials have been trained on the materials they describe\cite{xia2025largehamiltonian,haldar2025gearsh,li2026deephpack}, whereas a universal model would have to describe their optics and transport from crystal training alone.

The spectrum sets the second requirement. Errors in $H$ are amplified in its generalized eigenvalues when the overlap matrix $S$ is ill-conditioned, so a matrix error below 1~meV can still place a spurious (ghost) state in the band window\cite{li2025wanet,yin2025nextham}, where it distorts the properties computed from the spectrum. Band-energy and reciprocal-space terms in the training loss lower these errors\cite{zhong2023hamgnn,zhong2026unihamgnn,yin2025nextham}; a bound that holds beyond the training set has to come from the construction of the model. The third requirement is cost. Besides the network, the prior, the overlap and the spin-flip blocks of the spin--orbit Hamiltonian are built for every structure. For NextHAM, building $H^{(0)}$ requires the Hartree and exchange--correlation potentials on a grid that fills the cell, at a cost that grows as $N\log N$ for $N$ atoms at fixed density and with the cell volume at fixed $N$; this step has been reported to take most of the inference time of a model that corrects $H^{(0)}$\cite{yin2025nextham}.

Here we build a universal model, UniTB, on the structure of tight binding, which writes the Hamiltonian as a transferable atomic part plus a correction that depends on the environment and suggests a construction for each requirement. Like NextHAM, UniTB starts from $H^{(0)}$\cite{yin2025nextham}. For transfer across chemistry, it combines a mixture of experts with a charge-equilibration head. Tight-binding parametrizations give every element pair its own two-center tables\cite{slater1954koster,porezag1995dftb}, and a mixture of experts\cite{shazeer2017moe,fedus2022switch,wood2025uma,liu2026moemlip,zhou2026matrismoe} lets a shared network specialize in a comparable way: in UniTB, the bond type and the rotation invariants of the bond's prior block\cite{qiao2020orbnet,chen2020deepks}, which contains the potentials of the neighboring atoms, select the experts applied to that bond (PDQ-MoE)\cite{li2021dcd,chen2025mobe}. In self-consistent-charge density-functional tight binding (SCC-DFTB)\cite{elstner1998sccdftb}, the potentials of transferred charges shift each Hamiltonian block in proportion to the overlap of its two orbitals. UniTB adds a charge-equilibration head of this form; its site potentials come from local readouts and from latent charges that are solved as in QEq\cite{rappe1991qeq,ghasemi2015cent,ko2021fourthgeneration} and sum to zero over the cell.

For the spectrum, UniTB uses a property of $H^{(0)}$, the Hamiltonian of the Harris--Foulkes functional\cite{harris1985functional,foulkes1989tightbinding}: the self-consistent correction $H-H^{(0)}$ is a change of local potential, which shifts each ordered band by at most the largest magnitude of that change (Methods). A spectral projection imposes a bound of this kind on the predicted correction in the overlap metric, keeping each predicted band within a fixed distance of the band of $H^{(0)}$ with the same index. For the cost, the variant UniTB-Flash approximates $H^{(0)}$ from two-center integrals tabulated once per element pair and rotated into each bond by the Slater--Koster rules\cite{slater1954koster}, as tight-binding codes do\cite{sankey1989neutralatom,lewis2011fireball,soler2002siesta}. Because atomic orbitals and projectors have finite range, this grid-free prior, the overlap and the spin-flip blocks of the spin--orbit Hamiltonian are sums over a bounded number of neighbors, assembled at a cost linear in $N$ and independent of the cell volume\cite{goedecker1999linear,bowler2012onmethods,soler2002siesta}. With the prior at linear cost, the network and its SO(2) tensor products become the main cost of training and inference. We therefore provide fused GPU kernels for the forward and backward passes of these tensor products, SO2CUDA; they do not depend on UniTB and can be used by other equivariant models built on SO(2) convolutions, including interatomic potentials.

With 29,242 training crystals, UniTB reaches a mean absolute matrix-element error of 0.48~meV on nonmagnetic test crystals spanning 68 elements. After projection, none of the 200 band-test structures shows a spurious state in UniTB's bands under the strict detection criterion. On one graphics processing unit (GPU), UniTB-Flash generates $H$ and $S$ for a 3,456-atom cell in 1.65~s and UniTB in 5.11~s; the grid-free prior leaves a larger correction to learn and so trades some accuracy for speed. Without amorphous training data, both models reproduce the DFT-Hamiltonian optical constants and conductivity of amorphous Ge--Sb--Te supercells of 180--198 atoms as well as the computed valence-band photoemission spectra of amorphous Ge--Sb--Te and Si. On held-out thermoelectric crystals, they reproduce the Seebeck coefficients. We test both models on application domains chosen from common industrial needs: interconnect metals for chips, high-index photonic materials, photovoltaic absorbers, high-mobility semiconductors and thermoelectrics, together with band-gap design. In these six domains, UniTB-Flash reaches screening scores close to those of UniTB, with no significant difference, whereas UniTB gives the more accurate property values. In a two-stage screen of 16,195 public crystal structures for the first five domains, UniTB-Flash ranks the candidates, UniTB re-evaluates the highest ranked and DFT checks the final lists. The screen finds $n$-type thermoelectric candidates beyond the strongest reference material, among them compounds with measured large power factors, and in every domain the DFT checks confirm the accuracy of both models on crystals they have not seen.

\section{Results}
\label{sec:results}

\subsection{A universal Hamiltonian model with tight-binding structure}
\label{sec:model}
UniTB writes the Hamiltonian as a prior computed from atomic data plus learned corrections. In a basis of numerical atomic orbitals, it predicts each onsite block ($i=j$ in the same cell) and each hopping block (all other pairs, including periodic images) of the Hamiltonian $H$ as
\begin{equation}
 \widehat H_{ij}=\Pp_{ij}+\Delta_{\theta,ij}+\Delta^{\mathrm{pot}}_{ij},
 \label{eq:residual}
\end{equation}
where the prior Hamiltonian $\Pp$ is computed from the structure and tabulated atomic data, an equivariant network gives the correction $\Delta_\theta$, and site potentials add $\Delta^{\rm pot}$ through the overlap matrix $S$ (Fig.~\ref{fig:overview}). UniTB uses the zeroth-step Hamiltonian of superposed neutral-atom densities, $\Pp=H^{(0)}$, and UniTB-Flash the grid-free prior $\Pp=H_{\rm GF}$, assembled from tabulated two-center integrals as in tight binding (next section). The prior enters the model in three ways (orange paths in Fig.~\ref{fig:overview}a): it initializes the network features, its rotation invariants select the experts of each bond, and it is the baseline to which the corrections are added. With $H^{(0)}$ still given as input, learning the full Hamiltonian instead of the correction raises the mean matrix-element error of our dense model from $0.813$ to $5.692$~meV (Supplementary Table~\ref{S-tab:abl-reference}).

\begin{figure}[tbp]
\centering
\includegraphics[width=\textwidth]{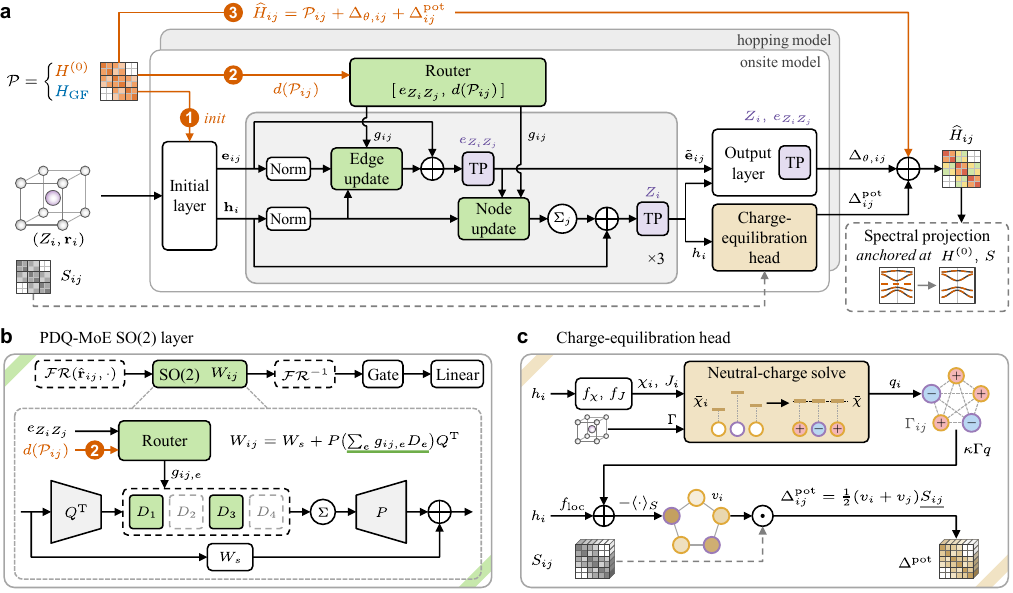}
\caption{\textbf{Architecture of UniTB.} \textbf{a},~UniTB uses the prior $\Pp=H^{(0)}$ and UniTB-Flash uses $\Pp=H_{\rm GF}$. Each model trains an onsite and a hopping network of the same architecture (stacked frames). The prior enters along the orange paths: it initializes the features (\raisebox{-0.15em}{\includegraphics[height=0.82em]{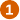}}), its rotation invariants $d(\Pp_{ij})$ feed the router (\raisebox{-0.15em}{\includegraphics[height=0.82em]{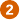}}), and it is the baseline of the residual sum in Eq.~\eqref{eq:residual} (\raisebox{-0.15em}{\includegraphics[height=0.82em]{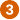}}). Three interaction layers update edge and node features with PDQ-MoE SO(2) maps (green) and element-dependent tensor products (TP). Only UniTB applies the spectral projection, shown removing a spurious in-gap level (inset). \textbf{b},~In each PDQ-MoE SO(2) layer, the router selects two of four expert cores $D_e$. The cores act between the shared bases $Q^{\mathrm T}$ and $P$ inside the bond-frame rotations, and $W_s$ is always active. \textbf{c},~In the charge-equilibration head, readouts of $h_i$ give electronegativities $\chi_i$ and hardnesses $J_i$. Charge equilibration at zero total charge (Eq.~\eqref{eq:qeq}) gives latent charges $q_i$, and their site potentials $v_i$ scale the overlap $S_{ij}$. Values are schematic.}
\label{fig:overview}
\end{figure}

UniTB shares parameters as a tight-binding parametrization does. Tight binding fits onsite and hopping blocks separately and indexes its tables by element and element pair. UniTB likewise predicts the two kinds of blocks with two separately trained networks of one architecture, and element-level tensor products make every update depend on the element or the ordered element pair (Fig.~\ref{fig:overview}a). Removing both choices raises the mean matrix-element error of the dense model from $0.813$ to $2.049$~meV (Supplementary Table~\ref{S-tab:abl-backbone}).

Within one element pair, the blocks still depend on the environment, the onsite blocks most of all: an onsite block contains the crystal field of the whole neighborhood, whereas a two-center hopping integral depends mainly on the element pair and the distance. The prior block $\Pp_{ij}$ already contains the potentials of all neighboring neutral atoms, so its rotation invariants $d(\Pp_{ij})$ describe the surroundings. Together with the ordered-bond embedding $e_{Z_iZ_j}$, they select two of four experts in each SO(2) map of the network (PDQ-MoE; Fig.~\ref{fig:overview}b),
\begin{equation}
 W_{ij}=W_s+P\Big(\sum_e g_{ij,e}D_e\Big)Q^{\mathrm T}.
 \label{eq:pdq}
\end{equation}
All interactions share the always-active map $W_s$ and the bases $Q$ and $P$; the gates $g_{ij,e}$ weight the selected expert cores $D_e$ that act between these bases, as in dynamic convolution decomposition\cite{li2021dcd}. Because the invariants do not depend on the orientation of $\Pp_{ij}$, the gates preserve equivariance (Supplementary Fig.~\ref{S-sfig:irreps}).

The prior contains no charge transfer between atoms, because it is evaluated at superposed neutral-atom densities. In SCC-DFTB, the potentials of transferred charges enter the onsite and hopping blocks through the overlap matrix. UniTB adds a charge-equilibration head (charge head) of this form (Fig.~\ref{fig:overview}c): latent charges are obtained by charge equilibration with a Gaussian-charge kernel $\Gamma$ under zero total charge, and together with local readouts they give site potentials that shift the blocks as in second-order SCC-DFTB\cite{elstner1998sccdftb},
\begin{equation}
 \Delta^{\mathrm{pot}}_{ii}=v^{\rm on}_iS_{ii},\qquad
 \Delta^{\mathrm{pot}}_{ij}=\tfrac12(v^{\rm hop}_i+v^{\rm hop}_j)S_{ij},
 \label{eq:potential-assembly}
\end{equation}
where the charges are latent variables learned from the Hamiltonian alone. Controlled comparisons of each component at a fixed training budget are collected in Supplementary Note~\ref{S-sn:ablations}.

\subsection{Grid-free construction of the prior Hamiltonian}
\label{sec:prior}
Besides the network output, every structure needs a prior, the overlap matrix and the spin-flip blocks of the spin--orbit Hamiltonian. The overlap, the spin-flip blocks and the grid-free prior are sums of finite-range atomic contributions, so their cost grows linearly with the number of atoms (Fig.~\ref{fig:prior}); $H^{(0)}$ also needs a potential on a grid that fills the cell. The zeroth-step prior
\begin{equation}
 H^{(0)}=T+V_{\rm nl}+V_{\rm ion,loc}+V_{\rm H}[\rho_0]+V_{\rm xc}[\rho_0+\rho_{\rm c}]
 \label{eq:hzero}
\end{equation}
is the Hamiltonian of the DFT code at the superposed neutral-atom valence density $\rho_0$, with the core density $\rho_{\rm c}$ of the nonlinear core correction in the exchange--correlation term; its Hartree and exchange--correlation potentials are evaluated on a real-space grid that fills the cell (Fig.~\ref{fig:prior}a, lower lane). The grid-free prior $H_{\rm GF}$ of UniTB-Flash replaces the grid by tables, as tight-binding codes do\cite{sankey1989neutralatom,soler2002siesta,lewis2011fireball}. The neutral-atom potential of each center is short-ranged because its ionic and electronic Coulomb tails cancel, so its matrix elements can be tabulated\cite{sankey1989neutralatom}. Two-center integrals of the kinetic energy, the overlap, the neutral-atom potential and the projector overlaps are tabulated once per element pair and rotated into each bond with the Slater--Koster rules\cite{slater1954koster}; the separable pseudopotential reduces its three-center terms to products of two-center projector overlaps\cite{kleinman1982bylander}, and a finite-rank factorization approximates the crystal field of the surrounding neutral atoms. Exchange--correlation enters as a tabulated pair term with an environmental correction from the multicenter weighted exchange--correlation density approximation (McWEDA) for hopping blocks\cite{horsfield1997multicenter,jelinek2005mcweda} and as an atom-centered quadrature for onsite blocks (Methods), so no grid spans the cell.

The overlap matrix comes from the same two-center tables and takes the form $S\otimes\mathds 1_2$ in the spinor basis. The local potentials of both priors are spin independent, so spin--orbit coupling enters only through the fully relativistic separable pseudopotential\cite{dalcorso2005soc}: the spin-flip blocks contract the projector overlaps already computed for the scalar term with the spin-resolved coefficients of each projector center (Methods; Fig.~\ref{fig:prior}a, magenta).

Each assembled term is a sum over neighbors within a fixed cutoff, and a sphere of fixed radius holds a bounded number of atoms, so the overlap, the spin-flip blocks and $H_{\rm GF}$ cost $O(N)$ operations for $N$ atoms, independent of the cell volume $\Omega$ (Methods; Supplementary Note~\ref{S-sn:cost}). $H^{(0)}$ adds its cell-wide Hartree and exchange--correlation potential on the $G\simeq\Omega/h^3$ points of a grid with spacing $h$, computed with fast Fourier transforms; its cost, $O(N+G\log G)$, becomes $O(N\log N)$ at fixed density and grows with the cell volume at fixed $N$ (Fig.~\ref{fig:prior}a, boxed labels), and the measured assembly times follow these forms (Fig.~\ref{fig:prior}c).

\begin{figure}[tbp]
\centering
\includegraphics[width=\textwidth]{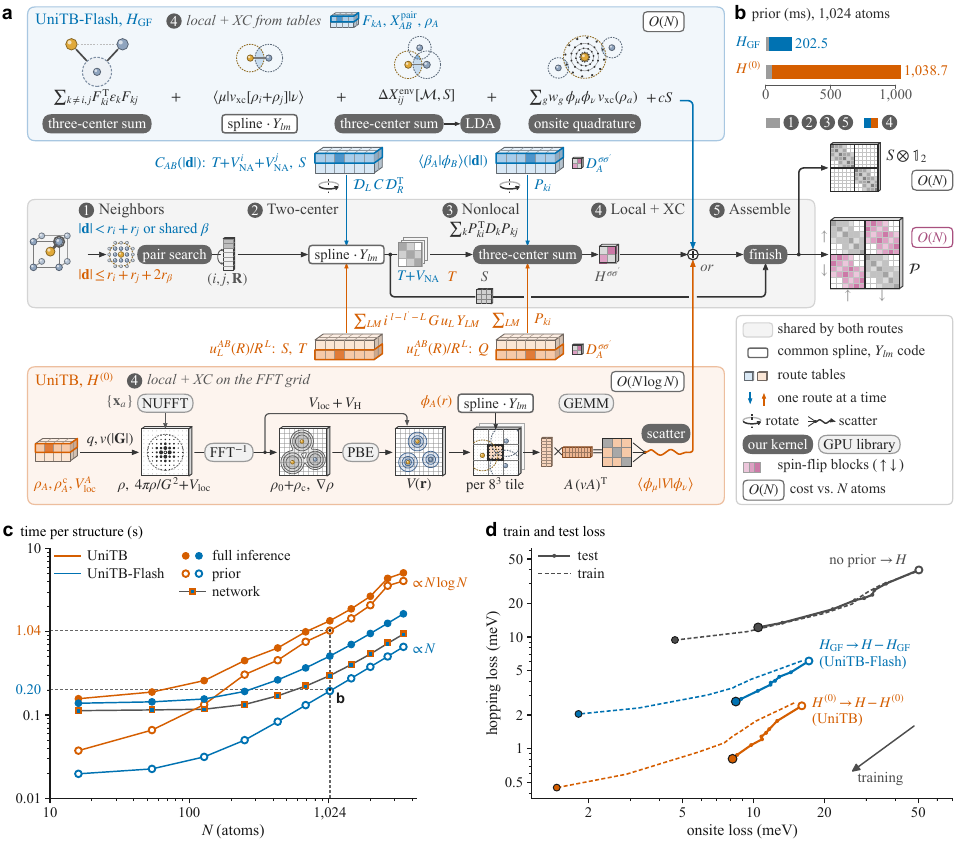}
\caption{\textbf{Assembly and cost of the two prior Hamiltonians and their effect on training.} \textbf{a},~The grid-free prior $H_{\rm GF}$ of UniTB-Flash and the zeroth-step Hamiltonian $H^{(0)}$ of UniTB are assembled in five steps (Methods). They differ in the two-center step and in step~4, where $H_{\rm GF}$ sums finite-range atomic terms and $H^{(0)}$ evaluates its potential on a grid that fills the cell. \textbf{b},~Assembly time of each prior for a 1,024-atom PbTe supercell. \textbf{c},~Inference time for PbTe supercells on one H200 GPU. Full inference runs from the structure to $H$ and $S$ and includes the prior assembly and the network. The labels $\propto N$ and $\propto N\log N$ give the derived cost forms, and dashed lines mark the prior times of \textbf{b}. \textbf{d},~Onsite and hopping losses, $\tfrac14$(MAE + RMSE), of three dense networks that are trained identically on MPHam-12K and differ only in the prior. Each trajectory runs from the first checkpoint (open circles) to the last (filled circles).}
\label{fig:prior}
\end{figure}

The two priors differ in cost mainly in the local-potential step, in which $H^{(0)}$ evaluates its cell-wide potential on the grid and integrates each orbital pair over it. For a 1,024-atom PbTe supercell, the five-step assembly (Methods) takes 0.20~s for $H_{\rm GF}$ and 1.04~s for $H^{(0)}$, of which this step takes 0.17 and 0.99~s (Fig.~\ref{fig:prior}b). Generating the sparse $H$ and $S$ of a 3,456-atom PbTe supercell from its structure takes $1.65$~s with UniTB-Flash and $5.11$~s with UniTB on one GPU (Fig.~\ref{fig:prior}c). The network forward passes of the two models, including charge equilibration, agree in time to within $4\%$, so the difference comes from the prior assembly.

The prior also sets the difficulty of the learning task. $H_{\rm GF}$ approximates $H^{(0)}$ without a cell-wide grid, and its learning target, $H-H_{\rm GF}$, equals $H-H^{(0)}$ plus the deviation of $H_{\rm GF}$ from $H^{(0)}$. Three dense networks, identical except for the prior, are trained on MPHam-12K to learn the full $H$ without a prior, $H-H_{\rm GF}$ and $H-H^{(0)}$; their combined test MAEs are $11.7$, $3.08$ and $0.813$~meV (Supplementary Note~\ref{S-sn:prior}). Fig.~\ref{fig:prior}d follows their onsite and hopping losses on the training and test structures through training. In the onsite blocks, $H_{\rm GF}$ uses an atom-centered quadrature and is already close to $H^{(0)}$. The onsite loss of the $H_{\rm GF}$ network therefore starts and ends close to that of the $H^{(0)}$ network; their final onsite test MAEs are $8.90$ and $8.49$~meV. In the hopping blocks, the multicenter exchange--correlation can only be approximated without the grid, here by a tabulated pair term with the McWEDA correction. The hopping loss of the $H_{\rm GF}$ network therefore starts well above that of the $H^{(0)}$ network and stays higher to the end; their final hopping test MAEs are $3.01$ and $0.726$~meV. UniTB-Flash thus gains speed from its cheaper prior, and UniTB keeps the higher accuracy (Table~\ref{tab:benchmark}).

With the prior at linear cost, most of the time of UniTB-Flash goes into the network: at 3,456 atoms, its forward passes take 0.92 of the 1.65~s needed to generate $H$ and $S$ (Supplementary Note~\ref{S-sn:screening}), and training runs the network alone. Most of the network's operations are equivariant tensor products evaluated as SO(2) maps in the frame of each bond. Our kernel library SO2CUDA fuses the rotations into and out of the bond frame with grouped matrix products over the experts, in both passes (Methods). For the SO(2) linear layer on one H200 GPU, it is faster than the official EquiformerV3 implementation, cuEquivariance and plain PyTorch in all 26 configurations tested, with median time ratios from 1.29 to 10.70, and it shortens a training step of UniTB's onsite network from 1.64~s with plain PyTorch to 0.79~s.

\subsection{Matrix-element accuracy across the periodic table}
\label{sec:benchmark}
With one onsite and one hopping network for all elements, UniTB recovers the self-consistent Hamiltonian to a mean matrix-element error below 1~meV (Table~\ref{tab:benchmark}). Trained on MPHam-12K, the de-duplicated NextHAM training list, at the published budget of about 100 epochs, it reaches $0.722$~meV on the 3,000 test structures of the NextHAM benchmark\cite{yin2025nextham}; trained on MPHam-29K, which adds MP-20 crystals\cite{xie2022cdvae}, for about 300 epochs, it reaches $0.493$~meV (Methods). The combined error weights every valid matrix entry equally, so it is set mostly by the far more numerous hopping entries (Table~\ref{tab:benchmark}).

\begin{table}[tbp]
\caption{\textbf{Matrix-element and band errors.} Matrix errors are evaluated on the 3,000 test structures and band errors on 200 of them drawn at random. Onsite, Hopping and Combined are mean absolute errors of the real spin-up--spin-up block. The band error $\mathrm{fw}_{10}$ is computed after the spectral projection with $b=10$~eV, or for UniTB-Flash after canonical orthogonalization (Methods). Indented rows add components cumulatively; UniTB also includes site channels and prior perturbations. UniTB-Flash uses the grid-free prior. LEM and SLEM\cite{zhouyin2025slem} are trained with the $H^{(0)}$ prior on the same data and budget as the MPHam-12K models. NextHAM reports no aggregate band errors\cite{yin2025nextham}; our labels are recomputed with its DFT settings (Supplementary Note~\ref{S-sn:data}).}
\label{tab:benchmark}
\centering\small
\setlength{\tabcolsep}{2pt}
\begin{tabular*}{\textwidth}{@{}l@{\extracolsep{\fill}}rrrrr@{}}
\toprule
 & \multicolumn{3}{c}{Matrix MAE (meV)} & \multicolumn{2}{c}{Band $\mathrm{fw}_{10}$ (meV)}\\
\cmidrule(lr){2-4}\cmidrule(l){5-6}
Model & Onsite & Hopping & Combined & Mean & Median\\
\midrule
\multicolumn{6}{@{}l@{}}{\textit{Published (Yin et al.)}\hfill\textit{12,000 structures, 100 epochs}}\\
NextHAM & & & 2.834 & & \\
\midrule
\multicolumn{6}{@{}l@{}}{\textit{Ours, MPHam-12K}\hfill\textit{11,953 structures, 100 epochs}}\\
LEM & 7.925 & 2.098 & 2.160 & 88.11 & 62.08\\
SLEM & 11.319 & 1.278 & 1.386 & 80.62 & 48.53\\
Dense (internal baseline) & 8.488 & 0.726 & 0.813 & 67.40 & 38.02\\
\quad + PDQ-MoE & 8.112 & 0.700 & 0.785 & 66.84 & 39.07\\
\quad + charge head & 7.977 & 0.670 & 0.753 & 64.26 & 33.41\\
\textbf{UniTB} & 8.098 & 0.638 & \textbf{0.722} & 63.14 & 35.96\\
\midrule
\multicolumn{6}{@{}l@{}}{\textit{Ours, MPHam-29K}\hfill\textit{29,242 structures, 300 epochs}}\\
Dense & 6.208 & 0.496 & 0.559 & 51.90 & 23.35\\
\textbf{UniTB} & 5.741 & 0.434 & \textbf{0.493} & 49.05 & 21.31\\
\addlinespace[2pt]
\textbf{UniTB-Flash} & 7.166 & 1.636 & \textbf{1.694} & 83.47 & 42.98\\
\bottomrule
\end{tabular*}
\end{table}

At the MPHam-12K budget, our dense model reaches $0.813$ meV; PDQ-MoE, the charge head, site channels and prior perturbations together lower the MAE by $11.3\%$ relative to it, and UniTB has the lower error on $81.8\%$ of the test structures. At the MPHam-29K budget, UniTB reaches $0.493$ meV, $11.9\%$ below the dense model at the same budget ($0.559$ meV). With the grid-free prior, UniTB-Flash reaches $1.694$ meV, $16.3\%$ below the dense model with the same prior ($2.023$ meV; Supplementary Table~\ref{S-tab:abl-recipe-amplitude}).

Averaged over the nonmagnetic test structures that contain a given element, the matrix error of UniTB trained on MPHam-29K is below 0.9 meV for each of the 68 elements in these structures, and its mean over all 2,431 structures is 0.48 meV (Fig.~\ref{fig:periodic}; UniTB-Flash in Supplementary Fig.~\ref{S-sfig:periodic-flash}).

\begin{figure}[tbp]
\centering
\includegraphics[width=\textwidth]{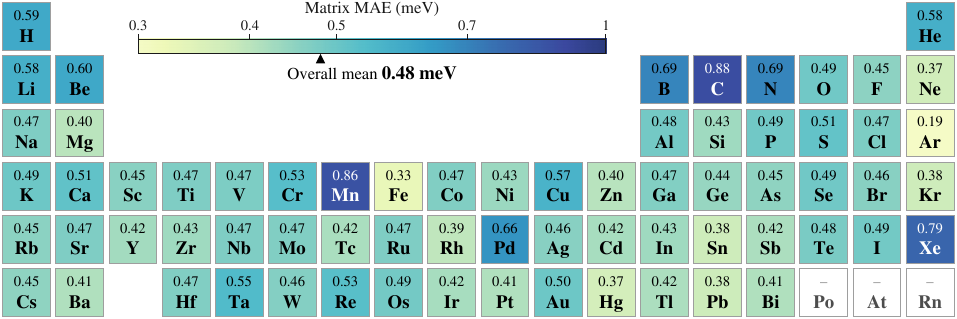}
\caption{\textbf{Matrix-element errors across the periodic table.} Values are for UniTB (MPHam-29K) on the 2,431 nonmagnetic test structures. Each tile averages the matrix-element MAE over the structures that contain the element, so a structure counts once for each of its elements. Dashes mark elements not in the test set. Logarithmic color scale. Values beyond the color range take the end color.}
\label{fig:periodic}
\end{figure}

Band energies test the effect of the learned correction on the spectrum. The band error $\mathrm{fw}_{10}$, the mean absolute eigenvalue error within $\pm10$ eV of the DFT valence-band maximum, averages $745.63$ meV over 200 test structures for $H^{(0)}$ alone and $63.14$ meV for UniTB at the MPHam-12K budget, and it decreases along the same sequence of models as the matrix error (Table~\ref{tab:benchmark}).

\subsection{Environment-aware routing of expert networks}
\label{sec:routing-results}
The prior descriptors matter most for the onsite blocks, as the crystal-field picture suggests. Relative to routing by bond type and prior descriptors, routing by bond type alone raises the onsite error by $7.17\%$ and the hopping error by $2.30\%$ (Supplementary Table~\ref{S-tab:moe}b). Routing by bond type and distance raises the onsite error by $8.82\%$ and lowers the hopping error by $1.63\%$. For the hopping blocks, distance thus describes the environment well enough; the onsite crystal field requires the prior descriptors. The trained router distinguishes environments within each bond type: about one edge in ten is assigned to an expert pair other than the most common pair for its type, and shuffling the gate vectors among edges of the same type raises the onsite and hopping errors by factors of $3.29$ and $2.13$ (Supplementary Table~\ref{S-tab:routing-statistics}). With this router, PDQ-MoE lowers the matrix error of the dense model by $3.49\%$, against $1.15\%$ for a wider dense model with a similar parameter count (Supplementary Table~\ref{S-tab:moe}a).

\subsection{Charge equilibration in the learned Hamiltonian}
\label{sec:charge-results}
The matrix error of a dense model without the charge head contains a component of overlap-weighted site potentials, the form in which SCC-DFTB adds the potentials of transferred charge. The RMS of this component is $6.4$ meV on training structures, $28.3$ meV on test structures whose formula occurs in the training set and $59.4$ meV on test structures with new formulas, and it increases across quartiles of distance to the training features, from $14.5$ to $97.2$ meV (Fig.~\ref{fig:charge}a). The charge head adds a correction of the same matrix form and lowers the matrix error on both backbones tested, by $2.08\%$ on the dense model and by $4.24\%$ on PDQ-MoE with site channels (Supplementary Table~\ref{S-tab:charge-controls}). A local-only head on PDQ-MoE, which omits the cell-wide charge equilibration, has a $1.97\%$ higher error than the charge head, so part of the gain comes from equilibrating charge over the whole cell. The paired gain from adding the head increases with cell size and reaches $6.98\%$ on PDQ-MoE and $3.42\%$ on the dense model for cells above 22 atoms (Fig.~\ref{fig:charge}b).

\begin{figure}[tbp]
\centering
\includegraphics[width=\textwidth]{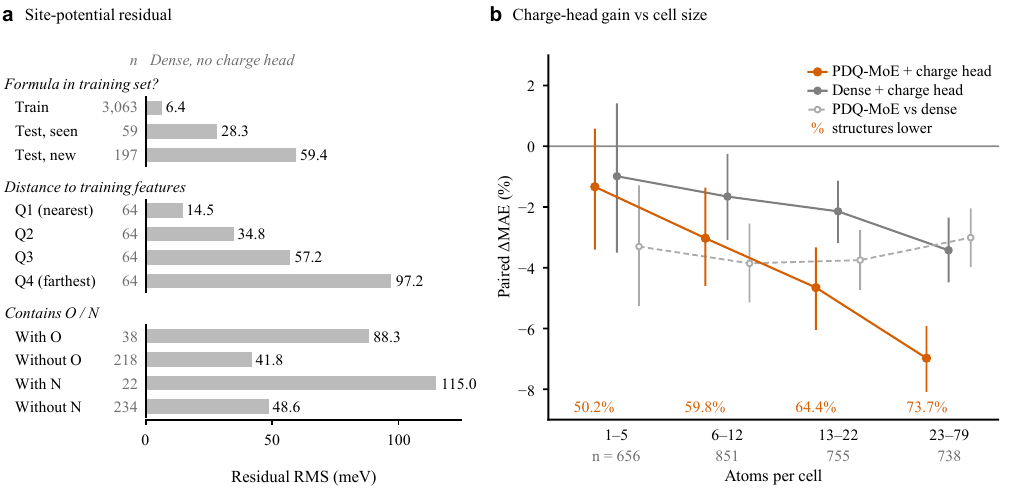}
\caption{\textbf{Site-potential errors and the charge-equilibration head.} \textbf{a},~A dense model without the charge head is trained for this diagnostic on its own split of 3,063 training and 256 test structures. Bars give the RMS of the component of its matrix error that has the overlap-weighted site-potential form of SCC-DFTB. \textbf{b},~Change in combined matrix MAE when the charge head is added to PDQ-MoE or to the dense model, for the 3,000 test structures binned by atoms per cell. Changes are paired over the same structures, with $95\%$ intervals. Percentages give the share of structures on which the head lowers the error of PDQ-MoE. The PDQ-MoE-versus-dense comparison is made without the head (Supplementary Note~\ref{S-sn:charge}).}
\label{fig:charge}
\end{figure}

\subsection{Spectral projection against spurious states}
\label{sec:band-results}
With the spectral projection, UniTB shows no spurious state under the strict detection criterion in any of the 200 band-test structures (Fig.~\ref{fig:projection}). The mean and median band errors of UniTB and UniTB-Flash are both below those of the dense models with the same prior and post-processing (Supplementary Fig.~\ref{S-sfig:bands}). Band calculations solve $H(\mathbf k)c_a=\varepsilon_aS(\mathbf k)c_a$ at each wavevector. The self-consistent correction $H-H^{(0)}$ is the change of local effective potential caused by charge redistribution. Its generalized eigenvalues therefore lie within the range of that potential, and each ordered band shifts by at most its largest magnitude (Methods); for the DFT labels of 3,063 training structures, these generalized eigenvalues stay below 8.5 eV in magnitude. A learned residual has no such bound, and an outlying generalized eigenvalue can place a spurious state in the band window. UniTB therefore clips the generalized spectrum of its predicted residual relative to $H^{(0)}$ at $b=10$ eV, above this range. The assembled residual $\widehat\Delta=\widehat H-H^{(0)}$ is whitened with the Cholesky factor of $S$ and diagonalized; eigenvalues outside $[-b,b]$ are moved to the nearest boundary, directions inside the interval are kept unchanged, and the result is added back to $H^{(0)}$. Every band of the projected Hamiltonian $\widehat H_b$ then obeys $\varepsilon_a(H^{(0)},S)-b\leq\varepsilon_a(\widehat H_b,S)\leq\varepsilon_a(H^{(0)},S)+b$ (Fig.~\ref{fig:projection}a; Methods).

In LiZrPd$_2$, a spurious band crosses the band window of the unprojected UniTB spectrum; projection removes it and lowers the full-path $\mathrm{fw}_{10}$ from 156.2 to 78.2 meV (Fig.~\ref{fig:projection}b). On the 200 band-test structures, projection reduces the mean $\mathrm{fw}_{10}$ of each of the four MPHam-12K models by 22 to 39 meV (Supplementary Fig.~\ref{S-sfig:bands}; Supplementary Note~\ref{S-sn:bands}).

The UniTB-Flash pipeline contains no $H^{(0)}$. Its residual relative to $H_{\rm GF}$ is $H-H_{\rm GF}=(H-H^{(0)})+(H^{(0)}-H_{\rm GF})$. The first term is the same bounded change of local potential. The second, the deviation of the grid-free prior from $H^{(0)}$, comes mainly from the approximate multicenter exchange--correlation of the hopping blocks and is not a local potential. It has no such bound, so a projection would also clip part of the correction the network has learned. UniTB-Flash is therefore not projected (Supplementary Table~\ref{S-tab:flash-bscan}).

\begin{figure}[tbp]
\centering
\includegraphics[width=\textwidth]{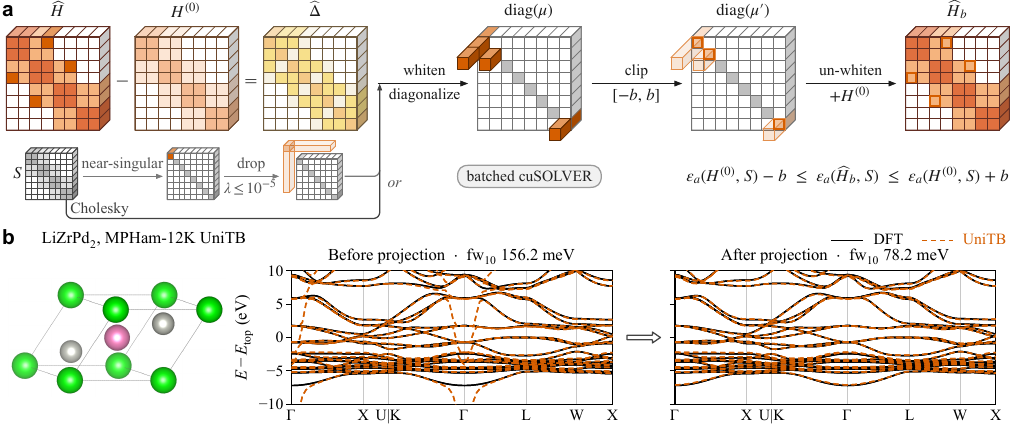}
\caption{\textbf{Spectral projection relative to $H^{(0)}$ and its effect on the bands.} \textbf{a},~The predicted residual $\widehat\Delta=\widehat H-H^{(0)}$ is whitened in the overlap metric and diagonalized. Eigenvalues outside $[-b,b]$ (orange) are clipped, and adding the result back to $H^{(0)}$ gives $\widehat H_b$, whose bands obey the bound shown (Methods). For a near-singular $S$, overlap eigenvalues below $10^{-5}$ are dropped first (gray branch). \textbf{b},~Bands of LiZrPd$_2$ (mp-1185423), a test structure in which a strict spurious state is detected before projection. In its primitive cell, Li is pink, Zr green and Pd grey. The projection uses $b=10$~eV, and each spectrum is aligned to its own valence-band top.}
\label{fig:projection}
\end{figure}

\subsection{Transferability from crystals to amorphous materials}
\label{sec:amorphous}
Amorphous solids test whether a correction learned from crystals remains accurate without long-range order. We evaluate both models on $18$ supercells with $180$--$216$ atoms: six Si cells from a-Si-24\cite{rosset2025asi24} and twelve cells of Ge--Sb--Te (GST) phase-change memory alloys of seven compositions from GST-ACE-24\cite{zhou2025gstace24} (Supplementary Note~\ref{S-sn:amorphous}). Trained only on crystals, both reproduce the DFT-Hamiltonian optical constants and conductivity of amorphous GST, their dependence on composition and the computed valence-band photoemission spectra of amorphous GST and Si (Fig.~\ref{fig:amorphous}). On all 18 cells, both models lower the matrix error relative to the dense model with the same prior, by $12.0\%$ for UniTB ($0.766$ meV) and by $10.9\%$ for UniTB-Flash ($1.818$ meV; Supplementary Fig.~\ref{S-fig:amorphous-l} and Supplementary Table~\ref{S-tab:abl-recipe-amplitude}), as on the crystal test set.

\begin{figure}[tbp]
\centering\includegraphics[width=\textwidth]{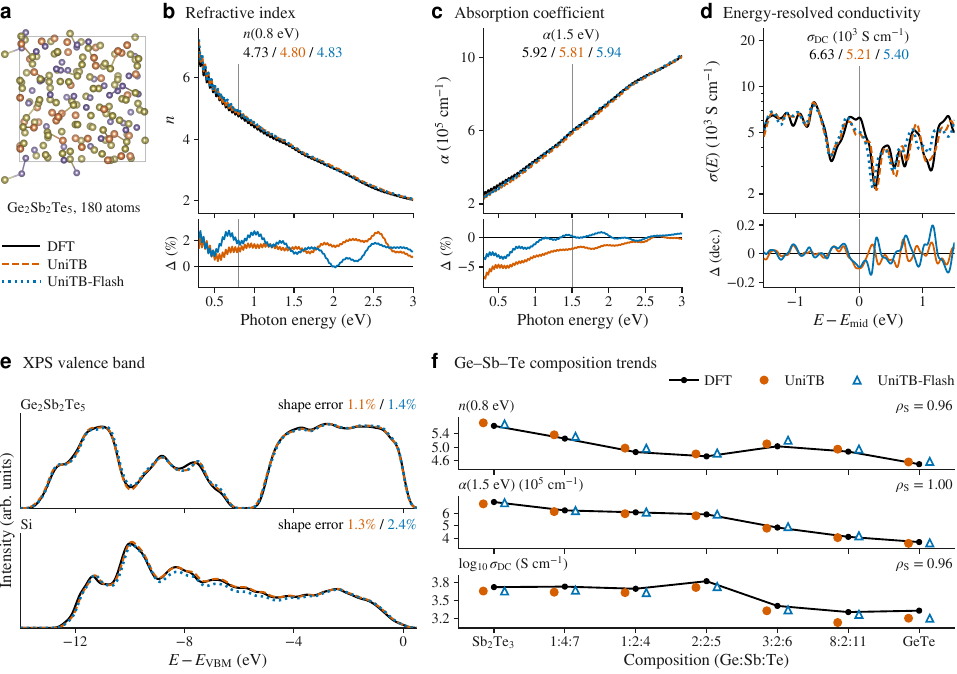}
\caption{\textbf{Properties of amorphous materials from models trained on crystals.} \textbf{a},~The amorphous Ge$_2$Sb$_2$Te$_5$ cell used in \textbf{b}--\textbf{e}, with Ge in purple, Sb in orange and Te in olive. \textbf{b}--\textbf{d},~Optical constants and conductivity of this cell, computed with the same Kubo--Greenwood protocol from the DFT, UniTB and UniTB-Flash Hamiltonians. The strips show model minus DFT, and energies in \textbf{d} are relative to the DFT midgap. \textbf{e},~Valence-band photoemission spectra computed for this cell and for a 216-atom amorphous Si network. Shape errors compare area-normalized spectra. \textbf{f},~Seven cells, one for each composition on the GeTe--Sb$_2$Te$_3$ line. $\rho_{\rm S}$ is the Spearman correlation with DFT across the compositions, equal for the two models. Protocols are in Methods.}
\label{fig:amorphous}
\end{figure}

\paragraph{GST optics and transport.}
GST alloys on the GeTe--Sb$_2$Te$_3$ line store information in the contrast between their amorphous and crystalline phases, which differ strongly in optical constants and electrical resistance. They are the active layers of rewritable optical discs and phase-change random-access memory\cite{wuttig2007pcm}, and they are used in non-volatile photonic devices and neuro-inspired computing\cite{wuttig2017photonic,zhang2019pcm}. The amorphous phase fixes one side of this contrast. Its refractive index near 0.8~eV, the photon energy of the 1,550~nm telecommunication band, is one end of the index contrast that photonic memories and switches use\cite{rios2015photonic}; its absorption coefficient sets how deeply light enters a film and how much of a switching pulse the film absorbs; and its DC conductivity sets the resistance of the amorphous state that a memory cell reads. Averaged over the twelve GST cells, both models agree with the DFT-Hamiltonian refractive index at 0.8 eV and absorption at 1.5 eV to within $2\%$ and with the DC conductivity to within $0.1$ decades; Fig.~\ref{fig:amorphous}b--d shows the 180-atom Ge$_2$Sb$_2$Te$_5$ cell, and Supplementary Table~\ref{S-tab:percell} lists every cell. Across seven compositions, the Spearman correlations of both models with DFT are $0.96$ for refractive index and conductivity and $1.00$ for absorption (Fig.~\ref{fig:amorphous}f), so both follow the composition dependence that alloy design uses\cite{lencer2008map,park2009gstoptical}.

\paragraph{Valence bands.}
Valence-band photoemission weights the occupied states by the photoionization cross-sections of their orbitals and is measured directly on amorphous films, so a computed spectrum tests the occupied states as a whole (Fig.~\ref{fig:amorphous}e). The shape error averages $1.00\%$ for UniTB and $1.15\%$ for UniTB-Flash over the twelve GST cells and $1.17\%$ and $2.77\%$ over the six cells of amorphous Si, which is used in thin-film transistors and solar cells; UniTB is closer to DFT in nine of the twelve GST cells and in all six Si cells (Supplementary Table~\ref{S-tab:percell}).

\subsection{Materials screening for industrial applications}
\label{sec:screening}
On the amorphous cells, UniTB-Flash comes close to UniTB in the optical and transport properties and orders the GST compositions in the same way, and it generates $H$ and $S$ of large cells in about a third of the time (Fig.~\ref{fig:prior}c). A materials screen must keep the best candidates of a large pool and needs accurate values only for those it keeps, which suggests two stages: UniTB-Flash ranks the pool at lower cost, and UniTB evaluates the retained candidates (Fig.~\ref{fig:screening}a). On nonmagnetic test crystals, we compare the two models in six application domains (Fig.~\ref{fig:screening}b; Methods). Each domain pairs a screening decision, such as the recall of the DFT top 20 averaged over shortlist sizes (maximum 0.893), with a property value; both are computed from the DFT Hamiltonian and from the Hamiltonian of each model. The screening scores of the two models differ by at most $0.051$, with no significant difference in any domain, whereas UniTB has the lower error for all six property values, by $42$--$59\%$ with paired $95\%$ intervals above zero (Supplementary Table~\ref{S-tab:domains}).

\begin{figure}[tbp]
\centering\includegraphics[width=\textwidth]{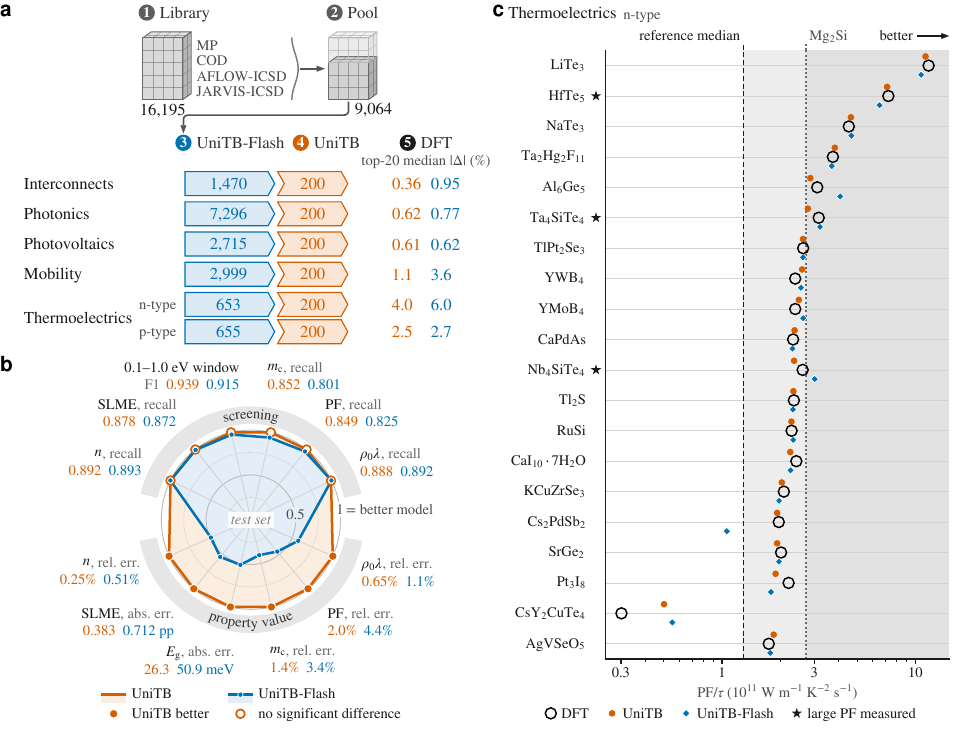}
\caption{\textbf{Model comparison and library screening.} \textbf{a},~Two-stage screen of a library of public crystal structures. The pools keep structures that match nonmagnetic Materials Project entries near the convex hull (rules in Methods), and translucent cubes are structures outside the pools. Blue cells count structures with valid UniTB-Flash values, and the pools overlap. Orange cells are the UniTB-Flash top 200, which UniTB evaluates. Median $|\Delta|$ is the median of $|$model $-$ DFT$|$/DFT over the UniTB top 20 with a DFT value. \textbf{b},~Comparison of the two models on nonmagnetic test crystals in six domains: photonics ($n$), photovoltaics (SLME), semiconductors ($E_{\rm g}$ and the window), mobility ($m_{\rm c}$), thermoelectrics (PF, n- and p-type combined) and interconnects ($\rho_0\lambda$). The radius is a model's score or inverse error relative to the better model. Screening is scored by the mean top-20 recall over $K=5$--60 (maximum 0.893), or by F1 for the band-gap window. Filled and open markers show whether the paired $95\%$ interval of the gain lies above zero or includes it. \textbf{c},~The final 20 $n$-type thermoelectric candidates in UniTB order, with values on the final $k$-point meshes. $\mathrm{PF}/\tau$ is the power factor per relaxation time at $10^{20}$~cm$^{-3}$ and 300~K. The reference median and Mg$_2$Si are the median and the strongest of the reference materials; grey marks the better side of these lines. Stars mark large power factors measured in pure or doped samples\cite{littleton1999pentatellurides,inohara2017ta4site4,okamoto2018nb4site4}.}
\label{fig:screening}
\end{figure}

We apply this two-stage screen to a library of 16,195 public crystal structures from the Materials Project (MP)\cite{jain2013materialsproject}, AFLOW-ICSD, COD and JARVIS-ICSD (Fig.~\ref{fig:screening}a), with two aims: to find promising materials and to test both models on crystals they have not seen. Each domain answers a common industrial need with one property: interconnect metals for chip wiring, ranked by the resistivity--mean-free-path product $\rho_0\lambda$\cite{gall2016mfp}, which sets the resistivity of narrow lines; high-index dielectrics for photonics, by the refractive index; thin-film photovoltaic absorbers, by the spectroscopic limited maximum efficiency (SLME)\cite{yu2012slme}; semiconductors for high-mobility channels, by the conduction-band mass; and thermoelectrics, by the power factor per relaxation time, $\mathrm{PF}/\tau$. After the pool rules (Methods), the five domain pools hold 9,064 distinct structures, 694 of them in the thermoelectric pool, where $n$- and $p$-type doping are ranked separately; the pools contain no training or test structures. In each pool, UniTB-Flash ranks every structure with a valid first-stage value, UniTB re-evaluates the 200 highest ranked, 3--14\% of these structures in the four larger pools, and DFT checks the 20 ranked highest by UniTB.

The thermoelectric screen finds strong $n$-type candidates (Fig.~\ref{fig:screening}c). On held-out test crystals, both models follow the DFT bands near the band edges and the DFT Seebeck coefficient from $10^{17}$ to $10^{21}$~cm$^{-3}$ at 300~K, to within about $4\%$ for three $n$-type crystals with high DFT power factors (Supplementary Fig.~\ref{S-sfig:te-n}; $p$-type crystals in Supplementary Fig.~\ref{S-sfig:te-p}). By DFT, 19 of the 20 exceed the median $\mathrm{PF}/\tau$ of classic reference materials computed with the same protocol (Supplementary Table~\ref{S-tab:hit-lines}), and six exceed Mg$_2$Si, the strongest $n$-type reference, led by LiTe$_3$ and HfTe$_5$ with 4.30 and 2.66 times its $\mathrm{PF}/\tau$. Three of the 20 have measured thermoelectric properties: HfTe$_5$ has been studied as a low-temperature thermoelectric\cite{littleton1999pentatellurides}, and whiskers of Ta$_4$SiTe$_4$ and of doped Nb$_4$SiTe$_4$ show large power factors\cite{inohara2017ta4site4,okamoto2018nb4site4}. LiTe$_3$ and NaTe$_3$, ranked first and third, have, to our knowledge, not been studied as thermoelectrics.

In the other four domains, UniTB's values differ from DFT by medians of 0.36--1.1\%, and UniTB-Flash's by 0.62--3.6\%; for $\mathrm{PF}/\tau$, the medians are 4.0\% and 6.0\% for $n$-type and 2.5\% and 2.7\% for $p$-type doping (Fig.~\ref{fig:screening}a). The second stage also corrects first-stage values: UniTB raises the conduction-band mass of Rb$_{14}$ZnGe$_{16}$ from 0.083 to 0.401~$m_{\rm e}$, which removes it from the final mobility list.

\skipsectionbarrier
\section{Discussion}
\label{sec:discussion}

Building on tight binding gives a construction for each requirement of a universal model. For transfer across chemistry, the invariants of the prior blocks, which describe each bond's surroundings, select the experts of each bond and help most in the onsite blocks, which contain the crystal field of the neighborhood; a charge-equilibration head of the SCC-DFTB form adds the site potentials of latent charges solved over the whole cell and lowers the matrix error more in larger cells. For the spectrum, the self-consistent correction is a change of local potential that shifts each band by a bounded amount, and the projection imposes a bound of the same kind on UniTB's bands relative to $H^{(0)}$, in the overlap metric that also scales the potentials of the charge head. For the cost, the finite range of atomic orbitals and projectors lets UniTB-Flash assemble its prior, the overlap and the spin-flip blocks from tabulated integrals at a cost linear in the number of atoms; the cheaper prior leaves a correction that is harder to learn, mostly in the hopping blocks (Fig.~\ref{fig:prior}d). The fused SO(2) kernels then shorten the training of the network, and other equivariant models built on SO(2) convolutions can use them unchanged.

Both models, trained only on crystals, also reproduce DFT-Hamiltonian properties of amorphous Ge--Sb--Te and Si (Fig.~\ref{fig:amorphous}). Which prior to use depends on the task. In the six test-set domains, the screening scores of UniTB-Flash stay close to those of UniTB despite its larger matrix errors, whereas UniTB gives the more accurate property values. A screen can therefore rank a large pool with the faster model and re-evaluate the shortlist with the more accurate one; in a screen of 16,195 public crystal structures for five application domains of industrial relevance, run this way, DFT confirms the values of the final candidates in every domain, and the $n$-type thermoelectric list holds candidates beyond the strongest reference material (Fig.~\ref{fig:screening}). These results are for Hamiltonians from nonmagnetic PBE calculations with spin--orbit coupling in one numerical atomic-orbital basis.

\section{Methods}
\label{sec:methods}

\subsection{Equivariant backbone}
The onsite and hopping models are separate networks of the same architecture, each trained on its own blocks. The backbone is equivariant to proper rotations, with reflection parity unconstrained. Scalar element or ordered-bond embeddings $z$ modulate features of degree $\ell$ as $x^{\ell}_{um}\leftarrow x^{\ell}_{um}(1+\sum_v a^{\ell}_{uv}z_v)$, with the same gain on every magnetic component $m$. At the output, a scalar-conditioned tensor product maps coupled features to orbital-pair components,
\begin{equation}
 \widehat\Delta^{\ell}_{wm}=\sum_uW^{\ell}_{0,wu}x^{\ell}_{um}+\sum_{uv}W^{\ell}_{wuv}z_vx^{\ell}_{um}.
 \label{eq:tp-readout}
\end{equation}
The model uses three interaction layers of bond-aligned SO(2) edge and node updates\cite{passaro2023escn,zhouyin2025slem}. Each layer first updates $\mathbf e_{ij}$ from the normalized $(\mathbf h_i,\mathbf e_{ij},\mathbf h_j)$ and then $\mathbf h_i$ from the normalized $\mathbf h_i$ and new $\mathbf e_{ij}$, with messages summed over the neighbors $j$ and scaled by the inverse square root of the average number of neighbors. Both updates are PDQ-MoE SO(2) maps (Eq.~\eqref{eq:pdq}) with the shared gates $g_{ij}$, and each is followed by a residual connection and the scalar modulation, with ordered-bond embeddings for edges and element embeddings for nodes; the three layers thus contain six routed SO(2) maps. Site-symmetry channel allocation sets the multiplicities at $\ell=0,\ldots,6$ to $(128,24,16,16,32,24,48)$, which preserve the hidden dimension and SO(2) weight budget of a uniform 32-channel allocation; training-set site-symmetry statistics guide the redistribution (Supplementary Table~\ref{S-tab:abl-backbone-allocation}). Prior perturbations add noise to the input prior during training only; for each element (onsite) or bond type (hopping) and each irreducible component, the noise scale is 0.5 (onsite) or 5 (hopping) times the batch RMS of the residual target of that type and component, with a random amplitude per structure. Residual targets and inference priors remain unperturbed.

\subsection{Prior descriptors and expert gates}
After a Clebsch--Gordan change of basis, scalar prior channels retain their signed values and higher-order channels contribute Gram entries $G^{\ell}_{uv}=\sum_mb^{\ell}_{um}b^{\ell}_{vm}$, formed within compatible representation blocks and passed through a signed logarithm. Concatenated with $e_{Z_iZ_j}$, these descriptors give cosine logits $\zeta_{ij,e}$ with scale 10, which determine the selected set and the gates,
\begin{equation}
 \mathcal S_{ij}=\TopK_e(\zeta_{ij,e}+\eta_e),\qquad
 g_{ij,e}=\frac{\exp \zeta_{ij,e}}{\sum_{a\in\mathcal S_{ij}}\exp \zeta_{ij,a}},\quad e\in\mathcal S_{ij}.
 \label{eq:route}
\end{equation}
Training adjusts the biases $\eta_e$ to balance selected-expert loads without an auxiliary loss\cite{wang2024lossfree}, and inference retains them. The final model selects two of four routed experts, has one always-active shared map and caps the shared-basis rank at 64. The cores $D_e$ are full matrices in the shared latent space, so the expert-dependent update in Eq.~\eqref{eq:pdq} has rank at most 64 and all experts act within one shared subspace. Selected experts are mixed before the activation, and each edge's gates are computed once and reused in all interaction layers.

\subsection{Charge-equilibration head}
Scalar features $h_i$ give $\chi_i=\chi_{Z_i}+f_\chi(h_i)$ and $J_i=J_{\min}+\softplus(j_{Z_i})+\softplus(f_J(h_i))$, and the QEq-inspired head\cite{rappe1991qeq,ko2021fourthgeneration} solves
\begin{equation}
 q=\argmin_{\one^{\mathrm T}q=0}\left[\chi^{\mathrm T}q+\tfrac12q^{\mathrm T}Aq\right],\qquad A=\diag(J)+\Gamma,
 \label{eq:qeq}
\end{equation}
with the Gaussian-charge kernel of a periodic cell of volume $\Omega$,
\begin{equation}
 \Gamma_{ij}=\frac{4\pi k_e}{\Omega}\sum_{0<|\mathbf G|\leq G_c}\frac{e^{-\sigma^2|\mathbf G|^2}}{|\mathbf G|^2}\cos[\mathbf G\cdot(\mathbf r_i-\mathbf r_j)].
 \label{eq:gamma}
\end{equation}
The sum includes both signs of each reciprocal vector, so $\Gamma$ is positive semidefinite, and positive hardnesses make $A$ positive definite. Stationarity gives $Aq+\chi+\lambda\one=0$; with $x=A^{-1}\chi$ and $y=A^{-1}\one$, the unique neutral solution is $q=-x+y\,(\one^{\mathrm T}x)/(\one^{\mathrm T}y)$, evaluated with differentiable double-precision Cholesky solves. The charges $q$ are latent variables learned through Hamiltonian supervision. The potential $\widetilde v_i=s[f_{\rm loc}(h_i)+\kappa(\Gamma q)_i]$ is centered as $v_i=\widetilde v_i-\sum_jd_j\widetilde v_j/\sum_jd_j$, with $d_i=\sum_{a\in\mathcal M_i}S_{ii,a}^2$ over the valid entries $\mathcal M_i$ of the onsite block of atom $i$, and enters Eq.~\eqref{eq:potential-assembly} as $v^{\rm on}_i$ or $v^{\rm hop}_i$. Here $\sigma=1.2$~\AA, $G_c=4$~\AA$^{-1}$, $J_{\min}=5$ in the kernel's energy convention and $s=0.1$. For inference, $\Gamma$ is evaluated by an Ewald decomposition of the same interaction and shared by the onsite and hopping models; its assembly scales as $N^2$ and the Cholesky solve as $N^3$, together taking 0.04~s for both models at 3,456 atoms.

\subsection{Spectral projection}
For a positive-definite overlap with $S=LL^{\dagger}$ and $X=L^{-\dagger}$, $X^{\dagger}SX=I$. For UniTB, the assembled residual $\widehat\Delta=\widehat H-H^{(0)}$ is diagonalized in the whitened basis, $X^{\dagger}\widehat\Delta X=U\diag(\mu)U^{\dagger}$. For a bound $b>0$, with $\mu'_a=\clip(\mu_a,-b,b)$ and $x_a=XU_{:a}$,
\begin{equation}
 \widehat\Delta_b=\widehat\Delta+\sum_{a:|\mu_a|>b}(\mu'_a-\mu_a)Sx_ax_a^{\dagger}S,\qquad
 \widehat H_b=H^{(0)}+\widehat\Delta_b .
 \label{eq:clip-update}
\end{equation}
The update preserves directions within the interval, moves outlying generalized eigenvalues to its boundary and retains the orbital dimension. Writing $f_b(A)=U\diag(\clip(\mu,-b,b))U^{\dagger}$ for $A=X^{\dagger}\widehat\Delta X$, the result is independent of the orthogonalizer and satisfies $-bI\preceq f_b(A)\preceq bI$, which gives the ordered-eigenvalue bound
\begin{equation}
 \varepsilon_a(H^{(0)},S)-b\leq\varepsilon_a(\widehat H_b,S)\leq\varepsilon_a(H^{(0)},S)+b
 \label{eq:weyl}
\end{equation}
on a fixed nonsingular overlap subspace (Supplementary Note~\ref{S-sn:spectral}). For a fixed basis and ionic nonlocal operator, the generalized Rayleigh quotient of a local effective-potential residual $\Delta V_{\rm Hxc}$ is the average of $\Delta V_{\rm Hxc}$ weighted by the density of the trial state and therefore lies in $[\inf\Delta V_{\rm Hxc},\sup\Delta V_{\rm Hxc}]$; this motivates a bound in the overlap metric. All reported projected spectra use $b=10$ eV. UniTB-Flash is not projected (Supplementary Table~\ref{S-tab:flash-bscan}); its only spectral post-processing is canonical orthogonalization, which removes overlap eigenvalues below $2\times10^{-3}$.

\subsection{Hamiltonian reconstruction}
Each onsite model is paired with the hopping model of the same architecture and prior. In the nonmagnetic calculations of the labels, the self-consistent change of the potential is spin independent, and spin--orbit coupling enters only through the fixed pseudopotential projectors. The predicted correction is therefore real and spin independent and is added to both diagonal spin blocks, as in NextHAM. Unless stated otherwise, spin-flip terms, imaginary components and blocks outside the learned graph come from the model's own prior, $H^{(0)}$ for UniTB and $H_{\rm GF}$ for UniTB-Flash; $S$ comes from the DFT output for UniTB and from the same two-center tables for UniTB-Flash. Directed network outputs $F$ are symmetrized as $H_{ij}=\tfrac12F_{i\to j}+\tfrac12F_{j\to i}^{\dagger}$.

\subsection{Prior construction}
$H^{(0)}$ (Eq.~\eqref{eq:hzero}) is reconstructed from two-center radial tables of the overlap, kinetic and projector integrals, together with the superposed density and local potentials on the cell's FFT grid (Fig.~\ref{fig:prior}). The tables are generated offline with ABACUS's numerical-orbital module on the 0.02 bohr radial grid with its projector sampling. For $H_{\rm GF}$, the neutral-atom potential of center $K$ is $V_{\mathrm{NA}}^K=V_{\mathrm{loc}}^K+V_{\mathrm H}[\rho_K^{\mathrm{val}}]$, and the separable pseudopotential contributes
\begin{equation}
 \langle\phi_a|V_{\mathrm{nl}}^K|\phi_b\rangle=\sum_{pq}\langle\phi_a|\beta_p^K\rangle D^K_{pq}\langle\beta_q^K|\phi_b\rangle ,
 \label{eq:kb}
\end{equation}
where $a,b$ combine atom and orbital indices; the factorization is exact for a separable pseudopotential\cite{kleinman1982bylander}. In both priors, the fully relativistic pseudopotential gives the spin blocks
\begin{equation}
 [V_{\rm nl}]^{\sigma\sigma'}_{ij}=\sum_K\langle\phi_i|\beta^K\rangle D^{K,\sigma\sigma'}\langle\beta^K|\phi_j\rangle,\qquad
 D^{K,\sigma\sigma'}=\bigoplus_{l,\pm}D^K_{l\pm}\otimes\bigl[P_{l\pm}\bigr]^{\sigma\sigma'},
 \label{eq:spinor-kb}
\end{equation}
where $D^K_{l\pm}$ couples the radial projectors of the channel with total angular momentum $j=l\pm\tfrac12$ and $P_{l\pm}=\bigl[(l+\tfrac12\pm\tfrac12)\mathds 1\pm\mathbf L\cdot\boldsymbol\sigma\bigr]/(2l+1)$ projects onto that channel\cite{dalcorso2005soc}. The base prior includes the neutral-atom crystal field in the onsite blocks and the potentials of the two bond ends in the hopping blocks,
\begin{align}
 (H^{\mathrm{base}}_{\mathrm{GF}})_{ii}&=T_{ii}+\sum_K(V_{\mathrm{NA}}^K)_{ii}+\sum_K(V_{\mathrm{nl}}^K)_{ii},\label{eq:nacf-onsite}\\
 (H^{\mathrm{base}}_{\mathrm{GF}})_{ij}&=T_{ij}+(V_{\mathrm{NA}}^i+V_{\mathrm{NA}}^j)_{ij}+\sum_K(V_{\mathrm{nl}}^K)_{ij},\label{eq:nacf-hopping}
\end{align}
where hopping includes distinct periodic centers and each sum includes the relevant periodic images; the crystal field uses a finite-rank factor approximation to the neutral-atom matrix elements. The prior of the trained UniTB-Flash models adds exchange--correlation and third-center local-potential terms,
\begin{align}
 (H_{\mathrm{GF}})_{ii}&=(H^{\mathrm{base}}_{\mathrm{GF}})_{ii}+X^{\mathrm{on}}_{ii}+cS_{ii},\label{eq:nacf-enhanced-on}\\
 (H_{\mathrm{GF}})_{ij}&=(H^{\mathrm{base}}_{\mathrm{GF}})_{ij}+V^{\mathrm{3c}}_{ij}+X^{\mathrm{pair}}_{ij}+\Delta X^{\mathrm{env}}_{ij}+cS_{ij}.\label{eq:nacf-enhanced-edge}
\end{align}
The onsite term integrates the local exchange--correlation potential of superposed atomic densities, including nonlinear core corrections\cite{louie1982nlcc}, by atom-centered quadrature. The hopping term combines a tabulated pair-density contribution with a McWEDA environmental correction\cite{horsfield1997multicenter,jelinek2005mcweda} built from unpolarized LDA exchange and PZ81 correlation\cite{perdew1981zunger}. The gauge $c$ is set by the cell volume and the valence-density second moments of the atoms and is defined only for fully periodic cells (Supplementary Note~\ref{S-sn:prior}). On 100 structures of 3--62 atoms, our assembly reproduces the ABACUS\cite{li2016abacus,zhou2025abacus} zeroth-step $H^{(0)}$, spin--orbit blocks included, and $S$ to within $0.08$~meV and $2.3\times10^{-9}$.

\subsection{Cost of the online assembly}
We count online arithmetic operations for $N$ atoms in a cell of volume $\Omega$, with fixed basis, projectors, tables, quadrature orders and grid spacing; offline tables are excluded. If all interatomic distances, including those to periodic images, are at least $d_{\min}$, a ball of radius $R$ holds at most $(1+2R/d_{\min})^3$ atoms, so the orbital-overlap partners of an atom, the atoms reached by one projector center and the neutral-atom centers that reach a bond form bounded lists, built using cell lists in $O(N)$ operations\cite{allen2017computer}. A two-center block costs one spline evaluation per Slater--Koster channel and one rotation. A separable contraction costs $O(m^2p+mp^2)$ per orbital pair and center, for $m$ orbitals per atom and $p$ projector functions per center (twice as many for each complex spin block of Eq.~\eqref{eq:spinor-kb}); a finite-rank term costs $O(rm^2)$ for rank $r$; and the onsite quadrature costs $O(q_{\rm on}(z_\rho+m^2))$ per atom for $q_{\rm on}$ points, where $z_\rho$ atoms contribute density at a point. Summing over the bounded lists gives $O(N)$ for $S$, the spin-flip blocks and $H_{\rm GF}$. For $H^{(0)}$, type-1 nonuniform FFTs give the structure factors in $O(N+G\log G)$ operations\cite{barnett2019finufft}, a fixed number of FFTs costs $O(G\log G)$\cite{cooley1965fft}, and the integration of each orbital pair over the $8^3$-point tiles of its common support costs $O(N)$. Supplementary Note~\ref{S-sn:cost} evaluates the counts at the parameters of our data.

\subsection{GPU assembly and band solves}
Both priors are assembled on the GPU in FP64 arithmetic with compact FP32 output, in five steps (Fig.~\ref{fig:prior}a): one pair search, two-center evaluation, one shared spinor-projector contraction, local-potential assembly and final block construction. The two priors differ in the two-center and local-potential steps. Spectral calculations batch wavevectors, test each residual's spectrum against $[-b,b]$ by Cholesky factorizations and diagonalize only the residuals that fail the test; final band solves use a batched cuSOLVER eigensolver with per-matrix status checks and an overlap fallback that retains overlap eigenvalues above $10^{-5}$ (timings in Supplementary Table~\ref{S-tab:spectral_path_timing}).

\subsection{SO2CUDA architecture and performance}
\label{sec:so2cuda}
Most of the network's operations are equivariant tensor products evaluated as SO(2) maps in the frame of each bond. In a bond-aligned frame, an SO(2) map separates into magnetic-order blocks; for $m>0$, $(y_m,y_{-m})=(A_mx_m-B_mx_{-m},\,B_mx_m+A_mx_{-m})$, with $m=0$ a real map. A plain implementation evaluates the resulting small matrix products in separate loops over orders and experts (Fig.~\ref{fig:so2cuda}a, lower band). SO2CUDA groups them in an $m$-major layout (Fig.~\ref{fig:so2cuda}a). Rotate-as-pack writes rotated features directly into the two selected-expert buffers; grouped products apply $W_s+PD_eQ^{\mathrm T}$, with the shared map folded in because the gates sum to one; rotate-as-gather applies gates and radial weights, rotates back and writes each output once without atomics. The backward pass reuses both kernels. The kernels implement the SO(2) linear layer for uniform and nonuniform channel allocations.

Operator benchmarks (Fig.~\ref{fig:so2cuda}b) time the forward-plus-backward pass of the SO(2) linear layer without radial weights on one H200 GPU in strict FP32, in 14 uniform-channel and 12 nonuniform configurations. Ratios divide the median time of each comparator (official EquiformerV3 with and without \texttt{torch.compile}, the fastest tested cuEquivariance configuration and our plain PyTorch implementation following upstream DeePTB\cite{gu2024deeptb}) by that of SO2CUDA for the same configuration; SO2CUDA has the lower median time in every configuration, with ratios from 1.29 to 10.70. Training steps (Fig.~\ref{fig:so2cuda}c) use real crystal batches of at most 32 crystals and on average 55,893 directed edges: a step of UniTB's onsite network takes 0.79~s with SO2CUDA, against 1.64~s with plain PyTorch and 1.66~s with cuEquivariance. In synthetic whole-model runs, outputs and parameter gradients agree with plain PyTorch to within $6.8\times10^{-6}$ and $1.52\times10^{-5}$ (Supplementary Note~\ref{S-sn:systems}).

\begin{figure}[tbp]
\centering
\includegraphics[width=\textwidth]{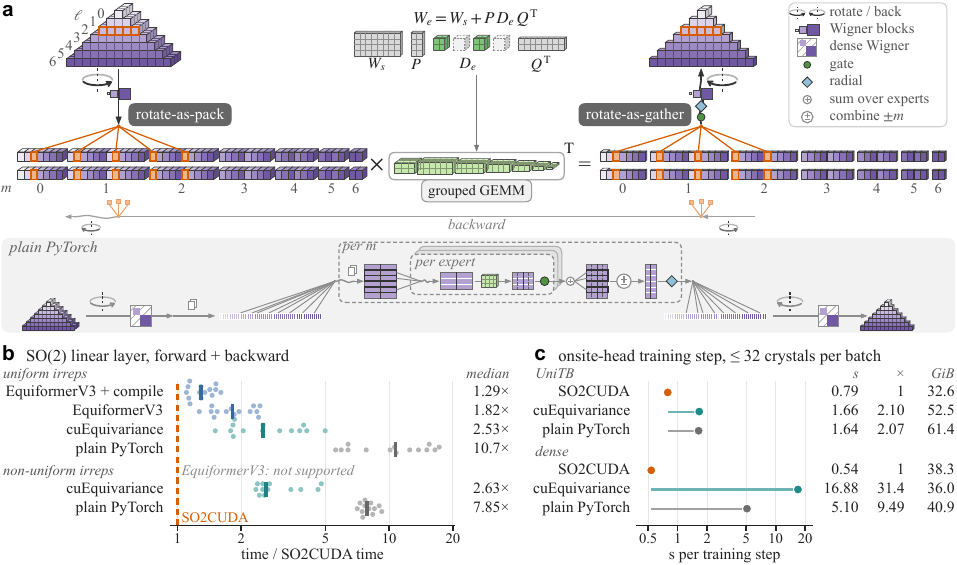}
\caption{\textbf{SO2CUDA architecture and performance.} \textbf{a},~Data flow of SO2CUDA for one edge of the UniTB edge update, with four experts and top-2 routing; depth shows channels. Rotate-as-pack rotates the features into the bond frame and writes expert-sorted copies. One grouped matrix multiplication per copy applies the expert maps, and rotate-as-gather rotates back and writes each output once. The backward pass reuses both kernels. The plain PyTorch band shows the same layer evaluated in separate loops over orders and experts. \textbf{b},~Forward-plus-backward time of the SO(2) linear layer relative to SO2CUDA. Points are 14 uniform and 12 nonuniform configurations, and bars are medians. \textbf{c},~Median time of one training step of the onsite network on real crystal batches, for UniTB and the dense model with three backends. Columns give the time, its ratio to SO2CUDA and the peak memory. All timings use one H200 GPU (Methods).}
\label{fig:so2cuda}
\end{figure}

\subsection{Datasets and labels}
MPHam-12K contains $11{,}953$ structures from the released NextHAM training list\cite{yin2025nextham}, after removal of 47 entries that duplicate another training structure or a test structure; the 3,000 released test structures are used unchanged (Supplementary Note~\ref{S-sn:data}). MPHam-29K adds $17{,}289$ structures selected from MP-20\cite{xie2022cdvae}, for $29{,}242$ structures; all their elements lie within the 68 elements covered by NextHAM, and none of these structures is in the NextHAM training or test sets. Labels were computed with ABACUS using the settings of NextHAM: the PBE functional\cite{perdew1996pbe}, PseudoDojo fully relativistic ONCV pseudopotentials, a 120 Ry wavefunction cutoff, a $\Gamma$-centered $6\times6\times6$ mesh and density convergence of $10^{-6}$. The learning target is $\operatorname{Re}H^{\uparrow\uparrow}$. Element-specific orbitals are embedded in a padded $4s2p2d1f$ set of 27 spatial orbitals, and the stored directed graph uses element-dependent cutoff radii (Supplementary Note~\ref{S-sn:protocol}).

\subsection{Training}
Models use HybridMuon, which combines Muon updates for selected matrix weights with AdamW updates for the remaining parameter groups. Training uses a learning rate of 0.01, weight decay of 0.01 and a gradient-norm cap of 0.3, with a warmup--stable--decay schedule and an absolute matrix loss without gauge shift. MPHam-12K training runs for 99.89 effective epochs (37,500 updates with dynamic batches capped at 32 structures, 31.84 structures on average); MPHam-29K training runs for about 300 effective epochs (91,800 updates with a 96-structure cap). UniTB-Flash changes the prior and residual target to $H_{\rm GF}$ and $H-H_{\rm GF}$ (Supplementary Tables~\ref{S-tab:config_s} and~\ref{S-tab:config_s_modules}). The final models are trained in strict FP32 with TF32 disabled, with the SO2CUDA kernels and data parallelism across the GPUs that train each network. The onsite and hopping models train concurrently; training takes about 14 h on one H200 GPU for MPHam-12K and about 37 h on two H200 GPUs for MPHam-29K (Supplementary Note~\ref{S-sn:training}).

\subsection{Matrix error}
For each structure $s$, onsite and hopping absolute-error sums are combined before division by the total number of valid matrix entries,
\begin{equation}
 e_s=\frac{L_{s,\mathrm{on}}+L_{s,\mathrm{hop}}}{N_{s,\mathrm{on}}+N_{s,\mathrm{hop}}},\qquad
 \mathrm{MAE}=\frac{1}{n}\sum_s e_s .
 \label{eq:matrix-metric}
\end{equation}
Onsite blocks use all valid spatial-orbital entries, hopping includes both directed orientations, and symmetry-enforced zero values inside valid blocks are included; the matrix results use the 3,000 released test structures and the real spin-up--spin-up block. Our entries use the final checkpoint after a fixed number of updates and the unaligned MAE; the NextHAM values use its selected checkpoint and published gauge convention (Supplementary Note~\ref{S-sn:protocol}).

\subsection{Band metrics}
Predicted and DFT spectra are aligned to their respective valence-band maxima. The $\mathrm{fw}_{10}$ metric averages absolute eigenvalue errors over reference levels in the DFT $[-10,+10]$ eV window, equally across bands and wavevectors and then equally across structures, with no additional fitted shift. A strict spurious-state detection requires an unmatched extra level at least 0.5 eV from every DFT level for at least 10\% of the evaluated wavevectors; the wide criterion omits the 0.5 eV separation. The band errors of Table~\ref{tab:benchmark} use the full band paths of 200 test structures drawn at random, and the projection-on/off comparison uses one double-precision solver on 15,459 sampled wavevectors (Supplementary Note~\ref{S-sn:bands}).

\subsection{Supercell properties}
For the optical and transport properties, the overlap, the spin--orbit components and the blocks outside the learned graph follow the reconstruction default above; DFT and model Hamiltonians share geometry and post-processing. At $\Gamma$, eigenpairs, projected for UniTB, are evaluated in a window of $\pm4$ eV about the DFT midgap. The velocity matrix is the $k$-derivative of the same Hamiltonian ($\widehat H_b$ for UniTB), $v_{nm}^{\alpha}=\hbar^{-1}c_n^\dagger(\partial_\alpha\widehat H-\bar\varepsilon_{nm}\partial_\alpha S)c_m$ with $\bar\varepsilon_{nm}=\tfrac12(\varepsilon_n+\varepsilon_m)$; $\partial_\alpha M=\mathrm i\,M\circ d^\alpha$ for unprojected matrices, with $d$ the orbital-center displacement including lattice translation, and Daleckii--Krein divided differences give the derivative of the clip (Supplementary Note~\ref{S-sn:amorphous}). Intra-atomic dipole terms are omitted for every Hamiltonian. Kubo--Greenwood response\cite{kubo1957response,greenwood1958conduction,calderin2017kubogreenwood} uses spinor states, including intraband and Kramers-pair contributions, at 300 K with a neutrality-determined chemical potential. The DC conductivity uses a Gaussian broadening of 50 meV and the optical response 100 meV; the refractive index is evaluated at 0.8 eV and the absorption at 1.5 eV after the Kramers--Kronig transform. Conductivity errors are absolute base-ten log ratios and optical errors absolute relative differences. Valence-band photoemission spectra weight each eigenstate with the Al K$\alpha$ photoionization cross-sections of its orbital components\cite{yeh1985subshell}, are broadened by Gaussians of 0.45 eV full width at half maximum and are aligned at each spectrum's valence-band maximum; the shape error is half the integrated absolute difference between the area-normalized model and DFT spectra.

\subsection{Application domains}
The six test-set domains use nonmagnetic crystals among the 3,000 test structures, identified by Materials Project labels\cite{jain2013materialsproject} (Supplementary Note~\ref{S-sn:screening}). Each property is computed with the same code from the DFT and model Hamiltonians. UniTB spectra are projected relative to $H^{(0)}$ with $b=10$ eV; UniTB-Flash spectra are not projected. The refractive index of a crystal is $n=(\operatorname{Tr}\varepsilon_\infty/3)^{1/2}$ from the electronic dielectric tensor. The SLME follows Yu and Zunger\cite{yu2012slme} for a 0.5~$\mu$m film under AM1.5G illumination at 300 K, without scissor correction. The conduction-band mass is the harmonic mean of the three principal masses from a central-difference Hessian at the band minimum. Power factors and Seebeck coefficients use a constant relaxation time at 300 K and a carrier density of $10^{20}$ cm$^{-3}$, or 25 densities from $10^{17}$ to $10^{21}$ cm$^{-3}$ for the Pisarenko curves; $\rho_0\lambda$ of metals is evaluated at 300 K, at the Fermi level of each Hamiltonian, with a constant mean free path\cite{gall2016mfp}. Screening decisions are scored by the recall of the DFT top 20 among the $K$ highest-ranked structures, averaged over $K=5$--60, and the band-gap window by the F1 of the 0.1--1.0 eV classification; property values are scored by median absolute errors or median absolute log ratios. Gains are $1-e_{\rm UniTB}/e_{\rm Flash}$ for values and score ratios minus one for decisions, with paired bootstrap intervals; a gain is called significant when its 95\% interval lies above zero.

\subsection{Library screen}
The library holds 16,195 crystal structures from MP, AFLOW-ICSD, COD and JARVIS-ICSD (Supplementary Note~\ref{S-sn:screening}). A structure enters a domain pool when its UniTB-Flash band gap lies in the domain window (at least 0.5 eV, at most 2.5 eV for photovoltaics and 3.0 eV for mobility; 0.1--1.0 eV for thermoelectrics; a metal for interconnects) and it has no unphysically short contacts; when it is, or matches, a Materials Project entry that is labeled nonmagnetic, lies within 0.1 eV per atom of the hull and contains no Tc, Pm, Po or heavier elements; and when neither it nor its entry matches a training or test structure, by Materials Project entry or by pymatgen StructureMatcher with default tolerances. Each entry keeps one structure. The five pools hold 9,064 distinct structures, 694 of them in the thermoelectric pool. The refractive index sums interband transitions over a $k$ mesh between states within 10 eV of the midgap of each Hamiltonian at $\Gamma$, without local-field effects. UniTB-Flash ranks every structure with a valid first-stage value, UniTB evaluates the 200 highest ranked, and DFT the 20 ranked highest by UniTB, using the labels' functional, basis and pseudopotentials; hit lines use the same codes, and the thermoelectric line is the median of the classic reference materials of each carrier type with defined values. Thermoelectric $\mathrm{PF}/\tau$ is ranked separately for $n$- and $p$-type doping on $k$ meshes chosen by cell size, and the final lists and reference materials are evaluated on dense meshes (Supplementary Note~\ref{S-sn:screening}).

\subsection{Inference timing}
Full inference runs from the atomic structure to the sparse $H$ and $S$ and includes the prior assembly and the forward passes of the onsite and hopping networks with charge equilibration. PbTe supercells of 16--3,456 atoms were timed on one H200 GPU; each point is the median of seven calls after three warm-up calls. The network runs in FP32 with TF32 disabled and the charge equilibration in FP64; model loading, compilation, file output and spectral post-processing are excluded (Supplementary Note~\ref{S-sn:screening}).

\subsection{Statistics}
For candidate and baseline errors $e^{(1)}_s$ and $e^{(0)}_s$, the relative change is $\overline e^{(1)}/\overline e^{(0)}-1$. Paired percentile intervals use 10,000 joint resamples of structure indices at the 95\% level and describe variation across test structures.

\section*{Data availability}
The public datasets used in this study are available from their
original sources\cite{yin2025nextham,xie2022cdvae,rosset2025asi24,
zhou2025gstace24,jain2013materialsproject}, as detailed in Methods
and the Supplementary Information.
The DFT Hamiltonians computed in this study, including those for MPHam-12K and MPHam-29K, together with the evaluation and screening records and the source data supporting the main and supplementary results, will be made publicly available upon publication.

\section*{Code availability}

The UniTB training and inference code, the GPU prior-assembly code and the SO2CUDA kernels will be made publicly available upon publication.

\section*{Acknowledgements}
We gratefully acknowledge fruitful discussions with Feitong Song, Kai Qi and Jijie Zou.
Computations were performed at the Supercomputing Center of the University of Science and Technology of China and on the High Performance Computing facilities at the National University of Singapore (NUS HPC) under Project CFP04-CF-019.

\section*{Funding}

This work was supported by the Advanced Materials-National Science and Technology Major Project (2025ZD0618401), the National Natural Science Foundation of China (12504285), and the Natural Science Foundation of Jiangsu Province (BK20250472). L.S. is supported by the Singapore Ministry of Education Tier 2 grant (No. MOE-T2EP50123-0012).

\section*{Competing interests}

The authors declare no competing interests.
\endgroup
\clearpage
\newgeometry{left=2cm,right=2cm,top=2cm,bottom=2cm}
\setstretch{1}
\fontsize{10}{12}\selectfont
\let\addcontentsline\savedaddcontentsline
\setcounter{figure}{0}
\setcounter{table}{0}
\setcounter{equation}{0}
\setcounter{footnote}{0}
\setcounter{page}{1}
\renewcommand{\thepage}{S\arabic{page}}
\renewcommand{\theequation}{S\arabic{equation}}
\renewcommand{\theHfigure}{si.\arabic{figure}}
\renewcommand{\theHtable}{si.\arabic{table}}
\renewcommand{\theHequation}{si.\arabic{equation}}
\renewcommand{\theHfootnote}{si.\arabic{footnote}}
\captionsetup{hypcap=false}
\captionsetup[table]{name=Supplementary Table}
\captionsetup[figure]{name=Supplementary Figure}
\renewcommand{\refname}{Supplementary References}
\pdfbookmark[0]{Supplementary Information}{si-part}
\begin{center}
{\Large\bfseries Supplementary Information}\\[0.8em]
{\large\bfseries Physical structure as an organizing principle for universal Hamiltonian learning}\\[0.8em]
\authorblock\\[1.2em]
{\small Supplementary Notes 1--14 $\cdot$ Supplementary Figures 1--\ref{S-sfig:te-p} $\cdot$ Supplementary Tables 1--\ref{S-tab:final-te} $\cdot$ Supplementary References}
\end{center}
\vspace{0.5em}
{\small\tableofcontents}

\snote{Extended related work}{sn:related}
\paragraph{Universal Hamiltonians and physical descriptors.}
DeepH and DeepH-E3 learn local electronic Hamiltonian blocks\cite{SI-li2022deeph,SI-gong2023deephe3}. Universal DeepH extends chemical coverage with a unified model\cite{SI-wang2024universaldeeph}. Uni-HamGNN models spin--orbit-coupled materials and demonstrates crystalline screening\cite{SI-zhong2026unihamgnn}. NextHAM uses zeroth-step blocks as features and residual baselines, assigns pairs to distance-based submodels, and applies reciprocal-space objectives to the spectrum\cite{SI-yin2025nextham}. LEM/SLEM supplies local equivariant operator updates\cite{SI-zhouyin2025slem}; UniTB builds its dense backbone on this representation, with separate onsite and hopping models and element-level modulation (Supplementary Note~\ref{S-sn:arch}). OrbNet uses symmetry-adapted atomic-orbital features\cite{SI-qiao2020orbnet}. DeePKS uses local operator descriptors in a learned energy functional\cite{SI-chen2020deepks}. UniTB uses prior-operator invariants to compute edge-wise expert coefficients.

\paragraph{Conditional computation.}
CondConv and DynamicConv condition convolutional kernels on inputs\cite{SI-yang2019condconv,SI-chen2020dynamicconv}. DCD expresses dynamic kernels through shared bases and an input-dependent core\cite{SI-li2021dcd}. Sparse MoE and Switch-style routing select subsets of experts\cite{SI-shazeer2017moe,SI-fedus2022switch}, while DeepSeekMoE combines specialized and shared experts\cite{SI-dai2024deepseekmoe}. Auxiliary-loss-free balancing controls expert selection through adaptive biases\cite{SI-wang2024lossfree}. Hash Layers use fixed side-information assignments\cite{SI-roller2021hashlayers}; Mobile V-MoE uses superclass information\cite{SI-daxberger2023mobilevmoe}. UMA uses structure-level expert mixtures\cite{SI-wood2025uma}; DPA3-MoE combines element-level routing with nonlinear experts on an invariant-feature backbone\cite{SI-liu2026moemlip}; MatRIS-MoE uses element-dependent expert selection\cite{SI-zhou2026matrismoe}. MoLoRA and LoRAMoE combine expert-specific low-rank adapters\cite{SI-zadouri2023molora,SI-dou2023loramoe}. MoBE studies shared bases for compressing language-model experts\cite{SI-chen2025mobe}. In UniTB, invariant prior blocks determine edge-wise gates over cores that share both bases.

\paragraph{Physical priors and charge response.}
Neutral-atom potentials, Slater--Koster tables and separable projectors underlie the grid-free prior\cite{SI-slater1954koster,SI-sankey1989neutralatom,SI-kleinman1982bylander,SI-soler2002siesta}, with multicenter exchange--correlation approximations from tight-binding methods\cite{SI-horsfield1997multicenter,SI-jelinek2005mcweda}. Harris and Foulkes--Haydock provide non-self-consistent prior constructions\cite{SI-harris1985functional,SI-foulkes1989tightbinding}. QEq and learned charge-equilibration models introduce constrained nonlocal responses\cite{SI-rappe1991qeq,SI-ghasemi2015cent,SI-ko2021fourthgeneration}. SCC-DFTB supplies the overlap-weighted potential form\cite{SI-elstner1998sccdftb}. HamGNN-LR uses effective-charge-based long-range Hamiltonian corrections\cite{SI-zhong2026hamgnnlr}. UniTB infers latent charges with zero total charge from equivariant features and adds their site potentials to the Hamiltonian in the overlap-weighted form of SCC-DFTB.
\snote{Architecture details}{sn:arch}
\begin{center}\begin{minipage}{\linewidth}
\centering\includegraphics[width=\textwidth]{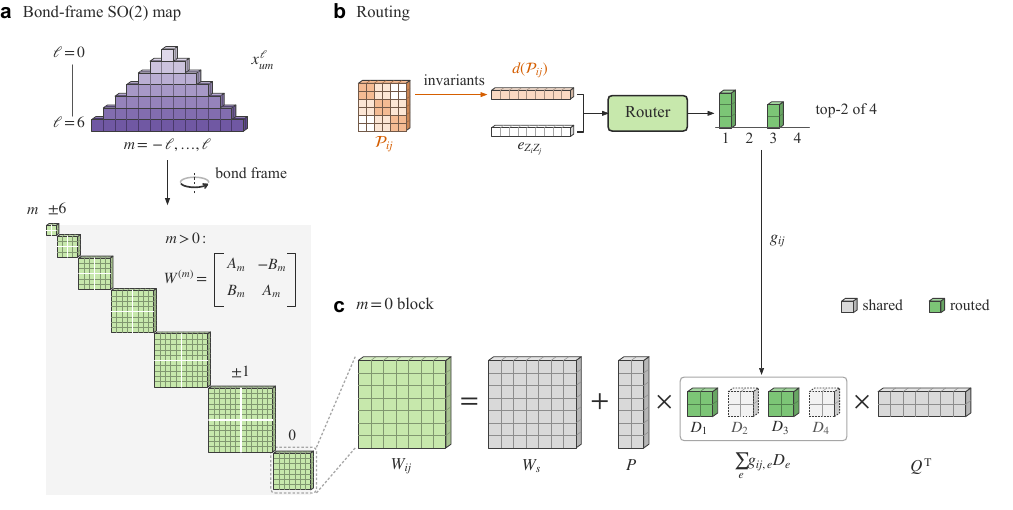}
\captionof{figure}{\textbf{PDQ-MoE maps on the bond-frame SO(2) blocks.} Prior-routed PDQ-MoE maps act on the bond-frame SO(2) blocks of the equivariant features. \textbf{a},~Depth is the channel index $u$, with multiplicities $128,24,16,16,32,24,48$ drawn schematically. In the bond frame the map is block-diagonal in $|m|$ (gray, zero). The $m=0$ block couples all degrees, and each $m>0$ block couples $x_m$ and $x_{-m}$ on degrees $\ell\ge|m|$. Block sides are drawn in degrees.
\textbf{b},~The invariants $d(\Pp_{ij})$ are signed scalar channels and Gram entries after a signed logarithm; bar heights and the selection are illustrative. Expert outputs mix before activation, and routing is reused across the three interaction layers.
\textbf{c},~The always-active $W_s$ and the bases $P$ and $Q$ are shared by all experts of a map, and the experts differ in their cores $D_e$ (Eq.~(\ref{eq:pdq})). The expert-dependent update has rank at most 64. Dashed cores are not selected.}
\label{S-sfig:irreps}
\end{minipage}\end{center}

\subsection*{Two features of the dense backbone}
The dense backbone and LEM\cite{SI-zhouyin2025slem} share the data, the $H^{(0)}$ prior and its feature initialization, the cutoffs, the basis, the optimizer and the training budget. They differ in two features. The onsite and hopping blocks are predicted by two separately trained models, each with 5,786,384 parameters, whereas LEM predicts both with one model of 3,643,141 parameters. After every edge and node update, and in the output layer, element-level tensor products modulate the features with the learned ordered-bond embedding $e_{Z_iZ_j}$ (edges) or the element one-hot encoding (nodes), 818,410 parameters per model; LEM uses linear readouts. Supplementary Note~\ref{S-sn:ablations} removes each feature from the dense backbone and retrains at the MPHam-12K budget, and describes the site-symmetry channel allocation and the prior perturbations of the training recipe.
\snote{Module ablations}{sn:ablations}
\label{S-sn:routing}%
Comparisons are grouped by design question. Unless stated otherwise, retrained comparisons use the fixed MPHam-12K budget (37,500 updates with dynamic batches of at most 32 structures, about 100 epochs) and are evaluated on the 3,000 test structures; a dagger marks evaluations on the 2,996-structure subset whose stored graphs close under the tested space-group operations. Within each group, the italic row is the baseline, and each row gives the onsite, hopping and combined MAE (meV, combined as in Methods) with their relative changes and the change of the mean $\mathrm{fw}_{10}$ on the 200 band structures after projection at $b=10$ eV, with the baseline mean in parentheses. Asterisks mark comparisons between configurations that differ in more than one setting. Unless stated otherwise, paired 95\% intervals use 10,000 structure resamples of the evaluated set of each comparison and describe variation across those structures; dashes and blank cells indicate that no value is reported.

\subsection*{Backbone}
Removing both the separate onsite and hopping networks and the element-level tensor products raises the dense combined MAE by 151.9\%; the jointly ablated model, at 2.049 meV, lies 1.236 meV above the dense backbone and 0.111 meV below LEM. A jointly trained network raises the combined MAE by 61.8\%, with half the parameters and training compute of the two separate networks, and removing the modulation raises it by 36.0\%; removing both raises the mean band error more than removing either one (Supplementary Table~\ref{S-tab:abl-backbone}). Channel allocations at equal feature width and site channels on PDQ-MoE are compared in Supplementary Table~\ref{S-tab:abl-backbone-allocation}; the allocation comparison refers to the dense backbone.

\providecommand{\tminus}{\textminus}
\begin{center}\begin{minipage}{\linewidth}
\singlespacing
\captionof{table}{\textbf{Ablations of the dense backbone.} The joint network has half the parameters and training compute of the two separate networks. LEM and SLEM are the upstream architectures with the same prior and data exposure.}
\label{S-tab:abl-backbone}\label{S-tab:dense-ablation}
\centering\footnotesize
\renewcommand{\arraystretch}{1.08}
\setlength{\tabcolsep}{2pt}
\begin{tabular}{@{}>{\raggedright\arraybackslash}p{4.0cm}rrr@{\hspace{3pt}}r@{\hspace{2pt}}l@{\hspace{3pt}}rr@{\hspace{3pt}}r@{\hspace{2pt}}l@{}}
\toprule
 & \multicolumn{3}{c}{MAE (meV)} & \multicolumn{2}{c}{Combined change (\%)} & \multicolumn{2}{c}{Change (\%)} & \multicolumn{2}{c}{$\Delta$ mean $\mathrm{fw}_{10}$ (meV)}\\
\cmidrule(lr){2-4}\cmidrule(lr){5-6}\cmidrule(lr){7-8}\cmidrule(lr){9-10}
Variant & Comb. & Onsite & Hopping & Value & 95\% CI & Onsite & Hopping & Value & 95\% CI\\
\midrule
\multicolumn{10}{@{}l@{}}{\textbf{a\enspace Separate onsite and hopping networks, and element-level tensor products (TP)}}\\
\textit{Dense backbone (separate networks, element-level TP)} & 0.8133 & 8.488 & 0.7261 &  &  &  &  & (67.4) & \\ %
One joint network, TP kept\textsuperscript{$\ast$} & 1.316 & 8.007 & 1.241 & +61.8 & {\scriptsize [+60.5, +63.1]} & \tminus5.7 & +70.9 & +1.3 & {\scriptsize [\tminus3.3, +6.0]}\\ %
No element-level TP (separate networks) & 1.106 & 8.744 & 1.020 & +36.0 & {\scriptsize [+35.0, +37.0]} & +3.0 & +40.4 & +4.2 & {\scriptsize [\tminus1.3, +11.2]}\\ %
Joint network, no element-level TP\textsuperscript{$\ast$} & 2.049 & 8.689 & 1.975 & +151.9 & {\scriptsize [+149.0, +154.9]} & +2.4 & +171.9 & +24.5 & {\scriptsize [+18.1, +31.4]}\\ %
\addlinespace[4pt]
\multicolumn{10}{@{}l@{}}{\textbf{b\enspace External baselines at the same prior and data exposure}}\\
\textit{Dense backbone} & 0.8133 & 8.488 & 0.7261 &  &  &  &  & (67.4) & \\ %
LEM\textsuperscript{$\ast$} & 2.1598 & 7.925 & 2.0983 & +165.5 & {\scriptsize [+162.3, +168.7]} & \tminus6.63 & +189.0 & +20.7 & {\scriptsize [+13.2, +28.7]}\\ %
SLEM\textsuperscript{$\ast$} & 1.3856 & 11.319 & 1.2782 & +70.37 & {\scriptsize [+68.72, +72.00]} & +33.36 & +76.02 & +13.2 & {\scriptsize [+3.8, +23.1]}\\ %
\textit{LEM} & 2.160 & 7.925 & 2.098 &  &  &  &  & (88.1) & \\ %
Dense, both features removed\textsuperscript{$\ast$} & 2.049 & 8.689 & 1.975 & \tminus5.1 & {\scriptsize [\tminus5.5, \tminus4.8]} & +9.6 & \tminus5.9 & +3.8 & {\scriptsize [\tminus3.4, +11.3]}\\ %
\bottomrule
\end{tabular}
\end{minipage}\end{center}

{\singlespacing\footnotesize\noindent $^\ast$ More than one setting differs from the baseline; LEM and SLEM keep their upstream architectures.\par}

\providecommand{\tminus}{\textminus}
\begin{center}\begin{minipage}{\linewidth}
\singlespacing
\captionof{table}{\textbf{Channel allocations at equal feature width.} Continuation of Supplementary Table~\ref{S-tab:abl-backbone}.}
\label{S-tab:abl-backbone-allocation}\label{S-tab:channel-allocation-controls}
\centering\footnotesize
\renewcommand{\arraystretch}{1.08}
\setlength{\tabcolsep}{2pt}
\begin{tabular}{@{}>{\raggedright\arraybackslash}p{4.0cm}rrr@{\hspace{3pt}}r@{\hspace{2pt}}l@{\hspace{3pt}}rr@{\hspace{3pt}}r@{\hspace{2pt}}l@{}}
\toprule
 & \multicolumn{3}{c}{MAE (meV)} & \multicolumn{2}{c}{Combined change (\%)} & \multicolumn{2}{c}{Change (\%)} & \multicolumn{2}{c}{$\Delta$ mean $\mathrm{fw}_{10}$ (meV)}\\
\cmidrule(lr){2-4}\cmidrule(lr){5-6}\cmidrule(lr){7-8}\cmidrule(lr){9-10}
Variant & Comb. & Onsite & Hopping & Value & 95\% CI & Onsite & Hopping & Value & 95\% CI\\
\midrule
\multicolumn{10}{@{}l@{}}{\textbf{c\enspace Channel allocation of the dense backbone (multiplicities at $\ell=0,\ldots,6$; every allocation has feature width 1,568)}}\\
\textit{Uniform (32 per degree), 5.79M} & 0.8133 & 8.488 & 0.7261 &  &  &  &  & (67.4) & \\ %
V-shaped (128,24,16,16,32,24,48), 6.05M & 0.7995 & 8.155 & 0.7155 & \tminus1.71 & {\scriptsize [\tminus2.21, \tminus1.18]} & \tminus3.92 & \tminus1.47 & \tminus4.4 & {\scriptsize [\tminus8.5, \tminus0.7]}\\ %
\addlinespace[4pt]
\multicolumn{10}{@{}l@{}}{\textbf{d\enspace Site-symmetry channels on PDQ-MoE (V-shaped allocation)}}\\
\textit{PDQ-MoE (4+1 experts, top 2), 7.73M}\textsuperscript{\dag} & 0.7851 & 8.111 & 0.7005 &  &  &  &  & (66.8) & \\ %
+ V-shaped channels (no charge head), 8.01M\textsuperscript{\dag}\textsuperscript{$\ast$} & 0.7684 & 8.230 & 0.6816 & \tminus2.12 & {\scriptsize [\tminus2.76, \tminus1.46]} & +1.46 & \tminus2.70 & \tminus0.7 & {\scriptsize [\tminus4.5, +3.1]}\\ %
\textit{PDQ-MoE + charge head, 7.74M} & 0.7531 & 7.977 & 0.6698 &  &  &  &  & (64.3) & \\ %
+ V-shaped channels, 8.03M & 0.7357 & 8.209 & 0.6491 & \tminus2.31 & {\scriptsize [\tminus2.86, \tminus1.73]} & +2.91 & \tminus3.09 & \tminus0.9 & {\scriptsize [\tminus5.0, +2.9]}\\ %
\bottomrule
\end{tabular}
\end{minipage}\end{center}

{\singlespacing\footnotesize\noindent $^\dagger$ 2,996-structure subset. $^\ast$ More than one setting differs from the baseline.\par}

\subsection*{Prior usage}
Learning the full Hamiltonian instead of the residual raises the dense combined MAE from 0.813 to 5.692 meV. Removing the input initialization raises the combined MAE by 1.54\% on PDQ-MoE and 0.77\% on UniTB, and passing the prior to the edge features only, instead of to nodes and edges, raises it on all three architectures (Supplementary Tables~\ref{S-tab:abl-reference} and~\ref{S-tab:abl-reference-edge}).

\providecommand{\tminus}{\textminus}
\begin{center}\begin{minipage}{\linewidth}
\singlespacing
\captionof{table}{\textbf{Training targets and prior inputs.} The input prior initializes node and edge features. The target is either the residual $H-H^{(0)}$, with $H^{(0)}$ added back to the output, or the full Hamiltonian $H$ with element- and bond-type shifts and scales estimated from the training set. In the $H^{(0)}$-alone rows, parentheses give the mean $\mathrm{fw}_{10}$ of $H^{(0)}$. $\ddagger$ marks unprojected band values (the two full-Hamiltonian-target comparisons and the baseline of the second); all other band values use projection at $b=10$ eV.}
\label{S-tab:abl-reference}\label{S-tab:reference-input-target-controls}\label{S-tab:reference-controls}
\centering\footnotesize
\renewcommand{\arraystretch}{1.08}
\setlength{\tabcolsep}{2pt}
\begin{tabular}{@{}>{\raggedright\arraybackslash}p{4.0cm}rrr@{\hspace{3pt}}r@{\hspace{2pt}}l@{\hspace{3pt}}rr@{\hspace{3pt}}r@{\hspace{2pt}}l@{}}
\toprule
 & \multicolumn{3}{c}{MAE (meV)} & \multicolumn{2}{c}{Combined change (\%)} & \multicolumn{2}{c}{Change (\%)} & \multicolumn{2}{c}{$\Delta$ mean $\mathrm{fw}_{10}$ (meV)}\\
\cmidrule(lr){2-4}\cmidrule(lr){5-6}\cmidrule(lr){7-8}\cmidrule(lr){9-10}
Variant & Comb. & Onsite & Hopping & Value & 95\% CI & Onsite & Hopping & Value & 95\% CI\\
\midrule
\multicolumn{10}{@{}l@{}}{\textbf{a\enspace Input initialization and training target (dense backbone)}}\\
$H^{(0)}$ alone, our labels & 4.717 &  &  &  &  &  &  & (745.6) & \\ %
$H^{(0)}$ alone, published labels\cite{SI-yin2025nextham} & 5.213 &  &  &  &  &  &  & -- & \\ %
\textit{Dense: $H^{(0)}$ input, residual target} & 0.8133 & 8.488 & 0.7261 &  &  &  &  & (67.4) & \\ %
No input prior & 0.7726 & 8.605 & 0.6834 & \tminus5.01 & {\scriptsize [\tminus5.89, \tminus4.15]} & +1.39 & \tminus5.88 & \tminus2.4 & {\scriptsize [\tminus7.4, +2.6]}\\ %
Full-Hamiltonian target $H$\textsuperscript{$\ast$} & 5.6921 & 13.186 & 5.5985 & +599.9 & {\scriptsize [+577.2, +623.5]} & +55.36 & +671.0 & +782.1\textsuperscript{\ddag} & {\scriptsize [+617.2, +971.8]}\\ %
\textit{No input prior, residual target} & 0.7726 & 8.605 & 0.6834 &  &  &  &  & (89.3)\textsuperscript{\ddag} & \\ %
Full-Hamiltonian target $H$\textsuperscript{$\ast$} & 11.6595 & 11.574 & 11.7046 & +1409 & {\scriptsize [+1372, +1448]} & +34.50 & +1613 & +1344\textsuperscript{\ddag} & {\scriptsize [+1145, +1571]}\\ %
\addlinespace[4pt]
\multicolumn{10}{@{}l@{}}{\textbf{b\enspace Input initialization on the expert and final models}}\\
\textit{PDQ-MoE with input initialization} & 0.7849 & 8.112 & 0.7003 &  &  &  &  & (66.8) & \\ %
Input initialization off (router still reads the prior)\textsuperscript{$\ast$} & 0.7970 & 8.128 & 0.7143 & +1.54 & {\scriptsize [+0.69, +2.44]} & +0.19 & +1.99 & \tminus7.3 & {\scriptsize [\tminus12.1, \tminus2.5]}\\ %
\textit{UniTB with input initialization} & 0.7215 & 8.098 & 0.6379 &  &  &  &  & (63.1) & \\ %
Input initialization off (prior perturbations kept)\textsuperscript{$\ast$} & 0.7270 & 7.881 & 0.6449 & +0.77 & {\scriptsize [\tminus0.05, +1.46]} & \tminus2.68 & +1.10 & \tminus1.5 & {\scriptsize [\tminus5.9, +2.8]}\\ %
\addlinespace[4pt]
\multicolumn{10}{@{}l@{}}{\textbf{c\enspace Layer-wise prior memory and prior as propagation operator}}\\
\textit{Dense + V-shaped channels} & 0.7995 & 8.155 & 0.7155 &  &  &  &  & (63.0) & \\ %
Layer-wise prior memory & 0.7951 & 8.077 & 0.7118 & \tminus0.54 & {\scriptsize [\tminus1.15, +0.03]} & \tminus0.96 & \tminus0.52 & +0.7 & {\scriptsize [\tminus2.3, +3.7]}\\ %
\textit{Dense backbone}\textsuperscript{\dag} & 0.8134 & 8.486 & 0.7262 &  &  &  &  & (67.4) & \\ %
Prior as propagation operator (one $H^{(0)}$ hop per layer)\textsuperscript{\dag} & 0.8179 & 8.398 & 0.7310 & +0.56 & {\scriptsize [+0.04, +1.07]} & \tminus1.04 & +0.67 & \tminus0.7 & {\scriptsize [\tminus4.2, +2.8]}\\ %
\addlinespace[4pt]
\multicolumn{10}{@{}l@{}}{\textbf{d\enspace Inference-time diagnostics on trained models (no retraining)}}\\
\textit{Propagation-operator model, unperturbed}\textsuperscript{\dag} & 0.8179 & 8.398 & 0.7310 &  &  &  &  &  & \\ %
Readings zeroed\textsuperscript{\dag} & 0.9060 & 8.822 & 0.8153 & +10.77 & {\scriptsize [+10.14, +11.43]} & +5.04 & +11.53 & -- & \\ %
Prior shuffled across edges\textsuperscript{\dag} & 0.8687 & 8.546 & 0.7808 & +6.21 & {\scriptsize [+5.90, +6.52]} & +1.75 & +6.82 & -- & \\ %
Layer-memory model, readings zeroed (against its unperturbed model) & -- & -- & -- & +114.5 & {\scriptsize [+110.5, +118.6]} & -- & -- & -- & \\ %
\bottomrule
\end{tabular}
\end{minipage}\end{center}

{\singlespacing\footnotesize\noindent $^\dagger$ 2,996-structure subset. $^\ast$ More than one setting differs from the baseline; the full-Hamiltonian target comes with its shift and scale readout. The unperturbed layer-memory model of the last row is the one trained in panel c. Dense models have 5,786,384 parameters per network, or 5,784,044 without input initialization. With the input prior held fixed, the full-Hamiltonian target has a higher combined MAE than the residual target by 4.879 meV (paired 95\% interval [4.695, 5.071]); without the input prior, the difference is 10.887 meV [10.597, 11.187].\par}

\providecommand{\tminus}{\textminus}
\begin{center}\begin{minipage}{\linewidth}
\singlespacing
\captionof{table}{\textbf{Prior in edge features only.} Continuation of Supplementary Table~\ref{S-tab:abl-reference}.}
\label{S-tab:abl-reference-edge}
\centering\footnotesize
\renewcommand{\arraystretch}{1.08}
\setlength{\tabcolsep}{2pt}
\begin{tabular}{@{}>{\raggedright\arraybackslash}p{4.0cm}rrr@{\hspace{3pt}}r@{\hspace{2pt}}l@{\hspace{3pt}}rr@{\hspace{3pt}}r@{\hspace{2pt}}l@{}}
\toprule
 & \multicolumn{3}{c}{MAE (meV)} & \multicolumn{2}{c}{Combined change (\%)} & \multicolumn{2}{c}{Change (\%)} & \multicolumn{2}{c}{$\Delta$ mean $\mathrm{fw}_{10}$ (meV)}\\
\cmidrule(lr){2-4}\cmidrule(lr){5-6}\cmidrule(lr){7-8}\cmidrule(lr){9-10}
Variant & Comb. & Onsite & Hopping & Value & 95\% CI & Onsite & Hopping & Value & 95\% CI\\
\midrule
\multicolumn{10}{@{}l@{}}{\textbf{a\enspace Prior in edge features only, against the node-and-edge baseline and against no input prior}}\\
\textit{Dense: prior in nodes and edges} & 0.8133 & 8.488 & 0.7261 &  &  &  &  & (67.4) & \\ %
Edge features only & 0.830 & 8.915 & 0.738 & +2.10 & {\scriptsize [+1.14, +3.07]} & +5.0 & +1.64 & -- & \\ %
\textit{PDQ-MoE: prior in nodes and edges} & 0.7849 & 8.112 & 0.7003 &  &  &  &  & (66.8) & \\ %
Edge features only & 0.796 & 8.741 & 0.706 & +1.46 & {\scriptsize [+0.66, +2.26]} & +7.7 & +0.81 & -- & \\ %
\textit{UniTB: prior in nodes and edges} & 0.7215 & 8.098 & 0.6379 &  &  &  &  & (63.1) & \\ %
Edge features only & 0.7349 & 8.176 & 0.6494 & +1.86 & {\scriptsize [+1.24, +2.46]} & +0.96 & +1.81 & \tminus1.0 & \\ %
\textit{Dense: no input prior} & 0.7726 & 8.605 & 0.6834 &  &  &  &  & (65.0) & \\ %
Edge features only & 0.830 & 8.915 & 0.738 & +7.49 &  & +3.60 & +7.99 & -- & \\ %
\textit{PDQ-MoE: no input prior} & 0.7970 & 8.128 & 0.7143 &  &  &  &  & (59.6) & \\ %
Edge features only & 0.796 & 8.741 & 0.706 & \tminus0.08 &  & +7.54 & \tminus1.16 & +8.3 & {\scriptsize [+3.7, +13.2]}\\ %
\textit{UniTB: no input prior} & 0.7270 & 7.881 & 0.6449 &  &  &  &  & (61.7) & \\ %
Edge features only & 0.7349 & 8.176 & 0.6494 & +1.08 & {\scriptsize [+0.32, +1.89]} & +3.74 & +0.70 & -- & \\ %
\bottomrule
\end{tabular}
\end{minipage}\end{center}

\subsection*{Expert routing}
PDQ-MoE lowers the dense combined MAE by 3.49\%, against 1.15\% for a wider dense backbone with a similar parameter count, and the other expert designs tested change it by $-1.62\%$ to $+29.2\%$ (panel a of Supplementary Table~\ref{S-tab:abl-routing}). Relative to the router on bond type and prior descriptors, a router on bond type alone raises the combined MAE by 2.74\% (onsite $+7.17\%$, hopping $+2.30\%$; panel b), and independent full-matrix experts and removal of the always-active shared map raise it by 12.46\% and 5.91\% (Supplementary Table~\ref{S-tab:abl-routing-shared}). Inference-time perturbations of the trained router and the routing statistics are in Supplementary Table~\ref{S-tab:abl-routing-diagnostics}.

\providecommand{\tminus}{\textminus}
\begin{center}\begin{minipage}{\linewidth}
\singlespacing
\captionof{table}{\textbf{Expert designs and router inputs.} PDQ-MoE has four routed experts and one always-active shared expert. It selects two experts per edge, routes on the bond type and prior descriptors, and its experts share low-rank bases. Shared bases, the shared map and routing without input initialization follow in Supplementary Table~\ref{S-tab:abl-routing-shared}.}
\label{S-tab:abl-routing}\label{S-tab:moe}
\centering\footnotesize
\renewcommand{\arraystretch}{1.08}
\setlength{\tabcolsep}{2pt}
\begin{tabular}{@{}>{\raggedright\arraybackslash}p{4.0cm}rrr@{\hspace{3pt}}r@{\hspace{2pt}}l@{\hspace{3pt}}rr@{\hspace{3pt}}r@{\hspace{2pt}}l@{}}
\toprule
 & \multicolumn{3}{c}{MAE (meV)} & \multicolumn{2}{c}{Combined change (\%)} & \multicolumn{2}{c}{Change (\%)} & \multicolumn{2}{c}{$\Delta$ mean $\mathrm{fw}_{10}$ (meV)}\\
\cmidrule(lr){2-4}\cmidrule(lr){5-6}\cmidrule(lr){7-8}\cmidrule(lr){9-10}
Variant & Comb. & Onsite & Hopping & Value & 95\% CI & Onsite & Hopping & Value & 95\% CI\\
\midrule
\multicolumn{10}{@{}l@{}}{\textbf{a\enspace Expert designs against the dense backbone}}\\
\textit{Dense backbone, 5.79M} & 0.8133 & 8.488 & 0.7261 &  &  &  &  & (67.4) & \\ %
PDQ-MoE, 4+1 experts, top 2, 7.73M\textsuperscript{$\ast$} & 0.7849 & 8.112 & 0.7003 & \tminus3.49 & {\scriptsize [\tminus4.17, \tminus2.83]} & \tminus4.42 & \tminus3.55 & \tminus0.6 & {\scriptsize [\tminus5.6, +4.4]}\\ %
Single shared low-rank core, no routing, 7.11M\textsuperscript{$\ast$} & 0.8001 & 8.332 & 0.7142 & \tminus1.62 & {\scriptsize [\tminus2.22, \tminus1.03]} & \tminus1.83 & \tminus1.65 & \tminus1.0 & {\scriptsize [\tminus4.9, +2.8]}\\ %
Whole-graph MoLE, 7.71M\textsuperscript{$\ast$} & 0.8080 & 8.758 & 0.7161 & \tminus0.65 & {\scriptsize [\tminus1.21, \tminus0.06]} & +3.19 & \tminus1.38 & +1.5 & {\scriptsize [\tminus3.1, +5.7]}\\ %
Switch-inspired, 256 full-matrix experts, top 1, 706.35M\textsuperscript{$\ast$} & 0.8568 & 8.209 & 0.7735 & +5.34 & {\scriptsize [+4.67, +6.01]} & \tminus3.29 & +6.52 & \tminus1.0 & {\scriptsize [\tminus5.5, +3.3]}\\ %
\textit{Dense backbone, 5.79M}\textsuperscript{\dag} & 0.8134 & 8.486 & 0.7262 &  &  &  &  & (67.4) & \\ %
Wider dense, 43 channels per degree, 7.63M\textsuperscript{\dag}\textsuperscript{$\ast$} & 0.8040 & 8.251 & 0.7190 & \tminus1.15 & {\scriptsize [\tminus1.69, \tminus0.62]} & \tminus2.77 & \tminus0.98 & \tminus2.0 & {\scriptsize [\tminus6.2, +2.0]}\\ %
DPA3-inspired, 24+1 experts, mixed after activation, 71.87M\textsuperscript{\dag}\textsuperscript{$\ast$} & 1.0510 & 8.329 & 0.9688 & +29.21 & {\scriptsize [+27.80, +30.60]} & \tminus1.85 & +33.42 & +1.6 & {\scriptsize [\tminus2.8, +5.8]}\\ %
\textit{Dense + V-shaped channels, 6.05M}\textsuperscript{\dag} & 0.7996 & 8.154 & 0.7156 &  &  &  &  & (63.0) & \\ %
PDQ-MoE + V-shaped channels, 8.01M\textsuperscript{\dag} & 0.7684 & 8.230 & 0.6816 & \tminus3.90 & {\scriptsize [\tminus4.57, \tminus3.20]} & +0.93 & \tminus4.75 & +3.2 & {\scriptsize [\tminus1.6, +7.8]}\\ %
\addlinespace[4pt]
\multicolumn{10}{@{}l@{}}{\textbf{b\enspace Routing inputs and expert selection (baseline: PDQ-MoE, router on bond type and prior descriptors)}}\\
\textit{PDQ-MoE, 4+1 experts, top 2, 7.73M} & 0.7849 & 8.112 & 0.7003 &  &  &  &  & (66.8) & \\ %
Router on bond type only & 0.8065 & 8.694 & 0.7164 & +2.74 & {\scriptsize [+2.09, +3.44]} & +7.17 & +2.30 & +4.9 & {\scriptsize [+0.0, +10.2]}\\ %
Router on bond type + distance (16 bond-length RBFs) & 0.7808 & 8.828 & 0.6890 & \tminus0.53 & {\scriptsize [\tminus1.19, +0.18]} & +8.82 & \tminus1.63 & +3.6 & {\scriptsize [\tminus2.0, +9.4]}\\ %
24+1 experts, top 2, 10.67M & 0.8265 & 8.475 & 0.7379 & +5.30 & {\scriptsize [+4.65, +5.96]} & +4.47 & +5.36 & +1.5 & {\scriptsize [\tminus2.8, +5.6]}\\ %
20+1 experts, top 4, 10.08M\textsuperscript{$\ast$} & 0.7788 & 8.149 & 0.6920 & \tminus0.78 & {\scriptsize [\tminus1.49, +0.05]} & +0.45 & \tminus1.19 & \tminus2.7 & {\scriptsize [\tminus7.6, +2.3]}\\ %
\textit{UniTB, 4+1 experts, top 2, 8.03M} & 0.7215 & 8.098 & 0.6379 &  &  &  &  & (63.1) & \\ %
UniTB with 20+1 experts, top 4, 10.40M & 0.7235 & 8.073 & 0.6397 & +0.28 & {\scriptsize [\tminus0.30, +0.86]} & \tminus0.31 & +0.28 & +3.6 & {\scriptsize [\tminus1.2, +8.7]}\\ %
\bottomrule
\end{tabular}
\end{minipage}\end{center}

{\singlespacing\footnotesize\noindent $^\dagger$ 2,996-structure subset. $^\ast$ More than one setting differs from the baseline: the expert designs of panel a replace or extend the dense update as a whole, the wider dense backbone grows the uniform width from 32 to 43 channels per degree (feature width 2,107), and the 20+1-expert row changes the pool and the selection together.\par}

\providecommand{\tminus}{\textminus}
\begin{center}\begin{minipage}{\linewidth}
\singlespacing
\captionof{table}{\textbf{Shared bases and the shared expert.} Continuation of Supplementary Table~\ref{S-tab:abl-routing}.}
\label{S-tab:abl-routing-shared}\label{S-tab:shared-basis-controls}\label{S-tab:routing-inputs}
\centering\footnotesize
\renewcommand{\arraystretch}{1.08}
\setlength{\tabcolsep}{2pt}
\begin{tabular}{@{}>{\raggedright\arraybackslash}p{4.0cm}rrr@{\hspace{3pt}}r@{\hspace{2pt}}l@{\hspace{3pt}}rr@{\hspace{3pt}}r@{\hspace{2pt}}l@{}}
\toprule
 & \multicolumn{3}{c}{MAE (meV)} & \multicolumn{2}{c}{Combined change (\%)} & \multicolumn{2}{c}{Change (\%)} & \multicolumn{2}{c}{$\Delta$ mean $\mathrm{fw}_{10}$ (meV)}\\
\cmidrule(lr){2-4}\cmidrule(lr){5-6}\cmidrule(lr){7-8}\cmidrule(lr){9-10}
Variant & Comb. & Onsite & Hopping & Value & 95\% CI & Onsite & Hopping & Value & 95\% CI\\
\midrule
\multicolumn{10}{@{}l@{}}{\textbf{c\enspace Shared bases and the shared map}}\\
\textit{PDQ-MoE, 4+1 experts, top 2, 7.73M}\textsuperscript{\dag} & 0.7851 & 8.111 & 0.7005 &  &  &  &  & (66.8) & \\ %
Independent full-matrix experts, 16.94M\textsuperscript{\dag}\textsuperscript{$\ast$} & 0.8829 & 8.495 & 0.7958 & +12.46 & {\scriptsize [+11.57, +13.34]} & +4.73 & +13.60 & \tminus0.8 & {\scriptsize [\tminus5.8, +4.6]}\\ %
Without the always-active shared expert, 4.99M\textsuperscript{\dag}\textsuperscript{$\ast$} & 0.8315 & 8.637 & 0.7386 & +5.91 & {\scriptsize [+5.03, +6.84]} & +6.47 & +5.45 & +2.4 & {\scriptsize [\tminus1.5, +6.3]}\\ %
\textit{Dense backbone, 5.79M}\textsuperscript{\dag} & 0.8134 & 8.486 & 0.7262 &  &  &  &  & (67.4) & \\ %
Independent full-matrix experts\textsuperscript{\dag}\textsuperscript{$\ast$} & 0.8829 & 8.495 & 0.7958 & +8.55 & {\scriptsize [+7.83, +9.27]} & +0.11 & +9.59 & \tminus1.4 & {\scriptsize [\tminus4.9, +2.1]}\\ %
Without the always-active shared expert\textsuperscript{\dag}\textsuperscript{$\ast$} & 0.8315 & 8.637 & 0.7386 & +2.22 & {\scriptsize [+1.35, +3.13]} & +1.77 & +1.72 & +1.8 & {\scriptsize [\tminus2.6, +6.5]}\\ %
\bottomrule
\end{tabular}
\end{minipage}\end{center}

{\singlespacing\footnotesize\noindent $^\dagger$ 2,996-structure subset. $^\ast$ More than one setting differs from the baseline: independent experts each have their own full matrix (parameters $\times$2.19), and the comparisons against dense add experts, router and shared map together.\par}

\providecommand{\tminus}{\textminus}
\begin{center}\begin{minipage}{\linewidth}
\singlespacing
\captionof{table}{\textbf{Routing diagnostics of trained models.} Panel a samples 1,878 ordered bond types and gives paired 95\% intervals. In panel c, nonmodal shares count edges whose expert pair or top-one expert differs from the most common one of their bond type. $H(E\mid T)/H(E)$ is the entropy of the top-one expert given the bond type, relative to its entropy.}
\label{S-tab:abl-routing-diagnostics}\label{S-tab:routing-statistics}
\centering\footnotesize
\renewcommand{\arraystretch}{1.08}
\setlength{\tabcolsep}{2pt}
\begin{tabular}{@{}>{\raggedright\arraybackslash}p{5.2cm}r@{\hspace{3pt}}lr@{\hspace{3pt}}lr@{\hspace{3pt}}l@{}}
\toprule
 & \multicolumn{2}{c}{Onsite} & \multicolumn{2}{c}{Hopping} & \multicolumn{2}{c}{Combined}\\
\cmidrule(lr){2-3}\cmidrule(lr){4-5}\cmidrule(lr){6-7}
 & Value & 95\% CI & Value & 95\% CI & Value & 95\% CI\\
\midrule
\multicolumn{7}{@{}l@{}}{\textbf{a\enspace Inference-time perturbations of one trained PDQ-MoE (800 structures; MAE ratio, perturbed/original)}}\\
Shuffle complete gate vectors within the bond type & 3.29 & {\scriptsize [3.13, 3.48]} & 2.13 & {\scriptsize [2.06, 2.21]} &  & \\
Replace prior descriptors by their bond-type mean & 3.03 & {\scriptsize [2.86, 3.21]} & 1.93 & {\scriptsize [1.86, 2.01]} &  & \\
Fully random routing & 3.16 & {\scriptsize [3.02, 3.32]} & 2.34 & {\scriptsize [2.25, 2.43]} &  & \\
Average over the four routed experts & 2.20 & {\scriptsize [2.11, 2.30]} & 1.84 & {\scriptsize [1.78, 1.91]} &  & \\
Keep only the shared expert & 2.49 & {\scriptsize [2.38, 2.61]} & 5.10 & {\scriptsize [4.89, 5.32]} &  & \\
Force one routed expert for all edges (range over the four experts) & \multicolumn{2}{l}{2.97--3.42} & \multicolumn{2}{l}{2.02--2.68} & & \\ %
\addlinespace[4pt]
\multicolumn{7}{@{}l@{}}{\textbf{b\enspace Retrained models without input initialization (3,000 structures; ratio = 1 + relative change)}}\\
PDQ-MoE: bond-type-mean descriptors & 5.23 &  & 4.09 &  & 4.17 & {\scriptsize [4.06, 4.28]}\\
PDQ-MoE: random edge of the same bond type & 5.72 &  & 4.39 &  & 4.50 & {\scriptsize [4.38, 4.61]}\\
PDQ-MoE: same bond type, length within 0.05~\AA & 1.19 &  & 1.063 &  & 1.074 & {\scriptsize [1.069, 1.080]}\\
UniTB recipe: bond-type-mean descriptors & 3.59 &  & 3.87 &  & 3.79 & {\scriptsize [3.70, 3.88]}\\
UniTB recipe: random edge of the same bond type & 4.13 &  & 3.86 &  & 3.85 & {\scriptsize [3.76, 3.94]}\\
UniTB recipe: same bond type, length within 0.05~\AA & 1.16 &  & 1.031 &  & 1.044 & {\scriptsize [1.040, 1.047]}\\
\addlinespace[4pt]
\multicolumn{7}{@{}l@{}}{\textbf{c\enspace Routing statistics of the same PDQ-MoE (800 structures; no interval)}}\\
Edges outside the modal expert pair (\%) & 8.90 &  & 10.50 &  &  & \\
Edges outside the modal top-one expert (\%) & 9.31 &  & 13.70 &  &  & \\
Effective top-one experts per bond type & 1.39 &  & 1.58 &  &  & \\
Entropy of the top-one expert given the bond type, relative to its entropy, $H(E\mid T)/H(E)$ & 0.25 &  & 0.33 &  &  & \\
\bottomrule
\end{tabular}
\end{minipage}\end{center}

{\singlespacing\footnotesize\noindent Shuffled gates permute complete gate vectors (selected experts and weights) within a bond type; type-mean descriptors replace the prior descriptors of each edge by the mean of its bond type.\par}

\subsection*{Charge head}
The charge head lowers the combined MAE by 2.08\% on dense and by 4.24\% on PDQ-MoE with site channels. On PDQ-MoE, the charge-head recipe lowers it by a further 2.25\% beyond an additive potential-shift head, and a local-only head without charge equilibration is 1.97\% above the charge head (Supplementary Table~\ref{S-tab:abl-charge}, which also gives the share of structures with lower error). The gain by number of atoms per cell is given in Supplementary Table~\ref{S-tab:abl-charge-size}.

\providecommand{\tminus}{\textminus}
\begin{center}\begin{minipage}{\linewidth}
\singlespacing
\captionof{table}{\textbf{Charge-head variants on two backbones.} The charge head combines a local readout with the long-range term $\kappa\Gamma q$ from charge equilibration. The additive potential-shift head is a local alternative. Lower (\%) is the share of test structures with a combined MAE below that of the row's baseline.}
\label{S-tab:abl-charge}\label{S-tab:charge-controls}
\centering\footnotesize
\renewcommand{\arraystretch}{1.08}
\setlength{\tabcolsep}{2pt}
\begin{tabular}{@{}>{\raggedright\arraybackslash}p{3.25cm}rrr@{\hspace{3pt}}r@{\hspace{2pt}}l@{\hspace{4pt}}r@{\hspace{3pt}}rr@{\hspace{3pt}}r@{\hspace{2pt}}l@{}}
\toprule
 & \multicolumn{3}{c}{MAE (meV)} & \multicolumn{3}{c}{Combined change} & \multicolumn{2}{c}{Change (\%)} & \multicolumn{2}{c}{$\Delta$ mean $\mathrm{fw}_{10}$ (meV)}\\
\cmidrule(lr){2-4}\cmidrule(lr){5-7}\cmidrule(lr){8-9}\cmidrule(lr){10-11}
Variant & Comb. & Onsite & Hopping & Value (\%) & 95\% CI & Lower (\%) & Onsite & Hopping & Value & 95\% CI\\
\midrule
\multicolumn{11}{@{}l@{}}{\textbf{a\enspace Dense backbone}}\\
\textit{Dense backbone} & 0.8133 & 8.488 & 0.7261 &  &  &  &  &  & (67.4) & \\ %
Charge head (local readout + $\kappa\Gamma q$) & 0.7964 & 8.283 & 0.7117 & \tminus2.08 & {\scriptsize [\tminus2.85, \tminus1.31]} & 54.5 & \tminus2.41 & \tminus1.99 & \tminus3.1 & {\scriptsize [\tminus6.8, +0.8]}\\ %
\addlinespace[4pt]
\multicolumn{11}{@{}l@{}}{\textbf{b\enspace PDQ-MoE}}\\
\textit{PDQ-MoE, 4+1 experts, top 2} & 0.7849 & 8.112 & 0.7003 &  &  &  &  &  & (66.8) & \\ %
Additive potential-shift head\textsuperscript{$\ast$} & 0.7704 & 7.914 & 0.6888 & \tminus1.85 & {\scriptsize [\tminus2.47, \tminus1.24]} & 59.3 & \tminus2.45 & \tminus1.64 & \tminus3.1 & {\scriptsize [\tminus7.0, +0.8]}\\ %
\textit{PDQ-MoE, 4+1 experts, top 2}\textsuperscript{\dag} & 0.7851 & 8.111 & 0.7005 &  &  &  &  &  & (66.8) & \\ %
Charge head\textsuperscript{\dag}\textsuperscript{$\ast$} & 0.7533 & 7.975 & 0.6700 & \tminus4.05 & {\scriptsize [\tminus4.83, \tminus3.26]} & 62.2 & \tminus1.69 & \tminus4.35 & \tminus2.6 & {\scriptsize [\tminus6.9, +2.0]}\\ %
\textit{PDQ-MoE + additive potential-shift head} & 0.7704 & 7.914 & 0.6888 &  &  &  &  &  & (63.8) & \\ %
Charge-head recipe\textsuperscript{$\ast$} & 0.7531 & 7.977 & 0.6698 & \tminus2.25 & {\scriptsize [\tminus3.18, \tminus1.33]} & 52.8 & +0.79 & \tminus2.76 & +0.5 & {\scriptsize [\tminus2.9, +4.1]}\\ %
\textit{PDQ-MoE + charge head} & 0.7531 & 7.977 & 0.6698 &  &  &  &  &  & (64.3) & \\ %
Local-only head (no charge equilibration, no $\kappa\Gamma q$)\textsuperscript{$\ast$} & 0.7679 & 7.972 & 0.6857 & +1.97 & {\scriptsize [+1.43, +2.54]} & 45.3 & \tminus0.05 & +2.37 & \tminus1.3 & {\scriptsize [\tminus5.4, +2.7]}\\ %
\textit{PDQ-MoE + V-shaped channels} & 0.7682 & 8.230 & 0.6814 &  &  &  &  &  & (66.1) & \\ %
Charge head & 0.7357 & 8.209 & 0.6491 & \tminus4.24 & {\scriptsize [\tminus4.98, \tminus3.52]} & 65.0 & \tminus0.27 & \tminus4.74 & \tminus2.8 & {\scriptsize [\tminus7.4, +2.1]}\\ %
\addlinespace[4pt]
\multicolumn{11}{@{}l@{}}{\textbf{c\enspace Heads on other expert configurations}}\\
\textit{All four routed experts (soft mixture)} & 0.7760 & 8.275 & 0.6899 &  &  &  &  &  & (68.4) & \\ %
Additive potential-shift head\textsuperscript{$\ast$} & 0.7536 & 8.115 & 0.6685 & \tminus2.89 & {\scriptsize [\tminus3.45, \tminus2.33]} & 62.8 & \tminus1.92 & \tminus3.10 & \tminus5.4 & {\scriptsize [\tminus9.0, \tminus2.1]}\\ %
\textit{20+1 experts, top 4} & 0.7788 & 8.149 & 0.6920 &  &  &  &  &  & (64.1) & \\ %
Additive potential-shift head & 0.7513 & 8.168 & 0.6650 & \tminus3.53 & {\scriptsize [\tminus4.21, \tminus2.87]} & 63.8 & +0.23 & \tminus3.90 & \tminus0.2 & {\scriptsize [\tminus3.9, +3.5]}\\ %
\bottomrule
\end{tabular}
\end{minipage}\end{center}

{\singlespacing\footnotesize\noindent $^\dagger$ 2,996-structure subset. $^\ast$ More than one setting differs from the baseline; the charge-head recipe changes the head as a whole (input embedding, MLP, readout branches), and the local-only head changes four response settings together.\par}

\providecommand{\tminus}{\textminus}
\begin{center}\begin{minipage}{\linewidth}
\singlespacing
\captionof{table}{\textbf{Charge head by cell size.} The 3,000 test structures are binned by atoms per cell. Paired 95\% intervals come from resamples within each bin. The PDQ-MoE and dense baselines have neither charge head nor site channels; MPHam-12K budget.}
\label{S-tab:abl-charge-size}
\centering\footnotesize
\renewcommand{\arraystretch}{1.08}
\setlength{\tabcolsep}{2pt}
\begin{tabular}{@{}>{\raggedright\arraybackslash}p{4.3cm}r@{\hspace{3pt}}lr@{\hspace{3pt}}lr@{\hspace{3pt}}l@{}}
\toprule
 & \multicolumn{2}{c}{PDQ-MoE + charge head} & \multicolumn{2}{c}{Dense + charge head} & \multicolumn{2}{c}{PDQ-MoE}\\
 & \multicolumn{2}{c}{versus PDQ-MoE} & \multicolumn{2}{c}{versus dense} & \multicolumn{2}{c}{versus dense}\\
\cmidrule(lr){2-3}\cmidrule(lr){4-5}\cmidrule(lr){6-7}
Atoms per cell (structures) & Change (\%) & 95\% CI & Change (\%) & 95\% CI & Change (\%) & 95\% CI\\
\midrule
1--5 (656) & \tminus1.33 & {\scriptsize [\tminus3.40, +0.59]} & \tminus0.99 & {\scriptsize [\tminus3.50, +1.42]} & \tminus3.30 & {\scriptsize [\tminus5.26, \tminus1.29]}\\
6--12 (851) & \tminus3.02 & {\scriptsize [\tminus4.59, \tminus1.36]} & \tminus1.66 & {\scriptsize [\tminus3.08, \tminus0.25]} & \tminus3.85 & {\scriptsize [\tminus5.14, \tminus2.55]}\\
13--22 (755) & \tminus4.65 & {\scriptsize [\tminus6.05, \tminus3.33]} & \tminus2.14 & {\scriptsize [\tminus3.18, \tminus1.13]} & \tminus3.75 & {\scriptsize [\tminus4.73, \tminus2.75]}\\
23--79 (738) & \tminus6.98 & {\scriptsize [\tminus8.08, \tminus5.91]} & \tminus3.42 & {\scriptsize [\tminus4.48, \tminus2.35]} & \tminus3.01 & {\scriptsize [\tminus3.97, \tminus2.04]}\\
\bottomrule
\end{tabular}
\end{minipage}\end{center}

\subsection*{Training recipe}
Along the site-channels-first path, site channels, PDQ-MoE, the charge head and prior perturbations with the constant $t=0$ embedding change the combined MAE by $-1.71\%$, $-3.90\%$, $-4.24\%$ and $-1.92\%$, and UniTB is 11.29\% below dense (Supplementary Table~\ref{S-tab:abl-recipe}); main-text Table~\ref{tab:benchmark} adds PDQ-MoE first. Perturbation forms and amplitudes and the full recipe at the MPHam-29K budget and on the 18 Si and Ge--Sb--Te supercells are compared in Supplementary Tables~\ref{S-tab:abl-recipe} and~\ref{S-tab:abl-recipe-amplitude}.

\providecommand{\tminus}{\textminus}
\begin{center}\begin{minipage}{\linewidth}
\singlespacing
\captionof{table}{\textbf{Steps of the recipe from the dense model to UniTB.} In panel a, each step is compared with the step above it; the last row follows the order of main-text Table~\ref{tab:benchmark}. Prior perturbations add noise to the input prior during training only (Methods).}
\label{S-tab:abl-recipe}\label{S-tab:roadmap}
\centering\footnotesize
\renewcommand{\arraystretch}{1.08}
\setlength{\tabcolsep}{2pt}
\begin{tabular}{@{}>{\raggedright\arraybackslash}p{4.0cm}rrr@{\hspace{3pt}}r@{\hspace{2pt}}l@{\hspace{3pt}}rr@{\hspace{3pt}}r@{\hspace{2pt}}l@{}}
\toprule
 & \multicolumn{3}{c}{MAE (meV)} & \multicolumn{2}{c}{Combined change (\%)} & \multicolumn{2}{c}{Change (\%)} & \multicolumn{2}{c}{$\Delta$ mean $\mathrm{fw}_{10}$ (meV)}\\
\cmidrule(lr){2-4}\cmidrule(lr){5-6}\cmidrule(lr){7-8}\cmidrule(lr){9-10}
Variant & Comb. & Onsite & Hopping & Value & 95\% CI & Onsite & Hopping & Value & 95\% CI\\
\midrule
\multicolumn{10}{@{}l@{}}{\textbf{a\enspace Recipe path from the dense backbone to UniTB}}\\
\textit{Dense backbone} & 0.8133 & 8.488 & 0.7261 &  &  &  &  & (67.4) & \\ %
+ site-symmetry channels (V-shaped) & 0.7995 & 8.155 & 0.7155 & \tminus1.71 & {\scriptsize [\tminus2.21, \tminus1.18]} & \tminus3.92 & \tminus1.47 & \tminus4.4 & {\scriptsize [\tminus8.5, \tminus0.7]}\\ %
+ PDQ-MoE (4+1 experts, top 2) & 0.7682 & 8.230 & 0.6814 & \tminus3.90 & {\scriptsize [\tminus4.59, \tminus3.22]} & +0.92 & \tminus4.76 & +3.2 & {\scriptsize [\tminus1.6, +7.8]}\\ %
+ charge head & 0.7357 & 8.209 & 0.6491 & \tminus4.24 & {\scriptsize [\tminus4.98, \tminus3.52]} & \tminus0.27 & \tminus4.74 & \tminus2.8 & {\scriptsize [\tminus7.4, +2.1]}\\ %
\textbf{+ prior perturbations and constant $t=0$ embedding (UniTB)}\textsuperscript{$\ast$} & \textbf{0.7215} & \textbf{8.098} & \textbf{0.6379} & \textbf{\tminus1.92} & \textbf{{\scriptsize [\tminus2.43, \tminus1.41]}} & \textbf{\tminus1.34} & \textbf{\tminus1.73} & \textbf{\tminus0.2} & \textbf{{\scriptsize [\tminus4.3, +3.8]}}\\ %
UniTB against dense (joint comparison)\textsuperscript{$\ast$} & 0.7215 & 8.098 & 0.6379 & \tminus11.29 & {\scriptsize [\tminus12.11, \tminus10.46]} & \tminus4.58 & \tminus12.15 & \tminus4.3 & {\scriptsize [\tminus9.6, +0.7]}\\ %
PDQ-MoE + charge head against dense\textsuperscript{$\ast$} & 0.7531 & 7.977 & 0.6698 & \tminus7.41 & {\scriptsize [\tminus8.27, \tminus6.57]} & \tminus6.02 & \tminus7.75 & \tminus3.1 & {\scriptsize [\tminus8.2, +2.2]}\\ %
\textit{PDQ-MoE + charge head (uniform channels)} & 0.7531 & 7.977 & 0.6698 &  &  &  &  & (64.3) & \\ %
UniTB (adds site channels and perturbations)\textsuperscript{$\ast$} & 0.7215 & 8.098 & 0.6379 & \tminus4.19 & {\scriptsize [\tminus4.81, \tminus3.53]} & +1.53 & \tminus4.77 & -- & \\ %
\addlinespace[4pt]
\multicolumn{10}{@{}l@{}}{\textbf{b\enspace Forms of prior perturbation on the 20+1-expert PDQ-MoE (top 4)}}\\
\textit{No perturbation} & 0.7788 & 8.149 & 0.6920 &  &  &  &  & (64.1) & \\ %
Pure prior perturbation, clean evaluation (adopted form)\textsuperscript{$\ast$} & 0.7665 & 8.508 & 0.6753 & \tminus1.6 & {\scriptsize [\tminus2.3, \tminus0.9]} & +4.4 & \tminus2.4 & +2.2 & {\scriptsize [\tminus2.5, +7.0]}\\ %
Flow matching with small perturbation (uniform $t$, endpoint loss)\textsuperscript{$\ast$} & 0.8748 & 9.965 & 0.7702 & +12.3 & {\scriptsize [+11.3, +13.3]} & +22.3 & +11.3 & -- & \\ %
Flow recipe (unit-noise scale, one-step evaluation from the clean start)\textsuperscript{$\ast$} & 1.6303 & 9.961 & 1.5445 & +109.3 & {\scriptsize [+106.2, +112.6]} & +22.2 & +123.2 & -- & \\ %
\addlinespace[4pt]
\multicolumn{10}{@{}l@{}}{\textbf{c\enspace Perturbation on uniform channels, and site channels with perturbation}}\\
\textit{PDQ-MoE + charge head (uniform channels)} & 0.7531 & 7.977 & 0.6698 &  &  &  &  & (64.3) & \\ %
+ prior perturbation\textsuperscript{$\ast$} & 0.7329 & 8.075 & 0.6494 & \tminus2.68 &  & +1.23 & \tminus3.06 & +1.3 & {\scriptsize [\tminus2.3, +4.9]}\\ %
\textit{PDQ-MoE + charge head + perturbation (uniform channels)} & 0.7329 & 8.075 & 0.6494 &  &  &  &  &  & \\ %
+ V-shaped channels (UniTB) & 0.7215 & 8.098 & 0.6379 & \tminus1.56 & {\scriptsize [\tminus2.14, \tminus0.97]} & +0.29 & \tminus1.77 & -- & \\ %
\bottomrule
\end{tabular}
\end{minipage}\end{center}

{\singlespacing\footnotesize\noindent $^\ast$ More than one setting differs from the baseline: UniTB against dense adds four modules (8.03M against 5.79M parameters), the other joint rows add two modules each, the perturbation steps come with the constant $t=0$ embedding, and the flow variants change trajectory, noise and loss together.\par}

\providecommand{\tminus}{\textminus}
\begin{center}\begin{minipage}{\linewidth}
\singlespacing
\captionof{table}{\textbf{Perturbation amplitudes and complete recipes.} Continuation of Supplementary Table~\ref{S-tab:abl-recipe}. The 18 supercells of panel e are absent from training.}
\label{S-tab:abl-recipe-amplitude}
\centering\footnotesize
\renewcommand{\arraystretch}{1.08}
\setlength{\tabcolsep}{2pt}
\begin{tabular}{@{}>{\raggedright\arraybackslash}p{4.0cm}rrr@{\hspace{3pt}}r@{\hspace{2pt}}l@{\hspace{3pt}}rr@{\hspace{3pt}}r@{\hspace{2pt}}l@{}}
\toprule
 & \multicolumn{3}{c}{MAE (meV)} & \multicolumn{2}{c}{Combined change (\%)} & \multicolumn{2}{c}{Change (\%)} & \multicolumn{2}{c}{$\Delta$ mean $\mathrm{fw}_{10}$ (meV)}\\
\cmidrule(lr){2-4}\cmidrule(lr){5-6}\cmidrule(lr){7-8}\cmidrule(lr){9-10}
Variant & Comb. & Onsite & Hopping & Value & 95\% CI & Onsite & Hopping & Value & 95\% CI\\
\midrule
\multicolumn{10}{@{}l@{}}{\textbf{d\enspace Perturbation amplitude (UniTB: onsite 0.5, hopping 5)}}\\
\textit{UniTB}\textsuperscript{\ddag} & 0.7215 & 8.197 & 0.6476 &  &  &  &  & (63.1) & \\ %
Onsite 5, hopping 20\textsuperscript{\ddag} & 0.7478 & 8.181 & 0.6731 & +3.64 & {\scriptsize [+2.97, +4.28]} & \tminus0.20 & +3.93 & +2.0 & {\scriptsize [\tminus1.5, +5.6]}\\ %
Onsite 5, hopping 50\textsuperscript{\ddag} & 0.7755 & 8.181 & 0.6992 & +7.48 & {\scriptsize [+6.81, +8.17]} & \tminus0.20 & +7.96 & +1.8 & {\scriptsize [\tminus1.7, +5.3]}\\ %
\addlinespace[4pt]
\multicolumn{10}{@{}l@{}}{\textbf{e\enspace Whole-recipe comparisons at the MPHam-29K budget (91,800 updates) and on 18 Si and Ge--Sb--Te supercells}}\\
\textit{Dense, $H^{(0)}$ prior} & 0.5592 & 6.208 & 0.4961 &  &  &  &  & (51.9) & \\ %
UniTB\textsuperscript{$\ast$} & 0.4928 & 5.741 & 0.4339 & \tminus11.89 & {\scriptsize [\tminus12.92, \tminus10.85]} & & & \tminus2.8 & {\scriptsize [\tminus7.8, +2.4]}\\ %
\textit{Dense, grid-free prior} & 2.0228 & 7.046 & 1.9709 &  &  &  &  & (76.2) & \\ %
UniTB-Flash\textsuperscript{$\ast$} & 1.6937 & 7.166 & 1.6361 & \tminus16.27 & {\scriptsize [\tminus16.68, \tminus15.82]} & & & \tminus5.4 & {\scriptsize [\tminus12.8, +2.1]}\\ %
\textit{Dense, $H^{(0)}$ prior, 18 supercells} & 0.8707 & 11.958 & 0.7032 &  &  &  &  &  & \\ %
UniTB\textsuperscript{$\ast$} & 0.7663 & 10.342 & 0.6203 & \tminus12.00 & {\scriptsize [\tminus13.81, \tminus10.23]} & & & -- & \\ %
\textit{Dense, grid-free prior, 18 supercells} & 2.0398 & 12.298 & 1.8878 &  &  &  &  &  & \\ %
UniTB-Flash\textsuperscript{$\ast$} & 1.8179 & 11.887 & 1.6694 & \tminus10.88 & {\scriptsize [\tminus12.98, \tminus8.46]} & & & -- & \\ %
\bottomrule
\end{tabular}
\end{minipage}\end{center}

{\singlespacing\footnotesize\noindent $^\ddagger$ Onsite and hopping values are pooled over matrix elements. $^\ast$ More than one setting differs from the baseline: UniTB and UniTB-Flash add four modules and differ from the dense models in further architectural and training settings. Supercell rows weight the 18 cells of 180 to 216 atoms equally; their intervals use 20,000 resamples, as in Supplementary Table~\ref{S-tab:amorphous-dense}.\par}

\subsection*{Spectral projection}
With one solver, projection at $b=10$ eV lowers the mean $\mathrm{fw}_{10}$ of five MPHam-12K configurations by 15.2--39.1 meV and removes all strict detections; full-path results and a threshold scan are given in Supplementary Tables~\ref{S-tab:abl-projection} and~\ref{S-tab:abl-projection-threshold}, and $b=10$ eV is used throughout.

\providecommand{\tminus}{\textminus}
\begin{center}\begin{minipage}{\linewidth}
\singlespacing
\captionof{table}{\textbf{Band errors with and without spectral projection.} Projection ($b=10$ eV) acts on the prediction at inference. Each row compares its own prediction before and after projection on 200 band structures. Panel a uses one solver, with paired 95\% intervals. Diagnostics are as in main-text Methods (Band metrics); counts are cells out of 200.}
\label{S-tab:abl-projection}\label{S-tab:clip_onoff_s_sampled_full}\label{S-tab:clip_onoff_s_production}\label{S-tab:clip_onoff_l_production}
\centering\footnotesize
\renewcommand{\arraystretch}{1.08}
\setlength{\tabcolsep}{2pt}
\begin{tabular}{@{}>{\raggedright\arraybackslash}p{5.5cm}rrr@{\hspace{2pt}}l@{\hspace{8pt}}rr@{\hspace{8pt}}rr@{}}
\toprule
 & \multicolumn{4}{c}{Mean $\mathrm{fw}_{10}$ (meV)} & \multicolumn{2}{c}{Wide cells} & \multicolumn{2}{c}{Strict cells}\\
\cmidrule(lr){2-5}\cmidrule(lr){6-7}\cmidrule(lr){8-9}
Configuration & Off & On & \multicolumn{2}{c}{Change [95\% CI]} & off & on & off & on\\
\midrule
\multicolumn{9}{@{}l@{}}{\textbf{a\enspace MPHam-12K, sampled path (15,459 wavevectors)}}\\
Dense & 95.02 & 67.76 & \tminus27.3 & {\scriptsize [\tminus52.2, \tminus8.5]} & 13 & 2 & 6 & 0\\
Dense + site channels & 86.55 & 63.39 & \tminus23.2 & {\scriptsize [\tminus46.3, \tminus5.7]} & 13 & 2 & 6 & 0\\
Dense + site channels + layer-wise prior memory & 78.69 & 63.54 & \tminus15.2 & {\scriptsize [\tminus36.7, \tminus2.5]} & 11 & 2 & 5 & 0\\
PDQ-MoE + charge head & 86.19 & 64.05 & \tminus22.1 & {\scriptsize [\tminus45.2, \tminus6.9]} & 11 & 1 & 8 & 0\\
UniTB & 102.40 & 63.30 & \tminus39.1 & {\scriptsize [\tminus76.8, \tminus10.8]} & 11 & 2 & 6 & 0\\
\addlinespace[4pt]
\multicolumn{9}{@{}l@{}}{\textbf{b\enspace MPHam-12K, full path (36,043 wavevectors)}}\\
Dense & 94.50 & 67.40 & \tminus27.1 & {\scriptsize [\tminus51.3, \tminus8.8]} & 13 & 2 &  & 0\\
Dense + site channels & 86.01 & 62.97 & \tminus23.0 & {\scriptsize [\tminus46.4, \tminus5.6]} & 13 & 1 &  & 0\\
Dense + site channels + layer-wise prior memory & 78.73 & 63.71 & \tminus15.0 & {\scriptsize [\tminus36.4, \tminus2.6]} & 11 & 2 &  & 0\\
PDQ-MoE + site channels &  & 66.13 &  &  &  & 3 &  & 0\\
PDQ-MoE + charge head + site channels &  & 63.38 &  &  &  & 2 &  & 0\\
PDQ-MoE + charge head & 85.72 & 64.26 & \tminus21.5 & {\scriptsize [\tminus44.3, \tminus6.7]} & 12 & 1 &  & 0\\
PDQ-MoE + additive shift head & 104.03 & 63.75 & \tminus40.3 &  & 14 & 4 &  & 0\\
All four experts + additive shift head & 90.63 & 62.92 & \tminus27.7 &  & 10 & 1 &  & 0\\
20+1 experts, top 4 + additive shift head & 80.84 & 63.89 & \tminus17.0 &  & 9 & 1 &  & 0\\
UniTB & 102.28 & 63.14 & \tminus39.1 & {\scriptsize [\tminus77.5, \tminus10.8]} & 11 & 2 &  & 0\\
UniTB with 20+1 experts, top 4 & 96.21 & 66.70 & \tminus29.5 &  & 13 & 3 &  & 0\\
20+1 experts + pure prior perturbation & 91.31 & 66.31 & \tminus25.0 &  & 17 & 4 &  & 0\\
UniTB with 20+1 experts, top 4 + frame encoding + mixed projection loss & 100.14 & 65.76 & \tminus34.4 &  & 13 & 1 &  & 0\\
Dense + frame encoding + mixed projection loss ($w=10$) & 76.76 & 67.63 & \tminus9.1 &  & 12 & 1 &  & 0\\
Dense + frame encoding + mixed projection loss ($w=1$) & 85.14 & 67.94 & \tminus17.2 &  & 13 & 1 &  & 0\\
\addlinespace[4pt]
\multicolumn{9}{@{}l@{}}{\textbf{c\enspace MPHam-29K, full path}}\\
Dense & 55.85 & 51.90 & \tminus4.0 &  & 3 & 1 &  & 0\\
UniTB &  & 49.05 &  &  &  & 2 &  & 0\\
\bottomrule
\end{tabular}
\end{minipage}\end{center}

{\singlespacing\footnotesize\noindent The last two rows of panel b use a mixed symmetry-projection loss in training (Supplementary Table~\ref{S-tab:abl-other-loss}); this table applies spectral projection at inference.\par}

\providecommand{\tminus}{\textminus}
\begin{center}\begin{minipage}{\linewidth}
\singlespacing
\captionof{table}{\textbf{Band errors at different projection bounds.} The solver and wavevectors are those of panel a of Supplementary Table~\ref{S-tab:abl-projection}. The label row projects the DFT label residual itself. All other results of the paper use $b=10$ eV.}
\label{S-tab:abl-projection-threshold}\label{S-tab:clip_threshold_s_sampled}
\centering\footnotesize
\renewcommand{\arraystretch}{1.08}
\setlength{\tabcolsep}{2pt}
\begin{tabular}{@{}>{\raggedright\arraybackslash}p{4.9cm}rrrrll@{}}
\toprule
 & \multicolumn{4}{c}{Mean $\mathrm{fw}_{10}$ (meV)} & \multicolumn{2}{c}{Paired change (meV) [95\% CI]}\\
\cmidrule(lr){2-5}\cmidrule(lr){6-7}
Configuration & Unprojected & $b=5$ & $b=7$ & $b=10$ & $b=5$ vs.\ $b=10$ & $b=7$ vs.\ $b=10$\\
\midrule
Label reconstruction reference & 0.86 & 5.29 & 0.86 & 0.86 &  & \\
\addlinespace[3pt]
Dense & 95.02 & 69.19 & 67.57 & 67.76 & +1.4 {\scriptsize [\tminus0.8, +4.0]} & \tminus0.2 {\scriptsize [\tminus0.6, +0.4]}\\
Dense + site channels & 86.55 & 66.20 & 63.59 & 63.39 & +2.8 {\scriptsize [+0.2, +6.0]} & +0.2 {\scriptsize [\tminus0.4, +1.1]}\\
Dense + site channels + layer-wise prior memory & 78.69 & 66.34 & 63.62 & 63.54 & +2.8 {\scriptsize [+0.3, +5.9]} & +0.1 {\scriptsize [\tminus0.2, +0.5]}\\
PDQ-MoE + charge head & 86.19 & 65.04 & 63.88 & 64.05 & +1.0 {\scriptsize [\tminus1.6, +3.6]} & \tminus0.2 {\scriptsize [\tminus0.4, +0.1]}\\
UniTB & 102.40 & 65.75 & 63.08 & 63.30 & +2.5 {\scriptsize [+0.1, +5.3]} & \tminus0.2 {\scriptsize [\tminus0.4, 0.0]}\\
\bottomrule
\end{tabular}
\end{minipage}\end{center}

\subsection*{Additional controls}
Additional controls compare frame encodings of the node features, a mixed symmetry-projection loss, a funnel last layer and two node-feature additions (Supplementary Tables~\ref{S-tab:abl-other} and~\ref{S-tab:abl-other-loss}); the retrained variants change the combined MAE by $-0.92\%$ to $+6.37\%$.

\providecommand{\tminus}{\textminus}
\begin{center}\begin{minipage}{\linewidth}
\singlespacing
\captionof{table}{\textbf{Further architectural variants.} The frame encoding injects an equivariant encoding of the crystal frame of each site at node initialization, optionally gated by site symmetry.}
\label{S-tab:abl-other}
\centering\footnotesize
\renewcommand{\arraystretch}{1.08}
\setlength{\tabcolsep}{2pt}
\begin{tabular}{@{}>{\raggedright\arraybackslash}p{4.0cm}rrr@{\hspace{3pt}}r@{\hspace{2pt}}l@{\hspace{3pt}}rr@{\hspace{3pt}}r@{\hspace{2pt}}l@{}}
\toprule
 & \multicolumn{3}{c}{MAE (meV)} & \multicolumn{2}{c}{Combined change (\%)} & \multicolumn{2}{c}{Change (\%)} & \multicolumn{2}{c}{$\Delta$ mean $\mathrm{fw}_{10}$ (meV)}\\
\cmidrule(lr){2-4}\cmidrule(lr){5-6}\cmidrule(lr){7-8}\cmidrule(lr){9-10}
Variant & Comb. & Onsite & Hopping & Value & 95\% CI & Onsite & Hopping & Value & 95\% CI\\
\midrule
\multicolumn{10}{@{}l@{}}{\textbf{a\enspace Frame encoding of node features (dense backbone)}}\\
\textit{Dense backbone}\textsuperscript{\dag} & 0.8134 & 8.486 & 0.7262 &  &  &  &  &  & \\ %
Node initialization\textsuperscript{\dag} & 0.8266 & 8.612 & 0.7369 & +1.62 & {\scriptsize [+0.99, +2.26]} & +1.49 & +1.48 & -- & \\ %
Node initialization + site gate\textsuperscript{\dag} & 0.8289 & 8.667 & 0.7390 & +1.91 & {\scriptsize [+1.33, +2.51]} & +2.13 & +1.78 & -- & \\ %
\bottomrule
\end{tabular}
\end{minipage}\end{center}

{\singlespacing\footnotesize\noindent $^\dagger$ 2,996-structure subset.\par}

\providecommand{\tminus}{\textminus}
\begin{center}\begin{minipage}{\linewidth}
\singlespacing
\captionof{table}{\textbf{Loss and layer variants.} Continuation of Supplementary Table~\ref{S-tab:abl-other}. The mixed projection loss trains on the space-group-projected output plus the raw output with weight $w$. The funnel layer is an equivariant nonlinear gate in the last layer. The per-order inner product and the low-rank bilinear residual are node-feature interaction modules.}
\label{S-tab:abl-other-loss}
\centering\footnotesize
\renewcommand{\arraystretch}{1.08}
\setlength{\tabcolsep}{2pt}
\begin{tabular}{@{}>{\raggedright\arraybackslash}p{4.0cm}rrr@{\hspace{3pt}}r@{\hspace{2pt}}l@{\hspace{3pt}}rr@{\hspace{3pt}}r@{\hspace{2pt}}l@{}}
\toprule
 & \multicolumn{3}{c}{MAE (meV)} & \multicolumn{2}{c}{Combined change (\%)} & \multicolumn{2}{c}{Change (\%)} & \multicolumn{2}{c}{$\Delta$ mean $\mathrm{fw}_{10}$ (meV)}\\
\cmidrule(lr){2-4}\cmidrule(lr){5-6}\cmidrule(lr){7-8}\cmidrule(lr){9-10}
Variant & Comb. & Onsite & Hopping & Value & 95\% CI & Onsite & Hopping & Value & 95\% CI\\
\midrule
\multicolumn{10}{@{}l@{}}{\textbf{b\enspace Mixed projection loss (dense + frame encoding; projected output)}}\\
\textit{Dense backbone} & 0.8133 & 8.488 & 0.7261 &  &  &  &  & (67.4) & \\ %
Raw-loss weight $w=1$\textsuperscript{$\ast$} & 0.8131 & 8.388 & 0.7268 & \tminus0.02 & {\scriptsize [\tminus0.61, +0.58]} & \tminus1.18 & +0.09 & +0.5 & {\scriptsize [\tminus2.9, +3.9]}\\ %
Raw-loss weight $w=10$\textsuperscript{$\ast$} & 0.8058 & 8.243 & 0.7217 & \tminus0.92 & {\scriptsize [\tminus1.47, \tminus0.37]} & \tminus2.88 & \tminus0.61 & +0.2 & {\scriptsize [\tminus3.1, +3.5]}\\ %
\textit{Mixed loss $w=1$} & 0.8131 & 8.388 & 0.7268 &  &  &  &  & (67.9) & \\ %
Mixed loss $w=10$ & 0.8058 & 8.243 & 0.7217 & \tminus0.90 & {\scriptsize [\tminus1.38, \tminus0.41]} & \tminus1.72 & \tminus0.70 & \tminus0.3 & {\scriptsize [\tminus3.4, +2.9]}\\ %
\addlinespace[4pt]
\multicolumn{10}{@{}l@{}}{\textbf{c\enspace Funnel last layer}}\\
\textit{Dense backbone} & 0.8133 & 8.488 & 0.7261 &  &  &  &  & (67.4) & \\ %
Funnel last layer (equivariant nonlinear gate) & 0.8651 & 8.403 & 0.7793 & +6.37 & {\scriptsize [+5.75, +7.00]} & \tminus0.99 & +7.33 & +3.5 & {\scriptsize [+0.2, +7.0]}\\ %
\textit{PDQ-MoE, 4+1 experts, top 2} & 0.7849 & 8.112 & 0.7003 &  &  &  &  & (66.8) & \\ %
Funnel last layer & 0.7991 & 8.284 & 0.7139 & +1.81 & {\scriptsize [+1.21, +2.41]} & +2.12 & +1.94 & +3.2 & {\scriptsize [\tminus1.9, +8.8]}\\ %
\addlinespace[4pt]
\multicolumn{10}{@{}l@{}}{\textbf{d\enspace Node-feature additions (dense backbone)}}\\
\textit{Dense backbone}\textsuperscript{\dag} & 0.8134 & 8.486 & 0.7262 &  &  &  &  &  & \\ %
Per-order inner product into the edge latent\textsuperscript{\dag} & 0.8085 & 8.345 & 0.7226 & \tminus0.60 & {\scriptsize [\tminus1.12, \tminus0.04]} & \tminus1.66 & \tminus0.49 & -- & \\ %
Low-rank bilinear node residual\textsuperscript{\dag} & 0.8171 & 8.347 & 0.7313 & +0.46 & {\scriptsize [\tminus0.14, +1.11]} & \tminus1.64 & +0.71 & -- & \\ %
\addlinespace[4pt]
\multicolumn{10}{@{}l@{}}{\textbf{e\enspace Frame encoding + mixed projection loss on UniTB with 20+1 experts (top 4)}}\\
\textit{UniTB with 20+1 experts, top 4} & 0.7235 & 8.073 & 0.6397 &  &  &  &  & (66.7) & \\ %
+ frame encoding + mixed loss $w=10$\textsuperscript{$\ast$} & 0.7239 & 8.024 & 0.6408 & +0.06 & {\scriptsize [\tminus0.59, +0.66]} & \tminus0.61 & +0.17 & \tminus0.9 & {\scriptsize [\tminus5.0, +2.7]}\\ %
\textit{UniTB: 4+1 experts, top 2} & 0.7215 & 8.098 & 0.6379 &  &  &  &  &  & \\ %
The same model against UniTB\textsuperscript{$\ast$} & 0.7239 & 8.024 & 0.6408 & +0.34 & {\scriptsize [\tminus0.24, +0.91]} & \tminus0.91 & +0.45 & -- & \\ %
\bottomrule
\end{tabular}
\end{minipage}\end{center}

{\singlespacing\footnotesize\noindent $^\dagger$ 2,996-structure subset. $^\ast$ More than one setting differs from the baseline: in the mixed-loss rows, frame encoding and loss change together and the output is the space-group-projected prediction.\par}
\snote{Datasets and labels}{sn:data}
\paragraph{Training set.}
The released NextHAM training list has 12,000 lines and contains three kinds of duplicates: the same ID listed twice (2 entries), truncated Materials Project IDs whose structures belong to a full ID that is also in the training set (32 entries), and truncated IDs whose structures are in the test set (13 entries). MPHam-12K removes these $2+32+13=47$ entries and keeps the test set unchanged, leaving 11,953 structures with no duplicated structure and no structure shared with the test set.

\paragraph{Test set.}
The 3,000 test entries contain no repeated ID; two entries are truncated IDs whose structures are also in the test set (``\texttt{mp-63}'' is \texttt{mp-631520} and ``\texttt{mp-16504}'' is \texttt{mp-1650463}), and the 13 structures shared with the training list were removed from the training side. Results use the full 3,000-entry test set, as in the published benchmark; Supplementary Table~\ref{S-tab:nodup} reports the 2,998-entry set without the two truncated copies.

\paragraph{Additional training data.}
We added 17,289 structures selected from MP-20. All their elements lie within the 68 elements covered by NextHAM (H to Bi, excluding La--Lu), and none of these structures is in the NextHAM training or test sets. Their spin--orbit-coupled Hamiltonians were computed with ABACUS. The resulting MPHam-29K training set contains 29,242 structures ($11{,}953+17{,}289$).

\paragraph{DFT labels.}
Labels were computed with ABACUS using the settings in Appendix H of ref.~\citenum{SI-yin2025nextham}: PBE, PseudoDojo fully relativistic ONCV pseudopotentials, a 120 Ry cutoff, a $\Gamma$-centered $6\times6\times6$ mesh and density convergence of $10^{-6}$.

\providecommand{\tminus}{\textminus}
\begin{center}\begin{minipage}{\linewidth}
\captionof{table}{\textbf{Matrix errors without the two truncated test copies.} Each column averages the real spin-up--spin-up matrix MAE equally across its stated test set. The 2,998-entry set removes \texttt{mp-63} and \texttt{mp-16504}.}
\label{S-tab:nodup}
\centering\small
\renewcommand{\arraystretch}{1.1}
\setlength{\tabcolsep}{10pt}
\begin{tabular}{@{}lrr@{}}
\toprule
 & \multicolumn{2}{c}{MAE (meV), test structures}\\
\cmidrule(lr){2-3}
Model & 3,000 & 2,998\\
\midrule
\multicolumn{3}{@{}l@{}}{\textit{MPHam-12K}}\\
Dense & 0.813 & 0.813\\
PDQ-MoE & 0.785 & 0.785\\
PDQ-MoE + charge head & 0.753 & 0.753\\
\textbf{UniTB} & \textbf{0.722} & \textbf{0.722}\\
\midrule
\multicolumn{3}{@{}l@{}}{\textit{MPHam-29K}}\\
Dense & 0.559 & 0.559\\
\textbf{UniTB} & \textbf{0.493} & \textbf{0.493}\\
\addlinespace[3pt]
Dense, grid-free prior & 2.023 & 2.023\\
\textbf{UniTB-Flash} & \textbf{1.694} & \textbf{1.694}\\
\bottomrule
\end{tabular}
\end{minipage}\end{center}
\snote{Gauge alignment and graph conventions}{sn:protocol}
Main-text matrix errors (Eq.~(\ref{eq:matrix-metric})) are unaligned. The released NextHAM code fits one L2 shift per structure, independently for each predicted Hamiltonian or prior:
\begin{equation}
 \lambda_{\rm L2}=
 \frac{\operatorname{Re}\langle\widehat H-H,S\rangle}
 {\|S\|_{\rm F}^2+10^{-6}},\qquad
 e^{\rm aligned}=\mathrm{MAE}(\widehat H,H+\lambda_{\rm L2}S).
\end{equation}
The inner product uses the spinor matrix. With real spin-diagonal $S$, identical diagonal-spin errors and negligible other errors, the real-block reduction uses regularizer $5\times10^{-7}$ for the 200-structure sensitivity check. L1 alignment instead uses the weighted median of $(\widehat H-H)_a/S_a$ with weights $|S_a|$; entries with $S_a=0$ retain their original error.

On the 200-structure sensitivity subset, alignment using the reduced NextHAM L2 convention changes UniTB MAE from $0.7113$ to $0.7018$ meV (Supplementary Table~\ref{S-tab:gauge}).

\providecommand{\tminus}{\textminus}
\begin{center}\begin{minipage}{\linewidth}
\captionof{table}{\textbf{Matrix errors under gauge alignment.} Real spin-up--spin-up MAE, averaged equally over structures. The 200-structure rows use the reduced L2 convention described above.}
\label{S-tab:gauge}
\centering\small
\renewcommand{\arraystretch}{1.1}
\setlength{\tabcolsep}{8pt}
\begin{tabular}{@{}lrrrr@{}}
\toprule
 & \multicolumn{3}{c}{MAE (meV)} & \\
\cmidrule(lr){2-4}
Model & Unaligned & L2 & L1 & \shortstack[r]{L2 change\\(\%)}\\
\midrule
\multicolumn{5}{@{}l@{}}{\textit{3,000 structures}}\\
\quad $H^{(0)}$ & 4.717 & 4.274 & 4.055 & \\
\midrule
\multicolumn{5}{@{}l@{}}{\textit{200 band structures}}\\
\quad $H^{(0)}$ & 4.827 & 4.391 & 4.158 & \\
\quad Dense, MPHam-12K & 0.818 & 0.787 & 0.774 & \tminus3.68\\
\quad \textbf{UniTB, MPHam-12K} & \textbf{0.711} & \textbf{0.702} & \textbf{0.687} & \textbf{\tminus1.34}\\
\bottomrule
\end{tabular}
\end{minipage}\end{center}

NextHAM reports $2.834$ meV for the real diagonal spin block and $1.417$ meV for its Overall metric. Overall averages the complex error across all four spin blocks. With equal diagonal-spin errors and negligible off-diagonal-spin errors, its value is approximately half the real diagonal-block value. 

\subsection*{Graph and orbital masks}
NextHAM uses an $8$ \AA{} atom-pair cutoff. Our stored directed graph retains $d_{ij}<(r_{Z_i}+r_{Z_j})/2$, with radii $7.4085$ \AA{} for $59$ elements, $8.4668$ \AA{} for Ba, Br, Cs, Ga, Ge, Kr, Sr and Y, and $9.5252$ \AA{} for Rb. The same edge mask acts on graph indices and all physical arrays. The learned radial envelope is applied within this stored graph. These radii are twice the orbital cutoffs (7, 8 and 9 bohr), so the graph holds every atom pair whose orbitals overlap.

Outside the graph, kinetic, overlap and local-potential elements vanish, so the out-of-graph blocks of the DFT Hamiltonian, $H^{(0)}$ and $H_{\rm GF}$ reduce to the same nonlocal pseudopotential terms. In these nonmagnetic calculations, spin-flip blocks and imaginary parts also come only from this term; Supplementary Table~\ref{S-tab:protocol-map} lists the matrices and spectral treatment used in each evaluation.

\begin{center}\begin{minipage}{\linewidth}
\captionof{table}{\textbf{Matrices and settings behind each evaluation.} Prior, the Hamiltonian that the network corrects; $S$, source of the overlap; fixed blocks, source of the spin-flip and imaginary components and of the blocks outside the learned graph; projection, spectral projection of the residual relative to $H^{(0)}$ with $b=10$ eV; removed overlap directions, overlap eigenvalues removed before the eigensolve. Own, assembled by our prior code from the structure (Fig.~\ref{fig:prior}); DFT, exported by ABACUS. $^\dagger$Reconstruction default of Methods (``Hamiltonian reconstruction''). $^\ddagger$Shared by the DFT, UniTB and UniTB-Flash Hamiltonians of each test crystal, which also omit blocks outside the learned graph. The fixed blocks of $H^{(0)}$ and of the DFT Hamiltonian coincide outside the learned graph and in their spin-flip and imaginary parts (Supplementary Note~\ref{S-sn:protocol}).}
\label{S-tab:protocol-map}
\centering\footnotesize
\renewcommand{\arraystretch}{1.12}
\setlength{\tabcolsep}{3pt}
\begin{tabular}{@{}>{\raggedright\arraybackslash}p{3.3cm}>{\raggedright\arraybackslash}p{1.45cm}>{\raggedright\arraybackslash}p{1.35cm}>{\raggedright\arraybackslash}p{1.75cm}>{\raggedright\arraybackslash}p{1.8cm}>{\raggedright\arraybackslash}p{2.0cm}>{\raggedright\arraybackslash}p{1.9cm}@{}}
\toprule
Evaluation & Prior & $S$ & Fixed blocks & Projection & Removed overlap directions & $k$ sampling\\
\midrule
Matrix errors (Table~\ref{tab:benchmark}, Fig.~\ref{fig:periodic}, Supplementary Fig.~\ref{S-sfig:periodic-flash}) & $H^{(0)}$ / $H_{\rm GF}$ & not used & not scored & none (matrix entries) & --- & --- \\
Band errors, UniTB and $H^{(0)}$-prior dense models (Fig.~\ref{fig:projection}, Supplementary Fig.~\ref{S-sfig:bands}) & $H^{(0)}$ & DFT$^\dagger$ & $H^{(0)}$, none outside the learned graph & yes & $\lambda\le10^{-5}$, only if $S$ is near-singular & full paths; sampled paths for on/off \\
Band errors, UniTB-Flash and grid-free dense model (Supplementary Fig.~\ref{S-sfig:bands}) & $H_{\rm GF}$ & own & $H_{\rm GF}$ & no & $\lambda<2\times10^{-3}$ & full paths \\
Supercell optics and DC conductivity, UniTB (Fig.~\ref{fig:amorphous}b--d) & $H^{(0)}$ & DFT & $H^{(0)}$ & yes; velocities include its derivative & none (Cholesky of $S$) & $\Gamma$ \\
Supercell optics and DC conductivity, UniTB-Flash (Fig.~\ref{fig:amorphous}b--d) & $H_{\rm GF}$ & own & $H_{\rm GF}$ & no & $\lambda<2\times10^{-3}$ & $\Gamma$ \\
Supercell photoemission, UniTB (Fig.~\ref{fig:amorphous}e) & $H^{(0)}$ & DFT & $H^{(0)}$ & yes & none (Cholesky of $S$) & $\Gamma$ \\
Supercell photoemission, UniTB-Flash (Fig.~\ref{fig:amorphous}e) & $H_{\rm GF}$ & own & $H_{\rm GF}$ & no & $\lambda<2\times10^{-3}$ & $\Gamma$ \\
Test-set domains and Seebeck curves (Fig.~\ref{fig:screening}b, Fig.~\ref{fig:amorphous}f) & $H^{(0)}$ / $H_{\rm GF}$ & DFT$^\ddagger$ & spin--orbit from $H_{\rm GF}$$^\ddagger$ & UniTB only & $\lambda<2\times10^{-3}$ & property meshes \\
Library screen, first stage (UniTB-Flash) & $H_{\rm GF}$ & own & own & no & $\lambda<2\times10^{-3}$ & property meshes \\
Library screen, second stage (UniTB) & $H^{(0)}$ & own & own & yes; velocities include its derivative & $\lambda<2\times10^{-3}$ & property meshes \\
Library screen, DFT check & --- & DFT & DFT & no & $\lambda<2\times10^{-3}$ & property meshes \\
Timing (Fig.~\ref{fig:prior}b,c) & $H^{(0)}$ / $H_{\rm GF}$ & own & own & not timed & --- & --- \\
\bottomrule
\end{tabular}
\end{minipage}\end{center}

Species-specific orbitals embed into a padded $4s2p2d1f$ vocabulary with $27$ spatial orbitals. Valid species slots define onsite and ordered-pair hopping masks, intersected with the real spin-up--spin-up target. The configured common orbital mask determines the effective training subspace.
\snote{Training settings, training time and errors by element}{sn:training}
The added MP-20 cells contain at most 20 atoms. MPHam-12K uses dynamic batches capped at 32 structures: batches average 31.84 structures, giving 99.89 effective epochs over 37,500 updates. MPHam-29K uses a 96-structure cap and 91,800 updates, about 300 effective epochs. Supplementary Tables~\ref{S-tab:config_s} and~\ref{S-tab:config_s_modules} give the corresponding configurations.

The dense backbone uses separate onsite and hopping models with element-level tensor-product modulation and prior initialization (Supplementary Note~\ref{S-sn:arch}). DPA3-inspired and Switch-inspired variants use this work's router.
For MPHam-12K, the learning rate rises linearly from $10^{-5}$ to $10^{-2}$ over 2,000 updates, stays at $10^{-2}$ until update 21,938, decays by a cosine to $10^{-5}$ at update 33,750 and stays at $10^{-5}$ for the last 3,750 updates; the MPHam-29K schedule has the same form with the update counts of Supplementary Table~\ref{S-tab:config_s}. HybridMuon applies Muon updates (momentum 0.95, update scale 0.2, automatic update clipping) to weight and tensor-product matrices and AdamW updates ($\beta=0.98,0.999$) to biases, normalization, router and the remaining parameters, with weight decay 0.01 and a gradient-norm cap of 0.3. Each head minimizes the mean absolute error of its predicted matrix elements, without a gauge shift. Reported results use the final checkpoint. The 200 band-test structures are a random sample of the 3,000 test structures.
\providecommand{\tminus}{\textminus}
\begin{center}\begin{minipage}{\linewidth}
\captionof{table}{\textbf{Optimizer and loss settings of the two budgets.} HybridMuon combines Muon updates for selected matrix weights with AdamW updates for the remaining parameter groups. The onsite and hopping heads are trained independently; UniTB-Flash changes the prior and residual target.}
\label{S-tab:config_s}\label{S-tab:config_l}
\centering\small
\renewcommand{\arraystretch}{1.1}
\setlength{\tabcolsep}{10pt}
\begin{tabular}{@{}lll@{}}
\toprule
Training setting & MPHam-12K & MPHam-29K\\
\midrule
Training structures & 11,953 & 29,242\\
Optimizer updates & 37,500 & 91,800\\
Effective epochs & 99.89 & about 300\\
Dynamic block-batch cap (structures) & 32 & 96\\
Warmup updates & 2,000 & 5,000\\
Schedule total / decay updates & 33,750 / 11,812 & 82,620 / 28,917\\
Prior & $H^{(0)}$ & $H^{(0)}$ or $H_{\rm GF}$\\
Residual target & $H-H^{(0)}$ & $H-H^{(0)}$ or $H-H_{\rm GF}$\\
\midrule
\multicolumn{3}{@{}l@{}}{\textit{Common to both budgets and both heads}}\\
Optimizer & \multicolumn{2}{l}{HybridMuon}\\
Learning rate & \multicolumn{2}{l}{0.01}\\
Weight decay & \multicolumn{2}{l}{0.01}\\
Gradient norm cap & \multicolumn{2}{l}{0.3}\\
Learning-rate schedule & \multicolumn{2}{l}{Warmup--stable--decay}\\
Explicit edge cap & \multicolumn{2}{l}{None}\\
Objective & \multicolumn{2}{l}{Absolute matrix loss; no gauge shift}\\
Labels / prediction mask & \multicolumn{2}{l}{SOC labels; common real $\uparrow\uparrow$ mask}\\
Readout and reconstruction & \multicolumn{2}{l}{Equivariant tensor product; direct blocks}\\
Readout hidden widths & \multicolumn{2}{l}{128, 128, 128}\\
Onsite symmetrization & \multicolumn{2}{l}{Enabled}\\
\bottomrule
\end{tabular}
\end{minipage}\end{center}
\providecommand{\tminus}{\textminus}
\begin{table}[htbp]
\caption{\textbf{Module settings of the trained models.} Settings of UniTB at both training budgets and of UniTB-Flash at MPHam-29K. Each independently trained model has its own response head. The onsite distance mask is $d<10^{-6}$ \AA. Hopping retains graph edges with $d\geq10^{-6}$ \AA{} and has no upper cutoff beyond the stored graph (Supplementary Note~\ref{S-sn:protocol}).}
\label{S-tab:config_s_modules}\label{S-tab:config_l_modules}
\centering\small
\renewcommand{\arraystretch}{1.05}
\setlength{\tabcolsep}{8pt}
\begin{tabular}{@{}ll@{}}
\toprule
Setting & Value (onsite and hopping models unless stated)\\
\midrule
\multicolumn{2}{@{}l@{}}{\textit{Backbone}}\\
Interaction layers & 3\\
Channels at angular degrees 0--6 & 128, 24, 16, 16, 32, 24, 48\\
Latent hidden widths; output width & 200, 128; 128\\
Radial basis size; readout channels & 128; 32\\
Element conditioning & Full tensor products in interaction layers and the output readout\\
\addlinespace[3pt]
\multicolumn{2}{@{}l@{}}{\textit{Prior-routed experts}}\\
Routed experts; selected; shared map & 4; 2; 1\\
Expert bases & Shared low-rank bases, rank capped at 64\\
Routing inputs & Ordered bond type and rotation-invariant prior descriptors\\
Router logits & Cosine similarity, scale 10\\
\addlinespace[3pt]
\multicolumn{2}{@{}l@{}}{\textit{Prior perturbations and conditioning}}\\
Prior perturbations & Sampled by type within each graph; clean residual targets\\
Residual-scale multiplier & 0.5 (onsite), 5 (hopping)\\
Time conditioning & Constant \textit{t}\,=\,0 embedding\\
Auxiliary supervision & No auxiliary context or loss\\
\addlinespace[3pt]
\multicolumn{2}{@{}l@{}}{\textit{Charge-response head}}\\
Potential readout & Per-atom, two layers of width 64\\
Charge response & Neutrality-constrained response plus local readout\\
Response hidden / element width & 64 / 8\\
Response scale; Gaussian width & 0.1; 1.2 \AA\\
Minimum hardness & 5 (kernel energy convention)\\
Reciprocal cutoff & 4 \AA\textsuperscript{\tminus1}\\
Kernel at inference & Ewald decomposition, shared by both models\\
\addlinespace[3pt]
\multicolumn{2}{@{}l@{}}{\textit{Predicted blocks}}\\
Onsite model & Onsite blocks\\
Hopping model & Non-onsite edges retained in the stored graph\\
\bottomrule
\end{tabular}
\end{table}

\subsection*{Training time}
Supplementary Table~\ref{S-tab:train_cost} gives epoch and total training times; an epoch is about 375 updates for MPHam-12K and 306 for MPHam-29K. Timings use strict FP32 with TF32 disabled, HybridMuon in fast mode and CUDA graphs. With two GPUs per head, each forms dynamic batches of at most half the global cap (16 and 48 structures), and gradients are averaged. Timed layouts are one H200 shared by both heads, one A100 per head and two A100 GPUs per head for MPHam-12K; MPHam-29K uses one H200 or two A100 GPUs per head. With two A100 GPUs per head, peak reserved memory per GPU is 20 GiB for MPHam-12K and 41 GiB for MPHam-29K, against 35.5 and 74 GiB with one GPU per head.
\providecommand{\tminus}{\textminus}
\begin{center}\begin{minipage}{\linewidth}
\captionof{table}{\textbf{Training time of UniTB.} Time is seconds per epoch multiplied by the number of epochs per head. Two GPUs per head use data parallelism; ranges span the onsite and hopping heads. GPUs are H200 (141 GB) and A100 (80 GB).}
\label{S-tab:train_cost}
\centering\small
\renewcommand{\arraystretch}{1.1}
\setlength{\tabcolsep}{10pt}
\begin{tabular}{@{}lrr@{}}
\toprule
Hardware and layout & \shortstack[r]{Seconds per\\epoch} & \shortstack[r]{All epochs\\(h)}\\
\midrule
\multicolumn{3}{@{}l@{}}{\textit{MPHam-12K, 100 epochs per head}}\\
1 H200, both heads on the GPU & 501 & 13.9\\
2 A100, one head per GPU & 512--519 & 14.2--14.4\\
\textbf{4 A100, two per head} & \textbf{353--356} & \textbf{9.8--9.9}\\
\midrule
\multicolumn{3}{@{}l@{}}{\textit{MPHam-29K, 300 epochs per head}}\\
2 H200, one head per GPU & 440--451 & 36.6--37.6\\
\textbf{4 A100, two per head} & \textbf{495--497} & \textbf{41.3--41.4}\\
\bottomrule
\end{tabular}
\end{minipage}\end{center}

\subsection*{Matrix error by element}
Main-text Fig.~\ref{fig:periodic} resolves the MAE of the MPHam-29K UniTB model by element. The attribution is structure-level, so a structure contributes to every element it contains, and the average uses the 2,431 test structures that the Materials Project labels as nonmagnetic, which contain all 68 elements of the dataset. Supplementary Fig.~\ref{S-sfig:periodic-flash} does the same for the MPHam-29K UniTB-Flash model, whose averages are higher than those of UniTB for every element.

\begin{center}\begin{minipage}{\linewidth}
\centering\includegraphics[width=\textwidth]{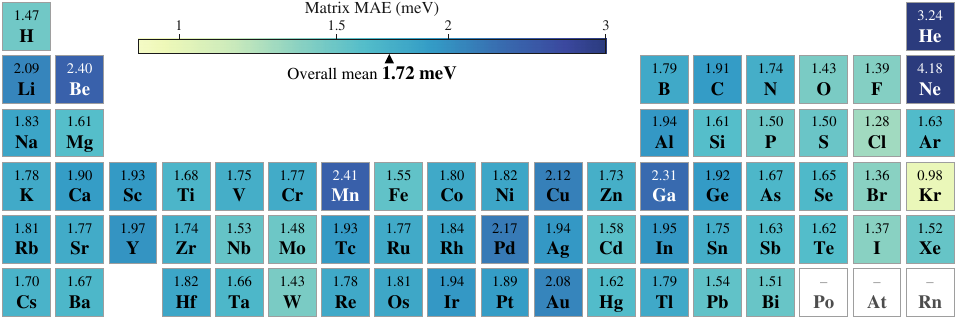}
\captionof{figure}{\textbf{Matrix-element errors of UniTB-Flash across the periodic table.} Values are for UniTB-Flash (MPHam-29K) on the 2,431 nonmagnetic test structures. Each tile averages the matrix-element MAE over the structures that contain the element, so a structure counts once for each of its elements. Dashes mark elements not in the test set. Logarithmic color scale, with limits three times those of main-text Fig.~\ref{fig:periodic}. Values beyond the color range take the end color.}
\label{S-sfig:periodic-flash}
\end{minipage}\end{center}
\snote{Charge-head diagnostics}{sn:charge}

The head variants and their controls are compared in Supplementary Table~\ref{S-tab:charge-controls}. The implemented head handles neutral, fully periodic or isolated structures; the local readout has width 64 and the element embedding dimension 8 (Supplementary Table~\ref{S-tab:config_s_modules}).

The diagnostic in Fig.~\ref{fig:charge}a uses a separately trained equivariant dense model without the charge head, with its own split of $3{,}063$ training and $256$ test structures. Its site-potential residual RMS is $6.4$ meV on training and $54.3$ meV on test structures; test errors are $28.3$ meV for formulas present in training and $59.4$ meV for new formulas. Quartiles of local-feature distance to training give $14.5$, $34.8$, $57.2$ and $97.2$ meV. Test structures containing O or N give $88.3$ and $115.0$ meV, against $41.8$ and $48.6$ meV for those without them. Figure~\ref{fig:charge}b groups the 3,000 test structures by atoms per cell into bins of 1--5, 6--12, 13--22 and 23--79 atoms (656, 851, 755 and 738 structures), with paired structure resampling within each bin.
\snote{Spectral projection: properties and protocols}{sn:spectral}
Main-text Eq.~(\ref{eq:clip-update}) defines the projection. For a full-rank positive-definite overlap, $S=LL^{\dagger}$, $X=L^{-\dagger}$ and $A=X^{\dagger}\widehat\Delta X=U\diag(\mu)U^{\dagger}$. Writing $f_b(A)=U\diag(\clip(\mu,-b,b))U^{\dagger}$, reconstruction in this case gives $\widehat H_b=H^{(0)}+X^{-\dagger}f_b(A)X^{-1}$. More generally, an orthogonalizer $X$ of a fixed retained overlap subspace satisfies $X^{\dagger}SX=I$, and the update of main-text Eq.~(\ref{eq:clip-update}) gives $X^{\dagger}\widehat H_bX=X^{\dagger}H^{(0)}X+f_b(A)$. Any other orthogonalizer of that same subspace is $Y=XQ$ for unitary $Q$. Functional calculus gives $f_b(Q^{\dagger}AQ)=Q^{\dagger}f_b(A)Q$, proving independence of orthogonalization within the subspace. When the input spectrum lies inside $[-b,b]$, the map is the identity. Otherwise, $-bI\preceq f_b(A)\preceq bI$ gives the ordered-eigenvalue bound
\begin{equation}
 \varepsilon_a(H^{(0)},S)-b\leq\varepsilon_a(\widehat H_b,S)
 \leq\varepsilon_a(H^{(0)},S)+b.
 \label{S-eq:weyl}
\end{equation}
The bound compares the two spectra on the same nonsingular overlap subspace, the retained subspace when the canonical solver removes directions.

For a fixed basis and ionic nonlocal operator, a local effective-potential residual has matrix elements $\Delta_{ab}=\langle\phi_a|\Delta V_{\rm Hxc}|\phi_b\rangle$. Its generalized Rayleigh quotient obeys
\begin{equation}
 \frac{c^{\dagger}\Delta c}{c^{\dagger}Sc}
 =\frac{\int\Delta V_{\rm Hxc}|\psi_c|^2\,\dd\mathbf r}{\int|\psi_c|^2\,\dd\mathbf r}
 \in[\inf\Delta V_{\rm Hxc},\sup\Delta V_{\rm Hxc}],
 \label{S-eq:potential-bound}
\end{equation}
where $\psi_c=\sum_a c_a\phi_a$. This motivates a bound in the overlap metric. All reported projected spectra use $b=10$ eV.

UniTB-Flash is not projected (main text). Supplementary Table~\ref{S-tab:flash-bscan} scans a projection relative to $H_{\rm GF}$ on 1,592 test structures.

\begin{center}\begin{minipage}{\linewidth}
\captionof{table}{\textbf{Projection relative to $H_{\rm GF}$ for UniTB-Flash.} Recall and window MAE use the 135 of these structures with DFT gaps in $[0.1,1.0]$ eV. Outliers (absolute gap error above 0.5 eV) and clipped structures (any residual eigenvalue moved to $\pm b$) are counted over all 1,592. Bounds of 20--500 eV lie between the rows shown.}
\label{S-tab:flash-bscan}
\centering\small
\begin{tabular*}{\linewidth}{@{}l@{\extracolsep{\fill}}rrrr@{}}
\toprule
$b$ (eV) & Recall & Outliers & Window MAE (meV) & Structures clipped\\
\midrule
No projection & 0.926 & 17 & 91 & 0\\
10 & 0.037 & $1{,}130$ & $1{,}730$ & $1{,}577$\\
$1{,}000$ & 0.696 & 169 & 329 & 642\\
\bottomrule
\end{tabular*}
\end{minipage}\end{center}

\subsection*{Band and state-count protocols}
Methods (``Band metrics'') defines $\mathrm{fw}_{10}$ and the strict and wide spurious-state criteria.

The overlap solver screens near-singular matrices and uses a canonical fallback retaining overlap eigenvalues above $10^{-5}$. Missing high-energy dimensions are padded at $10^4$ eV. The sampled comparison has $16$ fallback wavevectors per model and label among $15{,}459$, with zero removed directions. 

\subsection*{Label residual spectra}
For 3,063 training structures at 27 sampled $\mathbf k$ points, the stored real spin-up--spin-up label residual has generalized eigenvalues within $\pm10$ eV (maximum absolute value, $8.47$ eV).

The 200-structure sampled band paths give label residual eigenvalues in $[-5.81,+6.15]$ eV. With $b=10$ eV, the full-path DFT-label evaluation projects zero of $36{,}043$ wavevectors.
\snote{Band endpoints and projection comparisons}{sn:bands}
Supplementary Tables~\ref{S-tab:bands_s} and~\ref{S-tab:bands_l} report MPHam-12K and MPHam-29K separately and give the band columns of main-text Table~\ref{tab:benchmark}, and Supplementary Fig.~\ref{S-fig:band-sequence} shows the site-channels-first sequence. Supplementary Table~\ref{S-tab:bands_s} also lists the PDQ-MoE-first rows of that table, the prior $H^{(0)}$ without a network (mean 745.63 meV, median 617.60 meV) and LEM and SLEM, trained with the same $H^{(0)}$ prior and data exposure as the MPHam-12K models; the mean $\mathrm{fw}_{10}$ values of LEM and SLEM are 88.11 and 80.62 meV, with medians of 62.08 and 48.53 meV. For MPHam-29K, the UniTB-versus-dense paired median is $-2.23$ meV with interval $[-4.16,-0.11]$ meV; the paired median of UniTB-Flash against the grid-free dense model is $-4.13$ meV with interval $[-7.28,-1.59]$ meV.

\begin{center}\begin{minipage}{\linewidth}
\centering\includegraphics[width=\textwidth]{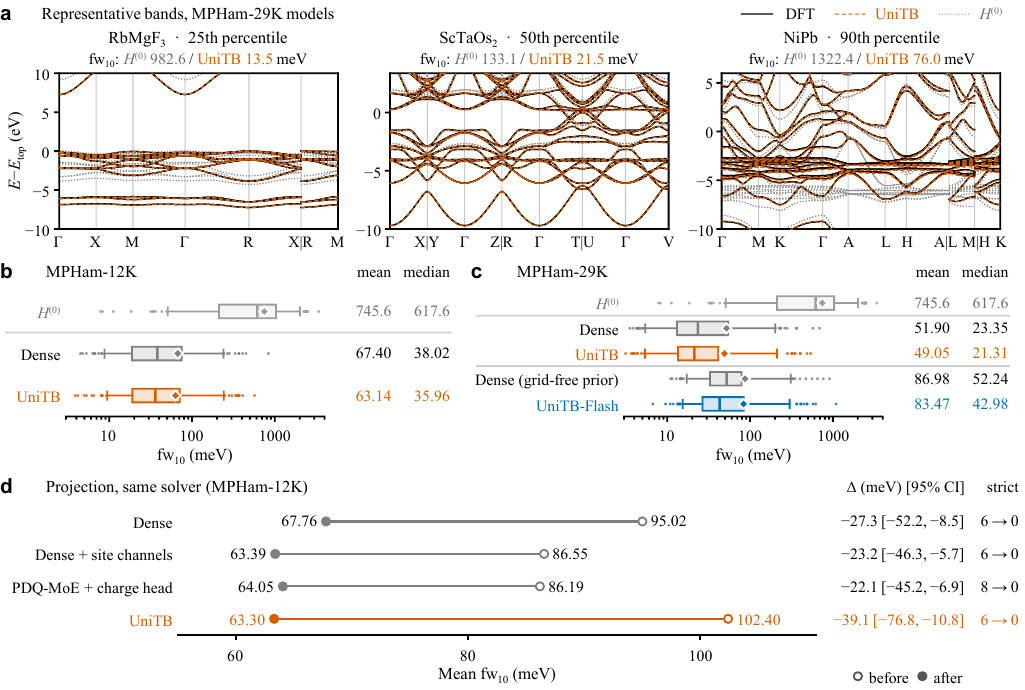}
\captionof{figure}{\textbf{Band structures and band errors.} \textbf{a},~Bands of three test cells with at most eight atoms, near the 25th, 50th and 90th percentiles of the UniTB $\mathrm{fw}_{10}$ over the 200 band-test structures. Each curve is aligned to its $N_{\rm occ}$-th band maximum. \textbf{b},\textbf{c},~Distributions of $\mathrm{fw}_{10}$ over the 200 structures for the MPHam-12K (\textbf{b}) and MPHam-29K (\textbf{c}) models. The $H^{(0)}$-prior models are projected at $b=10$~eV, and the grid-free models use overlap truncation. Boxes show quartiles and median, whiskers the 5th--95th percentiles, dots the remaining structures and diamonds the means. \textbf{d},~Mean $\mathrm{fw}_{10}$ of four MPHam-12K models before and after projection, from one solver on 15,459 sampled wavevectors of the 200 structures. Columns give the paired change with its $95\%$ interval and the number of structures with a strict spurious-state detection.}
\label{S-sfig:bands}
\end{minipage}\end{center}

\providecommand{\tminus}{\textminus}
\begin{center}\begin{minipage}{\linewidth}
\captionof{table}{\textbf{Band errors at MPHam-12K.} 200 structures after projection at \textit{b}\,=\,10 eV. Paired changes are from the preceding row, with 95\% intervals. Wide and strict diagnostics are defined in main-text Methods (Band metrics); strict detections and failed cells are 0 of 200 in every model row. LEM and SLEM\cite{SI-zhouyin2025slem} use the same $H^{(0)}$ prior and data exposure. The PDQ-MoE, LEM, SLEM and $H^{(0)}$ rows give endpoint statistics only; $H^{(0)}$ is the prior without a network.}
\label{S-tab:bands_s}
\centering\small
\renewcommand{\arraystretch}{1.1}
\setlength{\tabcolsep}{7pt}
\begin{tabular}{@{}lrrrrr@{\hspace{3pt}}r@{}}
\toprule
 & \multicolumn{2}{c}{$\mathrm{fw}_{10}$ (meV)} & Wide cells & \multicolumn{3}{c}{Paired change (meV)}\\
\cmidrule(lr){2-3}\cmidrule(lr){5-7}
MPHam-12K model & Mean & Median & (of 200) & Mean & \multicolumn{2}{c}{95\% CI}\\
\midrule
Dense & 67.40 & 38.02 & 2 & & &\\
Dense + site channels & 62.97 & 35.24 & 1 & \tminus4.42 & [\tminus8.46, & \tminus0.66]\\
PDQ-MoE + site channels & 66.13 & 37.08 & 3 & +3.16 & [\tminus1.57, & +7.80]\\
PDQ-MoE + charge head + site channels & 63.38 & 32.83 & 2 & \tminus2.76 & [\tminus7.37, & +2.11]\\
\textbf{UniTB} & \textbf{63.14} & \textbf{35.96} & \textbf{2} & \textbf{\tminus0.23} & \textbf{[\tminus4.31,} & \textbf{+3.79]}\\
\midrule
UniTB versus dense & & & & \tminus4.25 & [\tminus9.58, & +0.71]\\
\midrule
\multicolumn{7}{@{}l@{}}{\textit{Further rows of main-text Table~\ref{tab:benchmark} (PDQ-MoE first, without site channels)}}\\
PDQ-MoE & 66.84 & 39.07 & 3 & & &\\
PDQ-MoE + charge head & 64.26 & 33.41 & 1 & & &\\
\midrule
\multicolumn{7}{@{}l@{}}{\textit{Prior alone and external baselines, same prior and data exposure}}\\
$H^{(0)}$ (our labels) & 745.63 & 617.60 & & & &\\
LEM & 88.11 & 62.08 & 6 & & &\\
SLEM & 80.62 & 48.53 & 4 & & &\\
\bottomrule
\end{tabular}
\end{minipage}\end{center}
\providecommand{\tminus}{\textminus}
\begin{center}\begin{minipage}{\linewidth}
\captionof{table}{\textbf{Band errors at MPHam-29K.} 200 structures, after projection at \textit{b}\,=\,10 eV for the $H^{(0)}$-prior models and after overlap truncation for the grid-free models. Each model is paired with a dense model trained with the same data, budget and prior. Strict detections are 0 of 200 for the $H^{(0)}$-prior models, and no cell failed.}
\label{S-tab:bands_l}
\centering\small
\renewcommand{\arraystretch}{1.1}
\setlength{\tabcolsep}{9pt}
\begin{tabular}{@{}lrrrr@{\hspace{3pt}}r@{}}
\toprule
 & \multicolumn{2}{c}{$\mathrm{fw}_{10}$ (meV)} & \multicolumn{3}{c}{Paired change (meV)}\\
\cmidrule(lr){2-3}\cmidrule(lr){4-6}
MPHam-29K model & Mean & Median & Mean & \multicolumn{2}{c}{95\% CI}\\
\midrule
\multicolumn{6}{@{}l@{}}{\textit{Prior $H^{(0)}$}}\\
\quad Dense & 51.90 & 23.35 & & &\\
\quad \textbf{UniTB} & \textbf{49.05} & \textbf{21.31} & \textbf{\tminus2.85} & \textbf{[\tminus7.84,} & \textbf{+2.42]}\\
\addlinespace[3pt]
\multicolumn{6}{@{}l@{}}{\textit{Prior $H_{\rm GF}$ (grid-free)}}\\
\quad Dense & 86.98 & 52.24 & & &\\
\quad \textbf{UniTB-Flash} & \textbf{83.47} & \textbf{42.98} & \textbf{\tminus3.51} & \textbf{[\tminus15.29,} & \textbf{+10.36]}\\
\bottomrule
\end{tabular}
\end{minipage}\end{center}

\begin{center}\begin{minipage}{\linewidth}

\centering\includegraphics[width=\textwidth]{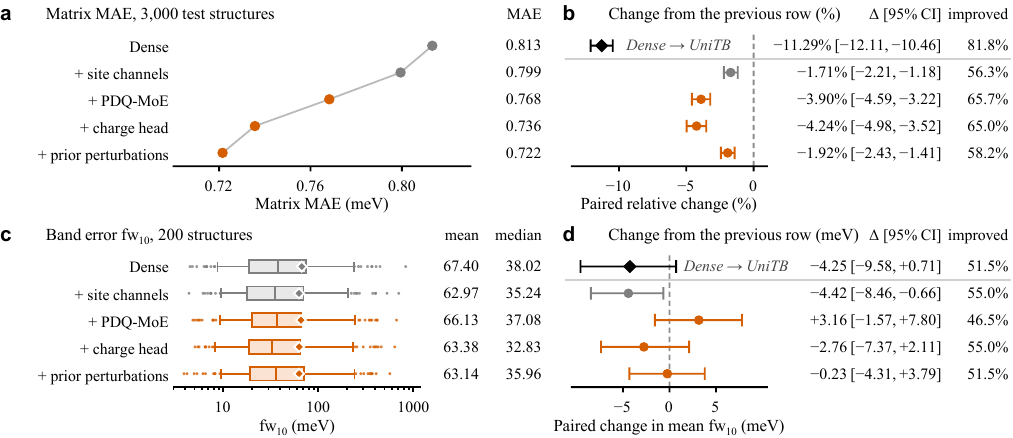}
\captionof{figure}{\textbf{From the dense model to UniTB.} MPHam-12K models. Each row adds one component to the row above, and the last row is UniTB. Panels \textbf{a} and \textbf{b} show matrix errors, and \textbf{c} and \textbf{d} band errors after projection. \textbf{b},\textbf{d},~Paired changes from the previous row, with $95\%$ intervals. The first row compares UniTB with the dense model, and improved is the share of structures with a lower error. \textbf{c},~Boxes show quartiles and median, whiskers the 5th--95th percentiles, dots the remaining structures and diamonds the means.}
\label{S-fig:band-sequence}

\end{minipage}\end{center}

\subsection*{Projection comparisons}
Supplementary Table~\ref{S-tab:abl-projection} compares each prediction with and without projection, under one solver on $15{,}459$ sampled wavevectors (panel a) and on the full paths (panels b and c); Supplementary Table~\ref{S-tab:abl-projection-threshold} gives the threshold sensitivity. Both tables are part of Supplementary Note~\ref{S-sn:ablations}.
\snote{Supercell selection, property protocol and results}{sn:amorphous}
\subsection*{Selection}
The a-Si-24 selections retain the source categories\cite{SI-rosset2025asi24}: four continuous random networks with cooling rates $10^{10}$--$10^{13}$ K/s, one paracrystal and one diamond control, all with $216$ atoms. Each random-network or paracrystalline selection is closest to its category's median density. The diamond control has the largest author-provided crystalline-similarity descriptor in its pool. GST-ACE-24 cells span $180$--$198$ atoms, multiple compositions and preparation categories, with density-median selections and two additional density-quartile selections from a GST pool\cite{SI-zhou2025gstace24}. The selection contains no repeated source frames. Supplementary Tables~\ref{S-tab:amorphous_selection_inventory} and~\ref{S-tab:amorphous_selection_inventory_gst} preserve frame identities and categories.

\providecommand{\tminus}{\textminus}
\begin{center}\begin{minipage}{\linewidth}
\captionof{table}{\textbf{Selection of the Si test cells.} CRN, continuous random network. Frame indices are zero-based within each source file, and the density rules describe the selection within each source category.}
\label{S-tab:amorphous_selection_inventory}
\centering\small
\renewcommand{\arraystretch}{1.15}
\setlength{\tabcolsep}{8pt}
\begin{tabular}{@{}lr>{\raggedright\arraybackslash}p{3.55in}@{}}
\toprule
Cell (source category) & Frame & Selection rule\\
\midrule
\multicolumn{3}{@{}l@{}}{\textit{Library a-Si-24; every cell is Si$_{216}$ with 216 atoms}}\\
\texttt{CRN}, $10^{10}$ K/s & 15 & Nearest density median of the author CRN pool; independent parent network relative to $10^{11}$ K/s\\
\texttt{CRN}, $10^{11}$ K/s & 190 & Nearest density median of the author CRN pool\\
\texttt{Paracrystalline} & 523 & Nearest density median of the author paracrystalline pool\\
\texttt{Diamond}, crystal control & 81 & Highest author \texttt{soap\_sim\_cSi} descriptor in the Diamond pool\\
\texttt{CRN}, $10^{12}$ K/s & 489 & Nearest density median of the author CRN pool (170 frames)\\
\texttt{CRN}, $10^{13}$ K/s & 723 & Nearest density median of the author CRN pool (198 frames)\\
\bottomrule
\end{tabular}
\end{minipage}\end{center}

\begin{center}\begin{minipage}{\linewidth}
\captionof{table}{\textbf{Selection of the Ge--Sb--Te test cells.} The source categories distinguish PBE, PBEsol and AIMD preparations: \texttt{standard\_pbe\_ace} is the ACE iterative MD and active-learning family, and \texttt{domain\_pbesol} the domain and phase-change training family. Quantiles are taken within the composition and category pool. Frame indices are zero-based within each source file.}
\label{S-tab:amorphous_selection_inventory_gst}
\centering\small
\renewcommand{\arraystretch}{1.15}
\setlength{\tabcolsep}{8pt}
\begin{tabular}{@{}lrlrlr@{}}
\toprule
Composition & $N_{\rm at}$ & Source category & Frame & Density selection & \shortstack[r]{Pool\\(frames)}\\
\midrule
\multicolumn{6}{@{}l@{}}{\textit{Library GST-ACE-24}}\\
$\mathrm{Ge}_{40}\mathrm{Sb}_{40}\mathrm{Te}_{100}$ & 180 & \texttt{standard\_pbe\_ace} & 3674 & nearest median &\\
$\mathrm{Ge}_{44}\mathrm{Sb}_{44}\mathrm{Te}_{110}$ & 198 & \texttt{standard\_pbesol} & 1746 & nearest median &\\
$\mathrm{Ge}_{44}\mathrm{Sb}_{44}\mathrm{Te}_{110}$ & 198 & \texttt{domain\_pbesol} & 2224 & nearest median &\\
$\mathrm{Ge}_{44}\mathrm{Sb}_{44}\mathrm{Te}_{110}$ & 198 & \texttt{aimd\_pbe} & 2738 & nearest median & 6\\
\addlinespace[3pt]
$\mathrm{Ge}_{90}\mathrm{Te}_{90}$ & 180 & \texttt{standard\_pbe\_ace} & 3367 & quantile 0.5 & 60\\
$\mathrm{Ge}_{72}\mathrm{Sb}_{18}\mathrm{Te}_{99}$ & 189 & \texttt{standard\_pbe\_ace} & 3342 & quantile 0.5 & 60\\
$\mathrm{Ge}_{51}\mathrm{Sb}_{34}\mathrm{Te}_{102}$ & 187 & \texttt{standard\_pbe\_ace} & 3322 & quantile 0.5 & 60\\
$\mathrm{Ge}_{26}\mathrm{Sb}_{52}\mathrm{Te}_{104}$ & 182 & \texttt{standard\_pbe\_ace} & 3654 & quantile 0.5 & 60\\
$\mathrm{Ge}_{15}\mathrm{Sb}_{60}\mathrm{Te}_{105}$ & 180 & \texttt{standard\_pbe\_ace} & 3267 & quantile 0.5 & 60\\
$\mathrm{Sb}_{72}\mathrm{Te}_{108}$ & 180 & \texttt{standard\_pbe\_ace} & 3604 & quantile 0.5 & 60\\
$\mathrm{Ge}_{40}\mathrm{Sb}_{40}\mathrm{Te}_{100}$ & 180 & \texttt{standard\_pbe\_ace} & 3668 & quantile 0.25 & 59\\
$\mathrm{Ge}_{40}\mathrm{Sb}_{40}\mathrm{Te}_{100}$ & 180 & \texttt{standard\_pbe\_ace} & 3670 & quantile 0.75 & 59\\
\bottomrule
\end{tabular}
\end{minipage}\end{center}

\subsection*{Localization descriptors}
Methods gives the shared property protocol. L\"owdin atomic weights are $w_{n,a}=\sum_{i\in a,\sigma}|(S^{1/2}c_n)_{i\sigma}|^2$, normalized per state and averaged within Kramers pairs. The inverse participation ratio is $\mathrm{IPR}_n=\sum_aw_{n,a}^2$. The mobility-gap criterion uses three times the median DFT IPR in the inner-band window $1$--$2$ eV from its edge, with the same threshold for each route on a given structure. Localized states are sought within $1$ eV of either band edge. The ISWO calculation uses overlap-squared orbital-pair contributions with L\"owdin weights and averages within $0.3$ eV of the band edges. The separately reported participation-atom count uses Mulliken weights. Conductivity error is the absolute base-ten log ratio, optical/localization errors are absolute relative differences, and energy-width errors are absolute differences.

\subsection*{Property errors and dense comparison}
Supplementary Table~\ref{S-tab:amorphous_l} summarizes the errors of the Ge--Sb--Te cells, Supplementary Table~\ref{S-tab:gst-properties} lists further Ge--Sb--Te observables, and Supplementary Table~\ref{S-tab:amorphous-dense} compares each model with the dense model of the same prior.

\subsection*{Per-cell errors and valence-band spectra}
Supplementary Table~\ref{S-tab:percell} lists the errors of every cell. UniTB has the lower shape error in all six Si cells and in nine of the twelve Ge--Sb--Te cells; for the optical constants and the DC conductivity of the Ge--Sb--Te cells, the mean errors of the two models lie within one percentage point and 0.03 decades of each other.

\providecommand{\tminus}{\textminus}
\begin{center}\begin{minipage}{\linewidth}
\captionof{table}{\textbf{Property errors of every supercell.} Errors are against same-cell DFT: absolute relative errors of $n(0.8\,\mathrm{eV})$ and $\alpha(1.5\,\mathrm{eV})$, absolute errors of $\log_{10}\sigma_{\rm DC}$ (Gaussian broadening) and XPS shape errors (half the integrated absolute difference between area-normalized valence-band spectra). U, UniTB; F, UniTB-Flash; CRN, continuous random network with its cooling rate. Ge--Sb--Te cells are identified as in Supplementary Table~\ref{S-tab:amorphous_selection_inventory_gst}; $^{a}$, the cell of main-text Fig.~\ref{fig:amorphous}a. For the Si cells, the XPS shape errors are listed. Protocols are in Methods.}
\label{S-tab:percell}
\centering\small
\renewcommand{\arraystretch}{1.08}
\setlength{\tabcolsep}{4pt}
\begin{tabular}{@{}lrrrrrrrr@{}}
\toprule
 & \multicolumn{2}{c}{$n(0.8\,\mathrm{eV})$ (\%)} & \multicolumn{2}{c}{$\alpha(1.5\,\mathrm{eV})$ (\%)} & \multicolumn{2}{c}{$\log_{10}\sigma_{\rm DC}$ (dec.)} & \multicolumn{2}{c}{XPS shape (\%)}\\
\cmidrule(lr){2-3}\cmidrule(lr){4-5}\cmidrule(lr){6-7}\cmidrule(lr){8-9}
Cell & U & F & U & F & U & F & U & F\\
\midrule
\multicolumn{9}{@{}l@{}}{\textit{Ge--Sb--Te, 180--198 atoms}}\\
\quad Ge$_{40}$Sb$_{40}$Te$_{100}$, median$^{a}$ & 1.50 & 2.01 & 1.85 & 0.29 & 0.105 & 0.089 & 1.12 & 1.36\\
\quad Ge$_{40}$Sb$_{40}$Te$_{100}$, quantile 0.25 & 2.32 & 2.92 & 2.14 & 0.15 & 0.157 & 0.133 & 1.34 & 1.42\\
\quad Ge$_{40}$Sb$_{40}$Te$_{100}$, quantile 0.75 & 1.83 & 1.33 & 1.84 & 0.58 & 0.075 & 0.011 & 0.60 & 0.87\\
\quad Ge$_{44}$Sb$_{44}$Te$_{110}$, standard PBEsol & 1.27 & 2.25 & 1.21 & 0.69 & 0.019 & 0.022 & 0.99 & 1.70\\
\quad Ge$_{44}$Sb$_{44}$Te$_{110}$, domain PBEsol & 1.23 & 1.57 & 0.81 & 0.94 & 0.005 & 0.061 & 0.91 & 0.95\\
\quad Ge$_{44}$Sb$_{44}$Te$_{110}$, AIMD PBE & 2.78 & 2.24 & 1.92 & 0.06 & 0.047 & 0.013 & 1.14 & 0.99\\
\quad Ge$_{90}$Te$_{90}$ & 1.43 & 1.81 & 2.85 & 1.77 & 0.126 & 0.125 & 1.62 & 1.21\\
\quad Ge$_{72}$Sb$_{18}$Te$_{99}$ & 1.48 & 1.52 & 1.51 & 1.99 & 0.175 & 0.037 & 1.20 & 1.93\\
\quad Ge$_{51}$Sb$_{34}$Te$_{102}$ & 1.40 & 3.53 & 1.11 & 1.46 & 0.082 & 0.066 & 0.90 & 1.15\\
\quad Ge$_{26}$Sb$_{52}$Te$_{104}$ & 2.38 & 2.31 & 1.78 & 0.42 & 0.064 & 0.064 & 0.80 & 0.91\\
\quad Ge$_{15}$Sb$_{60}$Te$_{105}$ & 2.08 & 1.23 & 1.58 & 0.31 & 0.093 & 0.059 & 0.62 & 0.68\\
\quad Sb$_{72}$Te$_{108}$ & 1.56 & 0.99 & 2.15 & 0.79 & 0.066 & 0.065 & 0.74 & 0.65\\
\quad \textit{Mean} & 1.77 & 1.98 & 1.73 & 0.79 & 0.084 & 0.062 & 1.00 & 1.15\\
\addlinespace[3pt]
\multicolumn{9}{@{}l@{}}{\textit{Si, 216 atoms}}\\
\quad CRN, $10^{10}$ K\,s$^{-1}$ & & & & & & & 1.43 & 3.70\\
\quad CRN, $10^{11}$ K\,s$^{-1}$ & & & & & & & 1.31 & 2.41\\
\quad CRN, $10^{12}$ K\,s$^{-1}$ & & & & & & & 0.72 & 2.08\\
\quad CRN, $10^{13}$ K\,s$^{-1}$ & & & & & & & 1.18 & 3.66\\
\quad Paracrystalline & & & & & & & 1.62 & 3.35\\
\quad Diamond control & & & & & & & 0.78 & 1.44\\
\quad \textit{Mean} & & & & & & & 1.17 & 2.77\\
\bottomrule
\end{tabular}
\end{minipage}\end{center}

\subsection*{Velocity operator}
The eigenpairs of UniTB come from the projected Hamiltonian $\widehat H_b$ of Methods, and its velocity matrix is the $k$-derivative of $\widehat H_b$ at $\Gamma$ in the atomic-center gauge, $v^{\alpha}_{nm}=\hbar^{-1}c_n^\dagger(\partial_\alpha\widehat H_b-\bar\varepsilon_{nm}\partial_\alpha S)c_m$. With $S=LL^\dagger$, $A=L^{-1}(\widehat H-H^{(0)})L^{-\dagger}=U\,\mathrm{diag}(\mu)\,U^\dagger$ and $g(x)=x-\mathrm{clip}(x,-b,b)$, the projection is $\widehat H_b=\widehat H-L\,g(A)\,L^\dagger$. Its derivative combines the derivative of the Cholesky factor with the Daleckii--Krein formula $\partial g(A)=U[G\circ(U^\dagger\partial A\,U)]U^\dagger$, where $G_{ab}=[g(\mu_a)-g(\mu_b)]/(\mu_a-\mu_b)$ and $G_{aa}=g'(\mu_a)$. $G$ vanishes unless a component is clipped; without clipped components, as for DFT in all 18 cells and for the unprojected UniTB-Flash, $v$ reduces to $(\mathrm i/\hbar)\,c_n^\dagger[(\widehat H-\bar\varepsilon_{nm}S)\circ d^\alpha]c_m$. Central finite differences agree with the analytic velocity to a relative Frobenius difference below $5\times10^{-6}$ in five cells, for DFT and UniTB. Over the twelve Ge--Sb--Te cells, the mean errors are 1.77\% for $n(0.8\,\mathrm{eV})$, 1.73\% for $\alpha(1.5\,\mathrm{eV})$ and 0.084 decades for $\sigma_{\rm DC}$ with UniTB and 1.98\%, 0.79\% and 0.062 decades with UniTB-Flash. Supplementary Table~\ref{S-tab:velocity-routes} compares the three combinations of eigenpairs and velocity for UniTB on the Ge$_2$Sb$_2$Te$_5$ cell.
\providecommand{\tminus}{\textminus}
\begin{center}\begin{minipage}{\linewidth}
\captionof{table}{\textbf{Eigenpairs and velocity for the optical and DC properties.} Errors against same-cell DFT are absolute relative for $n(0.8\,\mathrm{eV})$ and $\alpha(1.5\,\mathrm{eV})$ and absolute for $\log_{10}\sigma_{\rm DC}$ (Gaussian, 50 meV); the cell is the 180-atom Ge$_2$Sb$_2$Te$_5$ cell of Fig.~\ref{fig:amorphous}. UniTB uses (1) unprojected eigenpairs and velocities, (2) eigenpairs projected relative to $H^{(0)}$ ($b=10$ eV) with unprojected velocities, or (3) projected eigenpairs with the velocity of the projected Hamiltonian. UniTB-Flash uses the grid-free prior and overlap without projection, with canonical orthogonalization at $2\times10^{-3}$. Combination (3) is used for UniTB in main-text Fig.~\ref{fig:amorphous}b--d and Supplementary Tables~\ref{S-tab:percell}, \ref{S-tab:amorphous_l} and~\ref{S-tab:gst-properties}.}
\label{S-tab:velocity-routes}
\centering\small
\renewcommand{\arraystretch}{1.08}
\setlength{\tabcolsep}{5pt}
\begin{tabular}{@{}llrrr@{}}
\toprule
 & & $n(0.8\,\mathrm{eV})$ (\%) & $\alpha(1.5\,\mathrm{eV})$ (\%) & $\log_{10}\sigma_{\rm DC}$ (dec.)\\
\midrule
\quad \ModelH & (1) & 1.59 & 1.84 & 0.107\\
\quad \ModelH & (2) & 1.49 & 1.86 & 0.105\\
\quad \ModelH & (3) & 1.50 & 1.85 & 0.105\\
\quad \ModelGF & & 2.01 & 0.29 & 0.089\\
\bottomrule
\end{tabular}
\end{minipage}\end{center}

\providecommand{\tminus}{\textminus}
\begin{center}\begin{minipage}{\linewidth}
\captionof{table}{\textbf{Supercells against dense baselines.} Both pairs use MPHam-29K, matched training budgets and the same eighteen cells. Intervals use 20,000 paired structure resamples.}
\label{S-tab:amorphous-dense}
\centering\small
\renewcommand{\arraystretch}{1.1}
\setlength{\tabcolsep}{9pt}
\begin{tabular}{@{}lrr@{\hspace{3pt}}rr@{}}
\toprule
Same-prior comparison & \shortstack[r]{Combined matrix\\change (\%)} & \multicolumn{2}{c}{95\% CI} & \shortstack[r]{Cells improved\\(of 18)}\\
\cmidrule(lr){3-4}
\midrule
UniTB versus dense & \tminus12.00 & [\tminus13.81, & \tminus10.23] & 18\\
UniTB-Flash versus grid-free dense & \tminus10.88 & [\tminus12.98, & \tminus8.46] & 18\\
\bottomrule
\end{tabular}
\end{minipage}\end{center}

\begin{center}\begin{minipage}{\linewidth}

\centering\includegraphics[width=\textwidth]{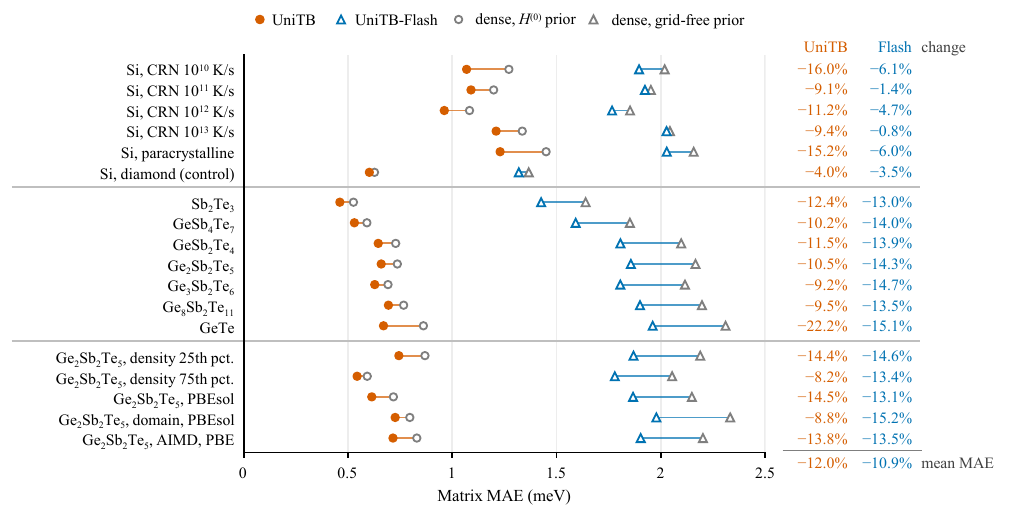}
\captionof{figure}{\textbf{Supercell matrix errors.} Matrix MAE of UniTB and UniTB-Flash (MPHam-29K) on the 18 supercells, each shown with the dense model of the same prior. Errors compare symmetrized blocks of the learned graph with DFT for the same cell. Changes are relative to the dense model.}
\label{S-fig:amorphous-l}

\end{minipage}\end{center}

\providecommand{\tminus}{\textminus}
\begin{center}\begin{minipage}{\linewidth}
\captionof{table}{\textbf{Property errors of the Ge--Sb--Te cells.} Matrix MAE averages symmetrized blocks of the learned graph equally across structures. UniTB properties use projected $\Gamma$ eigenpairs and the velocities of the same projected Hamiltonian; UniTB-Flash properties use its Hamiltonian after overlap truncation. DC errors use Gaussian broadening, and participation errors are absolute relative differences. Bold marks the lower error of the two models in each column.}
\label{S-tab:amorphous_l}
\centering\small
\renewcommand{\arraystretch}{1.1}
\setlength{\tabcolsep}{9pt}
\begin{tabular}{@{}lrrrr@{}}
\toprule
 & \multicolumn{4}{c}{Error against same-cell DFT}\\
\cmidrule(lr){2-5}
MPHam-29K model & \shortstack[r]{Matrix MAE\\(meV)} & \shortstack[r]{Mobility gap\\(eV)} & \shortstack[r]{DC conductivity\\(decades)} & \shortstack[r]{Conduction-edge\\participation (\%)}\\
\midrule
\multicolumn{5}{@{}l@{}}{\textit{Ge--Sb--Te family, $n=12$ cells}}\\
\ModelH & \textbf{0.636} & 0.007 & 0.084 & 3.29\\
\ModelGF & 1.813 & \textbf{0.005} & \textbf{0.062} & \textbf{2.89}\\
\bottomrule
\end{tabular}
\end{minipage}\end{center}
\providecommand{\tminus}{\textminus}
\begin{center}\begin{minipage}{\linewidth}
\captionof{table}{\textbf{Further Ge--Sb--Te properties.} Means over the twelve Ge--Sb--Te cells; percentages are mean absolute relative errors against same-cell DFT. Bold marks the lower error in each row.}
\label{S-tab:gst-properties}
\centering\small
\renewcommand{\arraystretch}{1.1}
\setlength{\tabcolsep}{12pt}
\begin{tabular}{@{}llrr@{}}
\toprule
Ge--Sb--Te observable & Error unit & \ModelH & \ModelGF\\
\midrule
DC conductivity, Gaussian & decades & 0.084 & \textbf{0.062}\\
DC conductivity, Lorentzian & decades & 0.072 & \textbf{0.048}\\
$n(0.8\,\mathrm{eV})$ & \% & \textbf{1.77} & 1.98\\
$\alpha(1.5\,\mathrm{eV})$ & \% & 1.73 & \textbf{0.79}\\
Conduction-edge ISWO & \% & 2.80 & \textbf{1.99}\\
Valence-edge ISWO & \% & \textbf{2.89} & 3.07\\
\bottomrule
\end{tabular}
\end{minipage}\end{center}

\subsection*{State counts}
At the evaluated $\Gamma$ points, the $\pm10$ eV projection leaves the DFT residual spectra unchanged in all 18 cells. In the six Si cells, it removes every UniTB level more than 1 eV below the DFT band bottom and all but two of the excess levels below the DFT midgap (Supplementary Fig.~\ref{S-fig:amorphous-si-clipping}).

\begin{center}\begin{minipage}{\linewidth}

\centering\includegraphics[width=\textwidth]{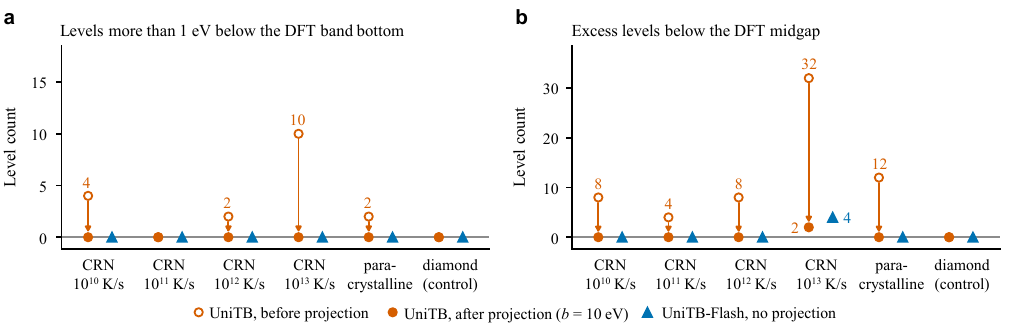}
\captionof{figure}{\textbf{Deep and excess levels in amorphous Si.} Level counts at $\Gamma$ in the six Si cells, from UniTB before and after projection and from UniTB-Flash. Panel \textbf{a} counts deep levels, and panel \textbf{b} counts excess levels relative to DFT.}
\label{S-fig:amorphous-si-clipping}

\end{minipage}\end{center}
\snote{Prior construction details}{sn:prior}
Main-text Eqs.~(\ref{eq:kb})--(\ref{eq:nacf-enhanced-edge}) define the grid-free prior. \paragraph{Gauge and stabilized environmental term.}
The environmental construction uses unpolarized LDA exchange and PZ81 correlation, normalized neutral-valence densities and unscaled nonlinear core corrections. For cell volume $\Omega$ in bohr$^3$, its overlap-proportional gauge is
\begin{equation}
 c=\frac{4\pi}{3\Omega}\sum_K M_{2,K}^{\mathrm{val}}\,\gamma_{\mathrm{Ry}\to\mathrm{eV}},
 \label{S-eq:gf-gauge}
\end{equation}
where $M_{2,K}^{\mathrm{val}}$ is the provider's valence-density second moment in bohr$^2$ and $\gamma_{\mathrm{Ry}\to\mathrm{eV}}$ is its energy conversion. The sum includes every atom in a fully periodic cell; this gauge is defined only for fully periodic cells. 

For an orbital-shell pair $(a,b)$, define $\kappa_{ab}=\sqrt{(2\ell_a+1)(2\ell_b+1)}$ and the Frobenius projection
\begin{equation}
 \mathcal C_B(M)=M\min\{1,B/\|M\|_{\mathrm F}\},\qquad \mathcal C_B(0)=0.
 \label{S-eq:gf-shell-project}
\end{equation}
Let $P,N\geq0$ be the pair and environment envelope moments, $D_p,D_e$ their signed AO moment blocks, $\rho_p$ the shell pair density, and $\bar\rho_e$ the environment density obtained from its envelope moment divided by the envelope overlap. With $\rho_t=\rho_p+\bar\rho_e$, $v_p=v_{\mathrm{xc}}(\rho_p)$, $v_t=v_{\mathrm{xc}}(\rho_t)$ and primes denoting density derivatives, the environmental correction is
\begin{align}
 \widetilde M_p&=\mathcal C_{2\kappa P}(D_p-\rho_p S),&
 \widetilde M_e&=\mathcal C_{2\kappa N}(D_e-\bar\rho_e S),\\
 \Delta X^{\mathrm{env}}&=(v_t-v_p)S+(v'_t-v'_p)\widetilde M_p+v'_t\widetilde M_e.
 \label{S-eq:gf-env-v2}
\end{align}
The projection rescales complete shell blocks while preserving their direction. The implementation suppresses unresolved signed environment moments, clamps negative envelope moments to zero, and applies the correction only on resolved shell overlaps with positive pair and total densities. These choices limit the growth of density derivatives at low density.

\subsection*{Online assembly on the GPU}
Radial integrals and factor tables are prepared offline. Online, both priors run through the same five GPU steps (main-text Fig.~\ref{fig:prior}a): (1) one pair search gives the block keys and their reverse blocks; (2) each route evaluates its own two-center tables in its own kernel on common spline, interval-search and real-harmonic code, with bond-frame rotation for $H_{\rm GF}$ and Gaunt coefficients for $H^{(0)}$; (3) one shared three-center path evaluates the projector contraction of Eq.~(\ref{eq:spinor-kb}) on FP64 tensor cores and gives the nonlocal term and the spin-flip blocks of both priors; (4) the local and exchange--correlation term is route specific, with the tabulated third-center, pair and environmental terms and the onsite quadrature for $H_{\rm GF}$, and the FFT-grid field, including the PBE potential, and its integration over $8^3$-point tiles for $H^{(0)}$; (5) one kernel completes the spinor blocks from their representative rows, applies the gauge and units, and returns $\Pp$ and $S\otimes\mathds 1_2$. Neighbor search and the PBE potential run on the GPU. The environmental moments and the third-center neutral-atom term of $H_{\rm GF}$ reuse the contraction of step 3. Global charge response and spectral solves are additional stages. Supplementary Note~\ref{S-sn:cost} derives the cost of the online assembly. For a 1,024-atom PbTe supercell on one H200 GPU, steps 1, 2, 3 and 5 take 28.5 and 46.7~ms and step 4 takes 173.7 and 991.8~ms for $H_{\rm GF}$ and $H^{(0)}$, totaling 202.5 and 1,038.7~ms (main-text Fig.~\ref{fig:prior}b; medians of three warm calls, taken separately).

\subsection*{Prior input and residual target}
UniTB-Flash takes $H_{\rm GF}$ as both its input prior and its baseline, so it learns the residual $H-H_{\rm GF}$ (Supplementary Table~\ref{S-tab:config_s}). Supplementary Table~\ref{S-tab:protocol-map} lists the matrices used in each evaluation. Main-text Fig.~\ref{fig:prior}d compares three dense networks trained identically on MPHam-12K (11,953 structures, 37,500 updates, batches of at most 32 structures) that differ only in the prior: $H^{(0)}$ with target $H-H^{(0)}$, $H_{\rm GF}$ with target $H-H_{\rm GF}$, and no prior with target $H$. Adding back the prior makes the three errors refer to the same Hamiltonian entries. On the 3,000 test structures, their combined, onsite and hopping MAEs are 0.813, 8.49 and 0.726~meV, 3.08, 8.90 and 3.01~meV, and 11.66, 11.57 and 11.70~meV. The panel plots the onsite loss against the hopping loss at ten checkpoints during training; each loss is $\tfrac14(\mathrm{MAE}+\mathrm{RMSE})$ of its head. The training loss is the mean over the minibatches of the epoch that ends at the checkpoint. The test loss is evaluated on the 3,000 test structures.
\snote{Operation counts of the online assembly}{sn:cost}
This note derives the online cost of each assembled matrix and evaluates the counts quoted in the main text; offline tables are excluded. $N$ is the number of atoms, $\Omega$ the cell volume, $m$ the orbitals per atom, $D=2\sum_im_i$ the spinor dimension, $p$ the projector functions of a center (magnetic components included), $r$ the rank of the neutral-atom factors, $q_{\rm on}$ the onsite quadrature points per atom, $h$ the spacing of the FFT grid and $G\simeq\Omega/h^3$ its number of points.

\paragraph{Bounded neighbor lists.}
Let all interatomic distances, including those to periodic images, be at least $d_{\min}>0$. A ball of radius $R$ then holds at most $n(R)=(1+2R/d_{\min})^3$ atoms: balls of radius $d_{\min}/2$ around these atoms are disjoint and lie in a ball of radius $R+d_{\min}/2$. With orbital cutoff $r_c$, projector cutoff $r_\beta$ and neutral-atom cutoff $r_{\rm NA}$, an atom has at most $z_2=n(2r_c)$ orbital-overlap partners, a projector center reaches at most $z_\beta=n(r_c+r_\beta)$ atoms, and at most $z_{\rm NA}=n(r_c+r_{\rm NA})$ neutral-atom centers reach an atom or a bond. The number of directed hopping blocks is $E\le Nz_2$. A cell list that stores only the occupied bins, of width $\ell$ at least the largest cutoff, examines 27 bins per atom, each holding at most $n(\sqrt3\,\ell/2)$ atoms, and builds every list in $O(N)$ operations\cite{SI-allen2017computer}.

\paragraph{Overlap and spin-flip blocks.}
A two-center block costs one spline evaluation on a uniform radial grid for each Slater--Koster channel, $\sum_{a,b}(\min(\ell_a,\ell_b)+1)$ channels over the shells $a$ and $b$ of the two atoms (116 for two 4s2p2d1f atoms), followed by one rotation. $S$ therefore costs $O(N)$, and its spinor form $S\otimes\mathds 1_2$ adds no arithmetic. In main-text Eq.~(\ref{eq:spinor-kb}), each center $K$ builds at most $z_\beta$ projector-overlap blocks $\langle\beta^K|\phi_i\rangle$ and contracts at most $z_\beta^2$ orbital pairs with the complex matrix $D^{K,\sigma\sigma'}$, each in $O(m^2p+mp^2)$ operations, so the spin-flip blocks cost $O(Nz_\beta^2(m^2p+mp^2))=O(N)$ and reuse the projector overlaps of the scalar term. Hermiticity, $H^{\downarrow\uparrow}_{ji}(-\mathbf T)=[H^{\uparrow\downarrow}_{ij}(\mathbf T)]^\dagger$ for lattice translation $\mathbf T$, and time reversal in a nonmagnetic prior, $H^{\downarrow\downarrow}=(H^{\uparrow\uparrow})^*$, leave two of the four spin blocks to compute.

\paragraph{Grid-free prior.}
Every item of $H_{\rm GF}$ belongs to a list of bounded length and takes a fixed amount of work (Supplementary Table~\ref{S-tab:cost-gf}), so $H_{\rm GF}$ costs $O(N)$. At fixed atom count and numerical settings, the operation count does not grow with $\Omega$ (the volume-dependent gauge $c$ is a single sum over atoms), and vacuum adds no work.

\begin{center}\begin{minipage}{\linewidth}
\captionof{table}{\textbf{Cost of the grid-free prior.} Each term sums over lists of bounded length with fixed work per item. Equation numbers refer to the main-text Methods. $z_{\rm env}$ counts the environment atoms of a bond, and $z_\rho$ the atoms that contribute density at a quadrature point.}
\label{S-tab:cost-gf}
\centering\small
\renewcommand{\arraystretch}{1.1}
\begin{tabular}{@{}llll@{}}
\toprule
Term & Item & Number of items & Work per item\\
\midrule
$T$, endpoint $V_{\rm NA}$, pair XC, $S$ & atom pair & $N(1+z_2)$ & one table block\\
Scalar separable term, Eq.~(\ref{eq:kb}) & center and orbital pair & $Nz_\beta^2$ & $O(m^2p+mp^2)$\\
Onsite crystal field, finite rank & atom and center & $Nz_{\rm NA}$ & $O(rm^2)$\\
Third-center term $V^{\rm 3c}$, finite rank & bond and center & $Nz_2z_{\rm NA}$ & $O(rm^2)$\\
McWEDA correction $\Delta X^{\rm env}$ & bond and environment atom & $Nz_2z_{\rm env}$ & $O(1)$, then $O(m^2)$ per bond\\
Onsite exchange--correlation & atom and quadrature point & $Nq_{\rm on}$ & $O(z_\rho+m^2)$\\
Gauge $cS$ & atom; block & $N$; $N(1+z_2)$ & $O(1)$; $O(m^2)$\\
\bottomrule
\end{tabular}
\end{minipage}\end{center}

\paragraph{Zeroth-step prior.}
$H^{(0)}$ shares the two-center and projector stages and adds the cell-wide potential (Supplementary Table~\ref{S-tab:cost-h0}). The structure factors $F_s(\mathbf G)=\sum_{A\in s}e^{-i\mathbf G\cdot\mathbf R_A}$ of each species $s$ come from type-1 nonuniform FFTs with a kernel width $w=O(\log\varepsilon^{-1})$ at tolerance $\varepsilon$\cite{SI-barnett2019finufft,SI-shih2021cufinufft}. With $G\propto\Omega$, the total is $O(N+G\log G)$, or $O(N\log N)$ at fixed density.

\begin{center}\begin{minipage}{\linewidth}
\captionof{table}{\textbf{Cost of the zeroth-step prior.} $s$ is the number of species, $U\le G$ the number of distinct reciprocal-vector lengths, $n_r$ the radial mesh of the pseudopotential and $q_\phi$ the grid points of one orbital support.}
\label{S-tab:cost-h0}
\centering\small
\renewcommand{\arraystretch}{1.1}
\begin{tabular}{@{}lll@{}}
\toprule
Stage & Operation & Work\\
\midrule
$T$, $S$, $V_{\rm nl}$ and spin-flip blocks & as for $H_{\rm GF}$ & $O(N)$\\
Structure factors & type-1 nonuniform FFT per species & $O(Nw^3+sG\log G)$\\
Radial form factors & radial quadrature at $U$ lengths & $O(sUn_r)\le O(sn_rG)$\\
Density, local and Hartree potentials & pointwise in reciprocal space & $O(sG)$\\
Real-space fields and PBE gradients & fixed number of three-dimensional FFTs & $O(G\log G)$\\
PBE potential & pointwise & $O(G)$\\
Orbital-pair integration & sum over the common support of each pair & $O(Nz_2q_\phi m^2)$\\
\bottomrule
\end{tabular}
\end{minipage}\end{center}

\paragraph{Counts at the parameters of our data.}
Supplementary Table~\ref{S-tab:cost-counts} uses median orbital cutoff $r_c=7$ bohr, 27 orbitals (4s2p2d1f), $p=34$ projector functions per atom with cutoff 1.7 bohr, rank $r=144$ with a 7-bohr factor cutoff, medium onsite quadrature $128\times24\times48$ and the labels' 480 Ry density cutoff ($h=0.143$ bohr). A table block costs ten operations per spline channel plus $8m^2$ for rotation, with four tables per hopping block. At uniform atomic density $\bar n$, list lengths are $z=\tfrac{4\pi}{3}\bar nR^3$. The common orbital support averaged over partners within $2r_c$ is one eighth of a support; with factor cutoff $r_c$, the mean third-center count is $z_{\rm NA}/8$, or 11 and 17 at the two densities. FFTs comprise twelve complex three-dimensional transforms at $5G\log_2G$ operations each, plus one transform per species on the nonuniform FFT grid upsampled twofold in each dimension. The radial-form-factor bound uses three species, all $|\mathbf G|$ distinct and the median fully relativistic radial mesh of 1,528 points.

\begin{center}\begin{minipage}{\linewidth}
\captionof{table}{\textbf{Operation counts of the priors.} The last row gives the number of atoms above which one dense spinor eigensolve, about $20D^3$ operations, exceeds the whole prior assembly.}
\label{S-tab:cost-counts}
\centering\small
\renewcommand{\arraystretch}{1.1}
\begin{tabular}{@{}lll@{}}
\toprule
Quantity & 0.05 atoms\,\AA$^{-3}$ & 0.08 atoms\,\AA$^{-3}$\\
\midrule
Grid points of one orbital support & $4.9\times10^5$ & $4.9\times10^5$\\
Grid points of a common support, mean & $6.1\times10^4$ & $6.1\times10^4$\\
Hopping partners per atom, $z_2$ & 85 & 136\\
\multicolumn{3}{@{}l@{}}{\textit{One hopping block}}\\
$H^{(0)}$, orbital-pair integration & $8.9\times10^7$ & $8.9\times10^7$\\
$H_{\rm GF}$, tables and third-center sum & $2.3\times10^6$ & $3.6\times10^6$\\
Ratio & 39 & 25\\
\multicolumn{3}{@{}l@{}}{\textit{Per atom}}\\
Scalar $H^{(0)}$, local stages & $8.3\times10^9$ & $1.3\times10^{10}$\\
Scalar $H_{\rm GF}$ & $4.5\times10^8$ & $8.0\times10^8$\\
Spin-flip blocks; overlap & $4.1\times10^7$; $6.0\times10^5$ & $1.1\times10^8$; $9.6\times10^5$\\
\multicolumn{3}{@{}l@{}}{\textit{Share of $H^{(0)}$, 20--2,000 atoms}}\\
FFTs, nonuniform transforms included & 2.1--2.7\% & 0.8--1.1\%\\
Radial form factors, bound & $\le22\%$ & $\le9\%$\\
\multicolumn{3}{@{}l@{}}{\textit{Band calculation}}\\
Atoms where one eigensolve exceeds the assembly, $H_{\rm GF}$; $H^{(0)}$ & 13; 51 & 17; 64\\
\bottomrule
\end{tabular}
\end{minipage}\end{center}

\paragraph{Band calculations.}
A dense generalized eigenproblem of dimension $D$ costs $O(D^3)$ per wavevector through Cholesky reduction, tridiagonalization and divide and conquer\cite{SI-golub2013matrix}; with complex spinor matrices, this is about $20D^3$ real operations. The whitening, screening and flagged residual decompositions of the spectral projection have the same order, and its update, of rank equal to the number $n_{\rm c}$ of clipped eigenvalues, costs $O(D^2n_{\rm c})$. A band path of $N_k$ wavevectors costs $O(N_kD^3)$. Above 13--17 atoms for $H_{\rm GF}$ and 51--64 atoms for $H^{(0)}$, one wavevector of the band calculation takes more operations than the whole assembly, overlap and spin-flip blocks included (Supplementary Table~\ref{S-tab:cost-counts}).
\snote{Implementation and timing}{sn:systems}
Figure~\ref{fig:so2cuda} in the Methods (``SO2CUDA architecture and performance'') summarizes the SO2CUDA mechanism and the operator and real-crystal benchmarks; the figures and tables of this note give the layouts, execution routes, operator costs, peak memory and synthetic whole-model timings.

Supplementary Table~\ref{S-tab:cuda-final} and Supplementary Figure~\ref{S-sfig:so2-model-speed} report synthetic whole-model timings of randomly initialized models on periodic structures, as medians of forward-plus-loss-backward times; the timing scope and baselines are defined below. Unless stated otherwise, the timings of this note use one H200 GPU in strict FP32 with TF32 disabled.

Across all six model/edge-count pairs, the maximum absolute SO2CUDA--PyTorch differences in training outputs, inference outputs and parameter gradients are $6.8\times10^{-6}$, $6.2\times10^{-6}$ and $1.5\times10^{-5}$, respectively, passing maximum- and RMS-error tolerances of \texttt{atol}+\texttt{rtol} times the corresponding baseline magnitude (\texttt{atol}/\texttt{rtol}: $10^{-5}$/$10^{-4}$ for outputs, $2\times10^{-6}$/$2\times10^{-4}$ for gradients).
\begin{center}\begin{minipage}{\linewidth}
\captionof{table}{\textbf{Whole-model timings with three backends.} Times are medians of ten synchronized forward-plus-loss-backward calls on synthetic workloads. Each speedup divides the baseline time by the SO2CUDA time for the same model and input. Setup, input cloning, buffer restoration and optimizer updates are excluded.}
\label{S-tab:cuda-final}
\centering\small
\renewcommand{\arraystretch}{1.1}
\setlength{\tabcolsep}{5pt}
\begin{tabular}{@{}lrrrrr@{}}
\toprule
Directed edges & \shortstack[r]{Plain PyTorch\\(ms)} & \shortstack[r]{cuEquivariance\\(ms)} & \shortstack[r]{SO2CUDA\\(ms)} & \shortstack[r]{vs. plain\\($\times$)} & \shortstack[r]{vs. cuEQ\\($\times$)}\\
\midrule
\multicolumn{6}{@{}l@{}}{\textit{Dense}}\\
\quad 20,000 & 549.78 & 554.32 & 251.16 & 2.19 & 2.21\\
\quad 50,000 & 1012.97 & 769.77 & 367.70 & 2.75 & 2.09\\
\quad 130,000 & 2232.17 & 1525.91 & 716.66 & 3.11 & 2.13\\
\addlinespace[3pt]
\multicolumn{6}{@{}l@{}}{\textbf{UniTB}}\\
\quad 20,000 & 663.28 & 771.85 & 290.67 & 2.28 & 2.66\\
\quad 50,000 & 1187.75 & 1232.56 & 466.19 & 2.55 & 2.64\\
\quad 130,000 & 2607.76 & 2657.88 & 953.76 & 2.73 & 2.79\\
\bottomrule
\end{tabular}
\end{minipage}\end{center}

\subsection*{Benchmark scope}
The operator, real-crystal and synthetic whole-model benchmarks are separate runs. The official EquiformerV3 comparator (\href{https://github.com/atomicarchitects/equiformer_v3/tree/a7300c58df683dc99cb48027d5bfd4c887486c48}{commit \texttt{a7300c58}}) is timed with and without \texttt{torch.compile}, and cuEquivariance 0.12.0 uses its fastest tested descriptor/rotation combinations. Plain PyTorch denotes our implementation following upstream DeePTB. The uniform comparison times the SO(2) linear layer without radial weights. Real-crystal training steps use the onsite network and a HybridMuon optimizer update, with four warm-up and twelve measured steps. Real-crystal step times use synchronized host wall time for the training iteration, including preparation inside the iteration, gradient clipping, metrics and the optimizer update; cloning each stored CPU batch is outside the timed interval. The cuEquivariance model uses the \texttt{escn\_tp\_compact} descriptor, \texttt{naive} method and PyTorch Wigner rotation, selected from a 16-configuration scan of numerical correctness and uniform-operator timing.

\subsection*{Synthetic whole-model protocol}
The implementation shares feature rotations across routed experts, groups matrix operations and mixes experts before the activation. The synthetic timings use the same randomly initialized models and periodic structures for all implementations, with three warm-up and ten measured calls. Each synthetic cell has 125 atoms and 80 neighbors per atom; cell replication gives 20,000, 50,000 and 130,000 directed edges. CUDA events cover model forward and loss backward, excluding setup, input cloning, buffer restoration and first JIT compilation, and no optimizer update is performed. Each total is the median of the combined forward-plus-backward calls. The plain PyTorch baseline uses uncached upstream SO(2) operations and selected-edge expert linears, with dense group-wise weight mixtures evaluated by the UMA graph formulation; model parameters, inputs, router, charge head and the remaining layers are shared across implementations. Training outputs, inference outputs and compared parameter gradients are checked against plain PyTorch with both scaled maximum-error and RMS-error criteria (the error criteria above). 

\subsection*{SO(2) layouts and execution routes}
Supplementary Figures~\ref{S-sfig:so2-layouts}--\ref{S-sfig:so2-routes} show the layouts, the corresponding implementation steps and the SO2CUDA execution routes of the benchmark scenarios.
\begin{center}\begin{minipage}{\linewidth}
\centering\includegraphics[width=\textwidth]{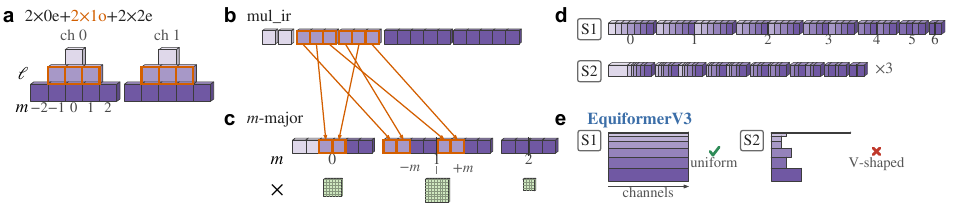}
\captionof{figure}{\textbf{The $m$-major feature layout.} \textbf{a},~An illustrative feature; orange marks the two $\ell=1$ vectors. \textbf{b},~The same entries in the e3nn order. \textbf{c},~Indexed placement into the $m=0$ and $[-m\,|\,+m]$ blocks of the $m$-major layout, each multiplied by its own GEMM weights. \textbf{d},~Block widths of the uniform scenario S1 (128 channels per degree) and the V-shaped scenario S2 (UniTB hidden irreps). \textbf{e},~The EquiformerV3 layout of 49 coefficients by $C$ channels, which needs the same channel count at every degree. Purple shades encode degree.}
\label{S-sfig:so2-layouts}
\end{minipage}\end{center}
\begin{center}\begin{minipage}{\linewidth}
\centering\includegraphics[width=\textwidth]{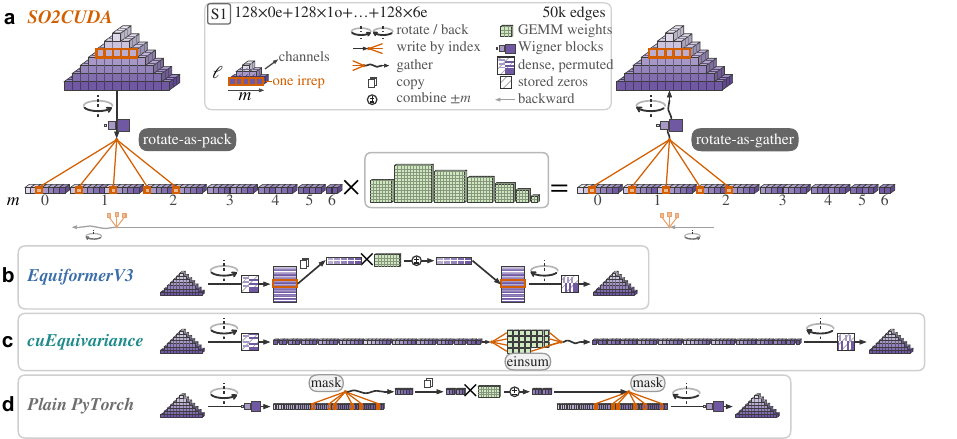}
\captionof{figure}{\textbf{Four SO(2) implementations.} All four perform the same linear multiply-adds and differ in rotation, layout and copy operations. \textbf{a},~SO2CUDA fuses rotation with indexed placement and collection around one GEMM per $m$; the backward pass reuses the same kernels. \textbf{b},~EquiformerV3 uses dense permuted Wigner matrices, per-$m$ slicing and contiguous copies, linear maps and the $\pm m$ combination. \textbf{c},~cuEquivariance uses a dense permuted rotation and a pathwise \texttt{einsum}. \textbf{d},~Plain PyTorch rotates per $\ell$, gathers by mask, copies, applies the linear map, combines $\pm m$ and writes back by mask. Autograd differentiates the composed operations in \textbf{b}--\textbf{d}.}
\label{S-sfig:so2-implementations}
\end{minipage}\end{center}
\begin{center}\begin{minipage}{\linewidth}
\centering\includegraphics[width=\textwidth]{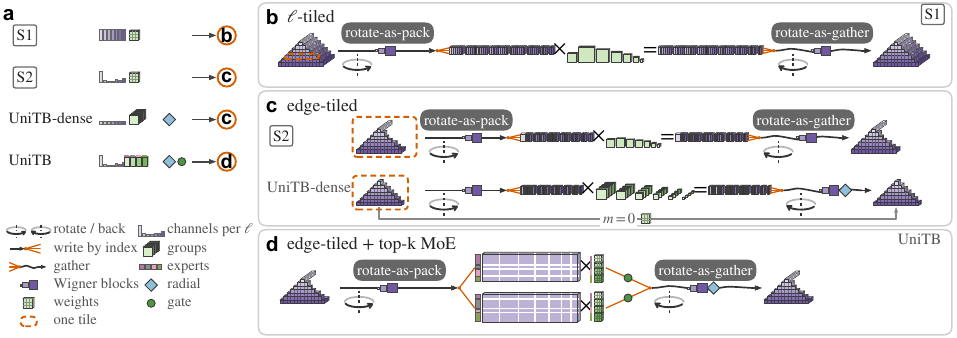}
\captionof{figure}{\textbf{SO2CUDA execution routes.} The channel layout and weight structure of a configuration select its route. \textbf{a},~Route of each benchmark configuration. UniTB-dense is the main-text dense model. One channel-depth voxel in the key represents 16 channels. \textbf{b},~One warp handles one edge, degree and 32-channel tile. \textbf{c},~An edge tile holds one edge in a thread block. S2 uses one weight set; the UniTB-dense hidden layer uses weights grouped by structure and radial weights on the gather. In the DeePTB dense layer the $m=0$ term is computed separately, so the fused path covers $m>0$ only. \textbf{d},~The UniTB routed layer covers $m=0$ and $m>0$. It packs top-two copies with expert-sorted rows, runs a grouped GEMM per copy and gathers with routing gates and radial weights. Edge tiling needs compact or identity Wigner data and enough shared memory; otherwise a general per-edge, per-channel kernel handles rotation and gathering.}
\label{S-sfig:so2-routes}
\end{minipage}\end{center}

\subsection*{Operator benchmarks and implementation costs}
Supplementary Figures~\ref{S-sfig:so2-operators} and~\ref{S-sfig:so2-costs} report operator-level timing, Wigner storage, kernel counts and peak allocated memory.
\begin{center}\begin{minipage}{\linewidth}
\centering\includegraphics[width=\textwidth]{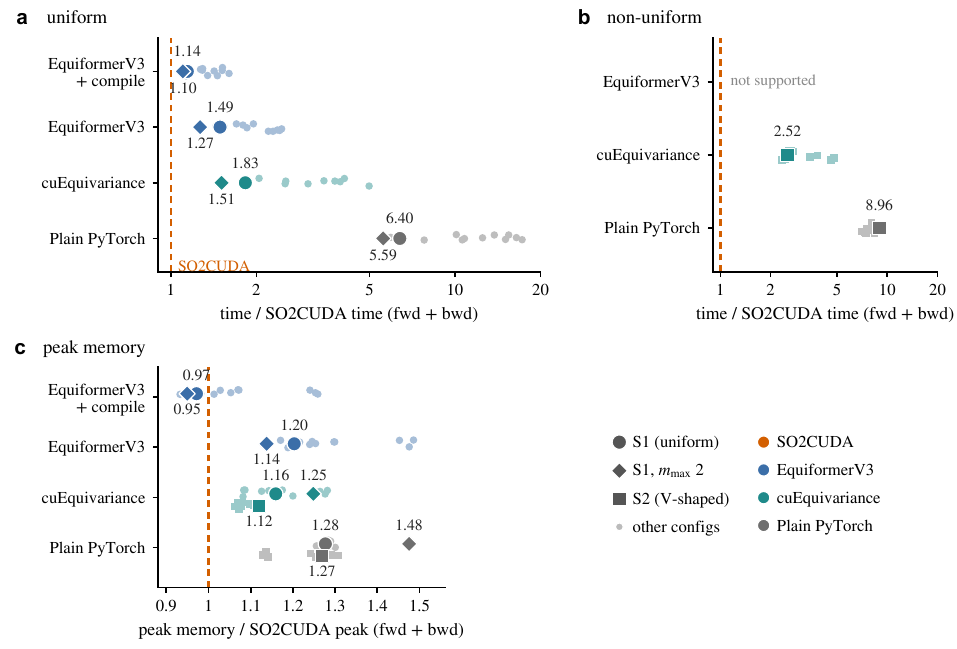}
\captionof{figure}{\textbf{Benchmarks of the SO(2) linear layer.} \textbf{a},\textbf{b},~Forward-plus-backward time relative to SO2CUDA for uniform (\textbf{a}) and nonuniform (\textbf{b}) channel allocations. \textbf{c},~Peak memory relative to SO2CUDA, with stored Wigner data counted. S1 has 128 channels per degree through $\ell=6$, and its variant marked $m_{\max}$ 2 keeps the orders $|m|\le2$. S2 uses the V-shaped UniTB hidden irreps. Geometry and Wigner data are precomputed, and each implementation starts from its native feature layout (protocol in Methods).}
\label{S-sfig:so2-operators}
\end{minipage}\end{center}
\begin{center}\begin{minipage}{\linewidth}
\centering\includegraphics[width=\textwidth]{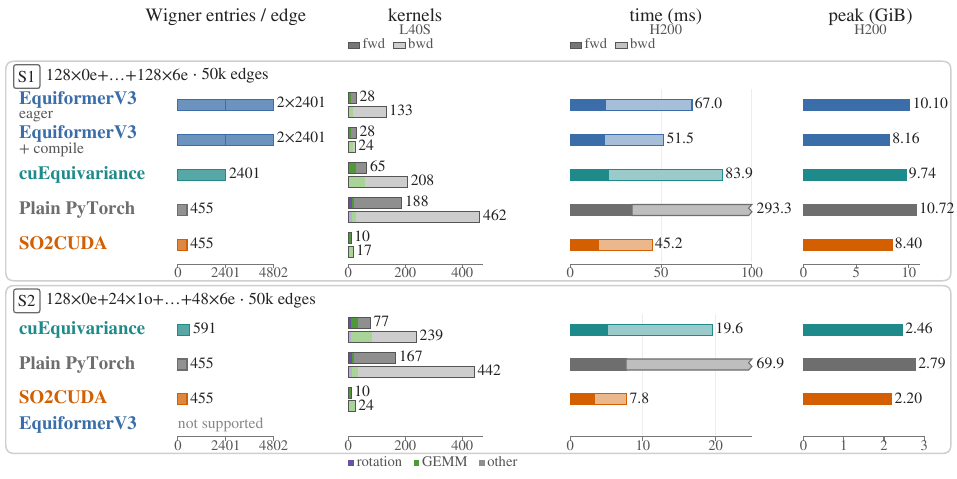}
\captionof{figure}{\textbf{Costs of the SO(2) implementations.} Wigner storage, kernel counts, time and peak memory of the SO(2) linear layer in scenarios S1 and S2. Dense $49\times49$ Wigner matrices store 2,401 entries per edge, and per-$\ell$ blocks store 455 (the $\ell=0$ entry is passed through). EquiformerV3 stores separate forward and backward rotations. cuEquivariance uses one matrix and its transpose, with 591 entries in five multiplicity groups for S2. Kernel counts come from PyTorch profiler runs and exclude memcpy and memset (forward: \texttt{no\_grad}; backward: combined minus forward with autograd). Times exclude Wigner construction and initial layout conversion. Peak memory is the allocated peak, including stored Wigner data.}
\label{S-sfig:so2-costs}
\end{minipage}\end{center}

\subsection*{Model benchmarks}
Supplementary Figure~\ref{S-sfig:so2-real-batch} supplements the real-crystal training steps of Fig.~\ref{fig:so2cuda}c with peak allocated memory. Supplementary Figure~\ref{S-sfig:so2-model-speed} gives the separate synthetic model timings split into forward and backward passes.
\begin{center}\begin{minipage}{\linewidth}
\centering\includegraphics[width=\textwidth]{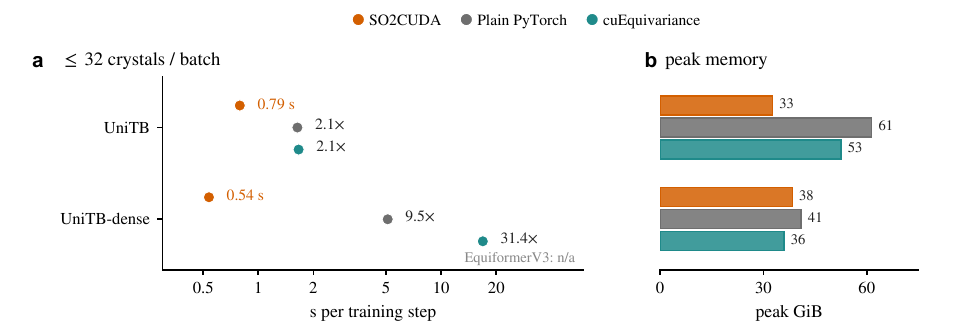}
\captionof{figure}{\textbf{Training steps on real crystals.} \textbf{a},~Median time of one training step of the onsite network, with the ratio to SO2CUDA. Batches hold at most 32 structures, on average 31.7 structures and 55,893 directed edges, and the step includes HybridMuon in its \texttt{fast} optimizer mode. \textbf{b},~Peak allocated memory of the same steps.}
\label{S-sfig:so2-real-batch}
\end{minipage}\end{center}
\begin{center}\begin{minipage}{\linewidth}
\centering\includegraphics[width=\textwidth]{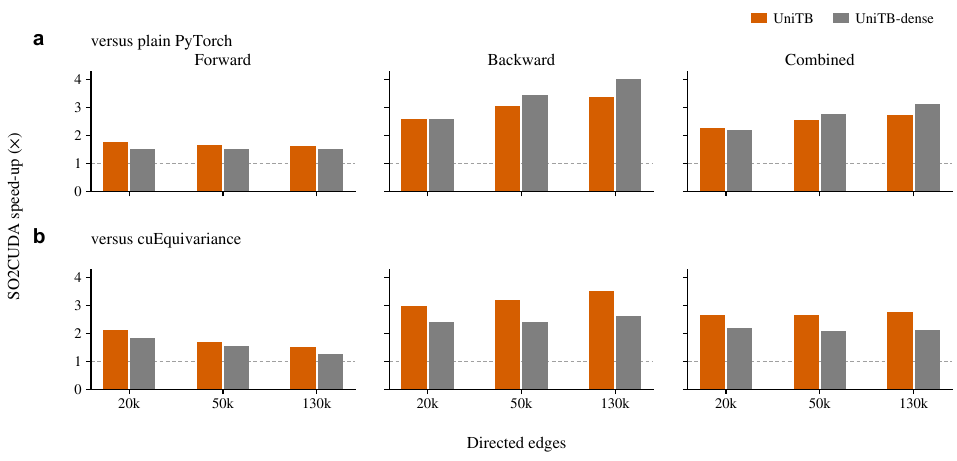}
\captionof{figure}{\textbf{Whole-model speedups of SO2CUDA.} Speedups over plain PyTorch (\textbf{a}) and cuEquivariance (\textbf{b}) in the forward pass, the loss backward and the combined call, which is measured directly. Speedup is the comparator median divided by the SO2CUDA median for the same model and directed-edge count. Randomly initialized models share parameters on periodic cells of 125 atoms and 80 neighbors per atom.}
\label{S-sfig:so2-model-speed}
\end{minipage}\end{center}

\subsection*{Spectral post-processing}
On one RTX PRO 6000 GPU, five MPHam-12K predictions on 200 structures give 1,000 calls, 77,295 sampled wavevectors and spinor dimensions 56--1,524. Mean solve times are 46.10 ms without projection and 56.64 ms with it ($22.85\%$ overhead), from one synchronized call per case/model/state after warmup on the first structure, excluding source reads, inference and solver setup. Eight full paths take 0.0112--1.1600 s/path (minimum of three calls, cached assembly indices), with maximum window-level differences of 0.219--0.969 meV from projected CPU double precision.
\providecommand{\tminus}{\textminus}
\begin{center}\begin{minipage}{\linewidth}
\captionof{table}{\textbf{Projected band-solve timing.} Each time is the minimum of three calls with cached assembly indices. The window maximum error is relative to the projected CPU double-precision spectrum.}
\label{S-tab:spectral_path_timing}
\centering\small
\renewcommand{\arraystretch}{1.1}
\setlength{\tabcolsep}{10pt}
\begin{tabular}{@{}rrrrr@{}}
\toprule
Atoms & $N_{\rm SOC}$ & $n_k$ & \shortstack[r]{Projected solve\\(s/path)} & \shortstack[r]{Window max error\\(meV)}\\
\midrule
3 & 106 & 122 & 0.011 & 0.219\\
4 & 168 & 122 & 0.018 & 0.969\\
4 & 208 & 183 & 0.029 & 0.317\\
12 & 328 & 184 & 0.076 & 0.402\\
16 & 584 & 224 & 0.202 & 0.520\\
24 & 608 & 244 & 0.189 & 0.258\\
46 & 1,004 & 184 & 0.626 & 0.319\\
29 & 1,342 & 144 & 1.160 & 0.400\\
\bottomrule
\end{tabular}
\end{minipage}\end{center}
\snote{Application domains, library screen and inference timing}{sn:screening}
\paragraph{Tasks.}
Each of the six domains of main-text Fig.~\ref{fig:screening}b pairs a screening decision with a property value (Supplementary Table~\ref{S-tab:domains}): photonics (high-index dielectrics), photovoltaics (thin-film absorbers), semiconductors (band-gap design), mobility (high-mobility channels), thermoelectrics and interconnect metals. A gain is significant when its paired 95\% interval lies above zero, and Methods defines the properties. A screening decision ranks the structures by the model value and scores the fraction of the DFT top 20 recovered among the $K$ highest-ranked structures, averaged over $K=5$--60 (a perfect ranking scores 0.893). Property values are scored by median absolute errors for the band gap and the SLME and by median absolute natural-log ratios for the other positive quantities, and the intervals use 10,000 paired bootstrap resamples. The pool contains 2,433 nonmagnetic test structures by Materials Project labels (the 2,431 of main-text Fig.~\ref{fig:periodic} plus two without a magnetic label). Each task uses a subset of these: 200 semiconductors for the refractive index and the SLME (SLME values for the 73 with DFT gaps of 0.5--2.5 eV), DFT gaps above 0.02 eV for band gaps, gaps of 0.5--3.0 eV for conduction masses, the thermoelectric subset of the test set for power factors, and metals for $\rho_0\lambda$.

\providecommand{\tminus}{\textminus}
\begin{center}\begin{minipage}{\linewidth}
\captionof{table}{\textbf{Screening scores and property errors in six domains.} Nonmagnetic test crystals; $n$, number of structures with values for all three Hamiltonians. A screen is scored by the recall of the DFT top 20 among the $K$ highest-ranked structures, averaged over $K=5$--60 (0.893 for a perfect ranking). Relative errors are median absolute natural-log ratios $e$, shown as $100\,[\exp(e)-1]$ percent. The gain is $1-e_{\rm UniTB}/e_{\rm Flash}$ for values and the score ratio minus one for screens, with paired bootstrap 95\% intervals (10,000 resamples). Bold, interval above zero.}
\label{S-tab:domains}
\centering\small
\renewcommand{\arraystretch}{1.08}
\setlength{\tabcolsep}{3pt}
\begin{tabular}{@{}llrrrr@{\hspace{3pt}}rr@{}}
\toprule
Domain and task & Metric & $n$ & UniTB & \shortstack[r]{UniTB-\\Flash} & \multicolumn{2}{c}{\shortstack{Gain (\%)\\{}[95\% CI]}} & \shortstack[r]{$R^2$\\U / F}\\
\midrule
\multicolumn{8}{@{}l@{}}{\textit{Photonics}}\\
\quad High-$n$ screen & mean top-20 recall & 200 & 0.892 & 0.893 & \tminus0.1 & [\tminus0.5, +1.2] & \\
\quad Crystal $n$ & median rel.\ error (\%) & 198 & 0.25 & 0.51 & \textbf{+51.0} & \textbf{[+31.1, +59.2]} & 0.998 / 0.996\\
\addlinespace[2pt]
\multicolumn{8}{@{}l@{}}{\textit{Photovoltaics}}\\
\quad High-SLME screen & mean top-20 recall & 200 & 0.878 & 0.872 & +0.6 & [\tminus0.2, +2.8] & \\
\quad SLME & median abs.\ error (pp) & 73 & 0.38 & 0.71 & \textbf{+46.1} & \textbf{[+8.1, +58.3]} & 0.949 / 0.940\\
\addlinespace[2pt]
\multicolumn{8}{@{}l@{}}{\textit{Semiconductors}}\\
\quad 0.1--1.0 eV window & F1 & 2,433 & 0.939 & 0.915 & +2.6 & [0.0, +5.4] & \\
\quad Band gap & median abs.\ error (meV) & 1,253 & 26.3 & 50.9 & \textbf{+48.3} & \textbf{[+43.5, +54.1]} & 0.998 / 0.962\\
\addlinespace[2pt]
\multicolumn{8}{@{}l@{}}{\textit{Mobility}}\\
\quad Light-$m_{\rm c}$ screen & mean top-20 recall & 680 & 0.852 & 0.801 & +6.4 & [\tminus2.6, +23.5] & \\
\quad Conduction $m_{\rm c}$ & median rel.\ error (\%) & 540 & 1.37 & 3.41 & \textbf{+59.4} & \textbf{[+53.8, +67.0]} & 0.995 / 0.951\\
\addlinespace[2pt]
\multicolumn{8}{@{}l@{}}{\textit{Thermoelectrics}}\\
\quad High-PF screen & mean top-20 recall, n/p & 261 & 0.849 & 0.825 & +2.9 & [\tminus5.5, +11.8] & \\
\quad Power factor & median rel.\ error (\%) & 226 & 2.01 & 4.41 & \textbf{+53.9} & \textbf{[+44.1, +62.5]} & 0.895 / 0.880\\
\addlinespace[2pt]
\multicolumn{8}{@{}l@{}}{\textit{Interconnects}}\\
\quad Low-$\rho_0\lambda$ screen & mean top-20 recall & 132 & 0.888 & 0.892 & \tminus0.5 & [\tminus1.7, +2.0] & \\
\quad $\rho_0\lambda$ & median rel.\ error (\%) & 132 & 0.65 & 1.13 & \textbf{+42.2} & \textbf{[+28.0, +61.0]} & 0.993 / 0.980\\
\bottomrule
\end{tabular}
\end{minipage}\end{center}

\paragraph{Photovoltaic efficiency.}
Supplementary Fig.~\ref{S-sfig:slme-sq} compares the SLME of the 200 semiconductors of the photovoltaic task with the Shockley--Queisser limit\cite{SI-shockley1961detailed}. The SLME implementation reproduces the limit, $33.68\%$ at 1.34 eV, for an ideal step absorber with radiative recombination only. The SLME of both models follows the DFT values, with $R^2$ of $0.986$ for UniTB and $0.982$ for UniTB-Flash on the 198 crystals with values from all three Hamiltonians, and among the 73 absorbers with DFT gaps of 0.5--2.5 eV, UniTB is closer to DFT for 67\%.

\begin{center}\begin{minipage}{\linewidth}
\centering\includegraphics[width=\textwidth]{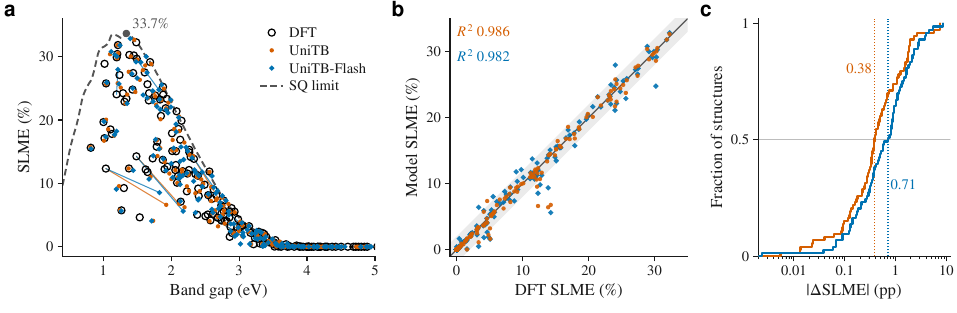}
\captionof{figure}{\textbf{SLME and the Shockley--Queisser limit.} \textbf{a},~200 nonmagnetic test crystals (UniTB-Flash: 198), each plotted at the band gap of its own Hamiltonian. Thin lines join model and DFT points more than 3 percentage points (pp) apart. The DFT points of the 40 crystals with gaps above 5 eV (SLME $\approx0$) lie outside the axis. \textbf{b},~The 198 crystals with all three values; gray band, $\pm2$ pp. \textbf{c},~The 73 crystals with DFT gaps of 0.5--2.5 eV; dotted lines, medians. Yu--Zunger SLME for a 0.5~$\mu$m film under AM1.5G at 300 K.}
\label{S-sfig:slme-sq}
\end{minipage}\end{center}

\paragraph{Seebeck coefficients.}
Supplementary Figs.~\ref{S-sfig:te-n} and \ref{S-sfig:te-p} show the structures, bands and Pisarenko curves of three n-type and three p-type test crystals with high DFT power factors at $10^{20}$ cm$^{-3}$; the bands and the Seebeck coefficients of each crystal come from the same three Hamiltonians. Each curve has 25 carrier densities spaced logarithmically from $10^{17}$ to $10^{21}$ cm$^{-3}$ at 300 K. Averaged over the 25 densities of the three n-type crystals, the magnitude of the Seebeck coefficient deviates from DFT by $3.8\%$ for UniTB and $4.2\%$ for UniTB-Flash, or by 11.4 and 10.8~$\mu$V\,K$^{-1}$.

\begin{center}\begin{minipage}{\linewidth}
\centering\includegraphics[width=\textwidth]{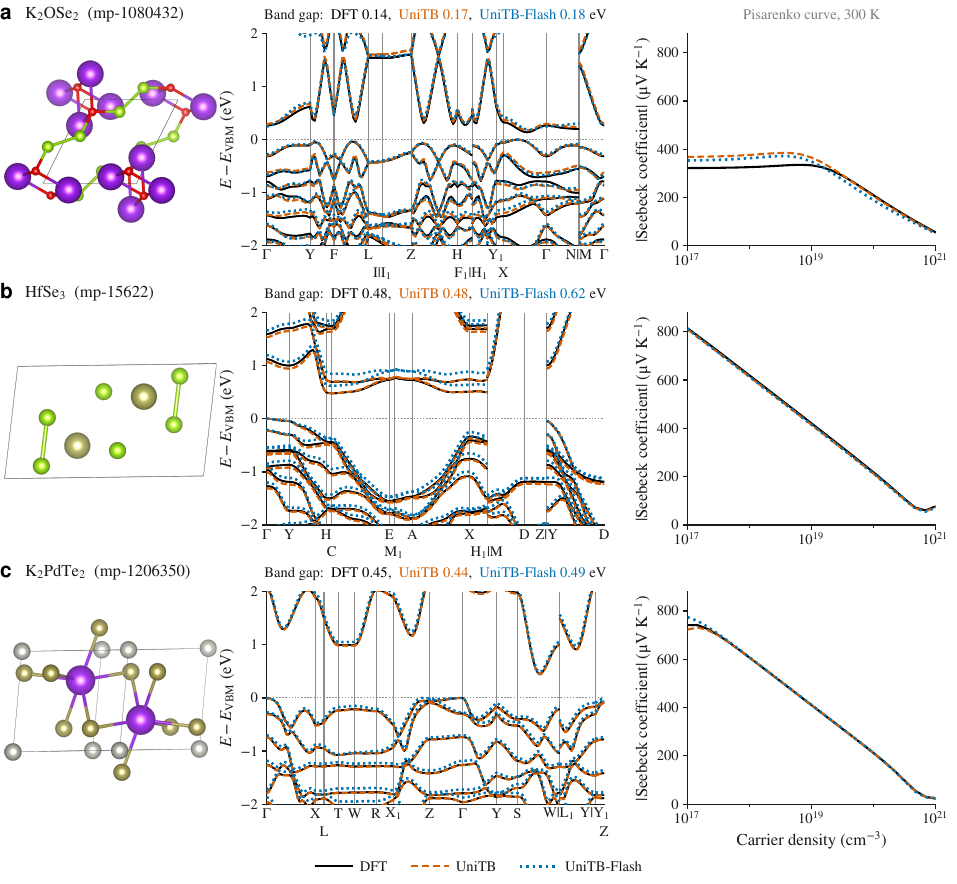}
\captionof{figure}{\textbf{Structures, bands and Seebeck coefficients of the n-type test crystals.} Three n-type test crystals with high DFT power factors at $10^{20}$ cm$^{-3}$. Each row shows the calculation cell, its bands and its Pisarenko curve, the magnitude of the Seebeck coefficient against carrier density at 300 K with a constant relaxation time. Bands follow standard paths and are relative to each Hamiltonian's valence-band maximum, and the band gaps above them are taken on the path. In the cells, K is purple, O red, Se green, Hf and Te olive, and Pd light grey.}
\label{S-sfig:te-n}
\end{minipage}\end{center}

\begin{center}\begin{minipage}{\linewidth}
\centering\includegraphics[width=\textwidth]{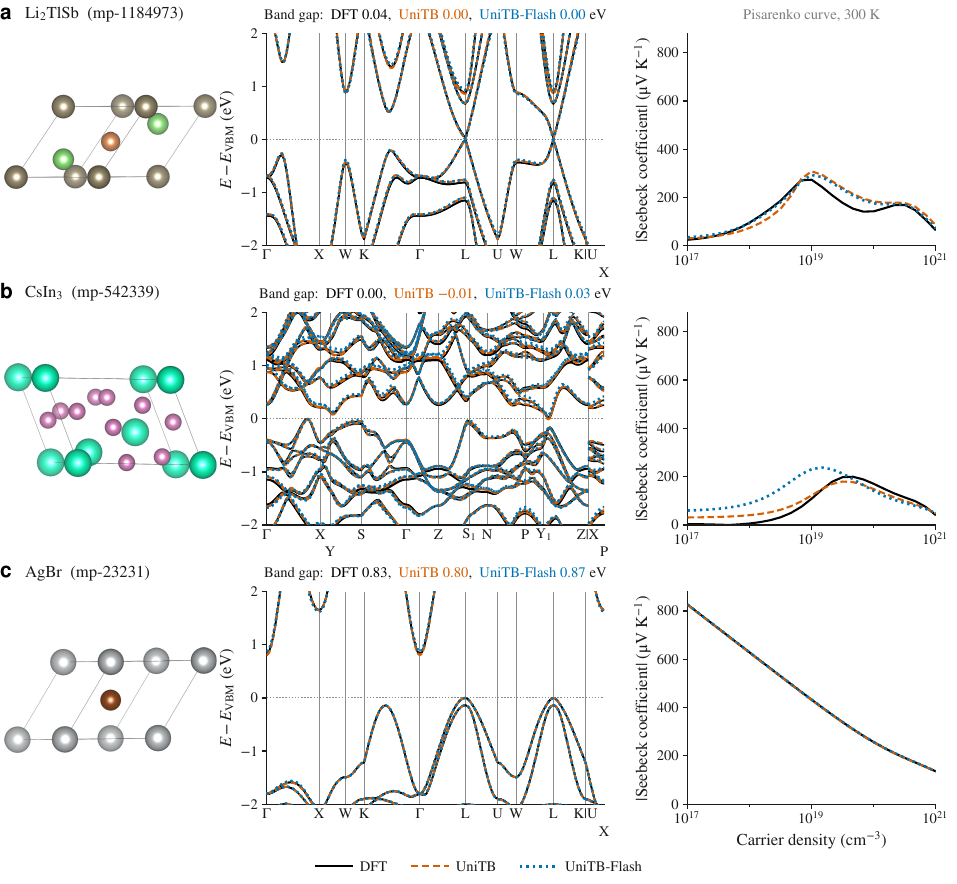}
\captionof{figure}{\textbf{Structures, bands and Seebeck coefficients of the p-type test crystals.} Three p-type test crystals with high DFT power factors at $10^{20}$ cm$^{-3}$. Each row shows the calculation cell, its bands and its Pisarenko curve, the magnitude of the Seebeck coefficient against carrier density at 300 K with a constant relaxation time. Bands follow standard paths and are relative to each Hamiltonian's valence-band maximum, and the band gaps above them are taken on the path. In the cells, Li is green, Tl grey-brown, Sb orange, Cs teal, In pink, Ag grey and Br brown.}
\label{S-sfig:te-p}
\end{minipage}\end{center}

\subsection*{Library screen}
\paragraph{Library and pool.}
The library holds 16,195 crystal structures: 11,384 from the Materials Project (MP)\cite{SI-jain2013materialsproject}, 2,103 from AFLOW-ICSD\cite{SI-curtarolo2012aflow}, 2,374 from COD\cite{SI-grazulis2012cod} and 334 from JARVIS-ICSD\cite{SI-choudhary2020jarvis}. The four domain pools contain 14,525 distinct structures before the pool rules and 8,785 after them, and the thermoelectric pool 1,249 and 694; the five pools together hold 9,064 distinct structures after the rules (Supplementary Table~\ref{S-tab:library-funnel}, which gives the per-rule removals and domain counts). Rules use the matched MP entry (release 2025-09-25; pymatgen\cite{SI-ong2013pymatgen} default structure-matching tolerances, choosing the lowest-$E_{\rm hull}$ entry when several match). They remove structures without an MP match, those with MP ordering other than NM or any matched ferromagnetic, antiferromagnetic or ferrimagnetic entry, those more than 0.1 eV per atom above the hull, those containing Tc, Pm or elements from Po on, and those that are, or match, a training or test structure, by MP entry or by StructureMatcher against every training and test structure with the same chemical system and reduced formula; a duplicated MP entry keeps one structure, the MP structure where there is one, and the same structure in every pool. Valid first-stage values must also pass the gap window or metal condition and the short-contact check.

\paragraph{Stages, hit lines and DFT.}
UniTB-Flash ranks each pool (thermoelectrics separately for $n$- and $p$-type doping), UniTB re-evaluates the 200 highest ranked, and the 20 best by UniTB are recomputed with DFT using the label settings with spin--orbit coupling and no spin polarization. The hit lines are evaluated with the same property codes for classic, widely used materials: W (mp-91) for interconnects, with Cu (mp-30) as a second line; Si (mp-149) for the refractive index and the conduction-band mass; CdTe (mp-406) for the SLME, which under this PBE protocol has a gap of 0.43 eV and an SLME of 3.4\%; and, for thermoelectrics, the median $\mathrm{PF}/\tau$ of the classic reference materials of each carrier type on the dense meshes of the final lists, taken over the reference materials with defined values (Supplementary Table~\ref{S-tab:hit-lines}). The five final lists hold 120 entries and 114 distinct structures; three thermoelectric structures appear for both carrier types and three also in another domain. Supplementary Tables~\ref{S-tab:final-metal}--\ref{S-tab:final-te} list the final candidates with the values of the three Hamiltonians.

\paragraph{Property protocols.}
The three Hamiltonians go through the same property codes. The refractive index is $n=\varepsilon_\infty^{1/2}$ from the independent-particle interband sum, without local-field, excitonic or ionic terms, over the states within $\pm10$ eV of the midgap of each Hamiltonian at $\Gamma$, with velocity matrix elements that include the overlap term, overlap eigen-directions below $2\times10^{-3}$ removed and $k$ meshes with $n_i=\lceil|\mathbf b_i|/0.16\,\text{\AA}^{-1}\rceil$ points along each reciprocal vector. The SLME uses a 0.5~$\mu$m film under AM1.5G illumination at 300 K, absorption from the same optics code with 0.025 eV broadening, the allowed gap over all retained states and no scissor correction. The conduction-band mass is the conductivity mass of the valleys at the conduction-band minimum from finite-difference Hessians with steps of 0.04 and 0.02 \AA$^{-1}$, averaged over the star of each valley, with a 2 meV degeneracy tolerance and a 0.05 eV valley window. $\rho_0\lambda$ uses Hellmann--Feynman velocities on the Fermi surface with full spin--orbit coupling, Fermi--Dirac occupations at 300 K, a Fermi window widened until the weights converge and $k$-point densities set by the number of atoms per cell. Thermoelectric $\mathrm{PF}/\tau$ and the Seebeck coefficient use constant-relaxation-time Boltzmann transport in the rigid-band approximation with spinor states at 300 K and $10^{20}$ cm$^{-3}$; in the screen, they are left undefined when the gap on the mesh is 0.02 eV or less. Their $\Gamma$-centered meshes take the smallest even $n_i\geq\max(6,\lceil L/|\mathbf a_i|\rceil)$, with $L=600$ \AA{} for the final lists and the reference materials and, in the ranking stages, $L=600$, 450 or 300 \AA{} for cells of up to 4, 5--16 and 17 or more atoms. UniTB spectra are projected with $b=10$ eV, and their velocities include the derivative of the projection; UniTB-Flash and DFT spectra are not projected. The DFT checks use ABACUS 3.10.1 with PBE, full spin--orbit coupling and no spin polarization, the pseudopotentials and orbitals of the training labels, a 120 Ry cutoff, $\Gamma$-centered meshes with $n_i=\min(6,\max(1,\lceil|\mathbf b_i|/0.199\,\text{\AA}^{-1}\rceil))$, 0.001 Ry Gaussian smearing and a density convergence of $10^{-6}$; the exported $H(\mathbf R)$ and $S(\mathbf R)$ pass through the same property codes.

\paragraph{Thermoelectrics.}
The thermoelectric pool takes the library structures with a UniTB-Flash band gap of 0.1--1.0 eV on a coarse mesh (spacing 0.35 \AA$^{-1}$) and a nonmagnetic label in the library record, 1,249 in all, and the pool rules leave 694. UniTB-Flash ranks them by $\mathrm{PF}/\tau$ separately for $n$- and $p$-type doping (653 and 655 valid values), UniTB re-evaluates the 200 highest ranked of each type, and the 20 best by UniTB are recomputed with DFT. By DFT, 19 $n$-type and 13 $p$-type candidates exceed the median line ($1.291\times10^{11}$ and $1.859\times10^{11}$ W\,m$^{-1}$\,K$^{-2}$\,s$^{-1}$), and six $n$-type candidates exceed the strongest $n$-type reference material, Mg$_2$Si ($2.72\times10^{11}$); K$_6$CdPb$_8$ ($p$ type) has no defined DFT value and counts as below the line (Supplementary Table~\ref{S-tab:final-te}). Against DFT, the values of UniTB differ by medians of 3.99\% ($n$) and 2.50\% ($p$), against 6.04\% and 2.68\% for UniTB-Flash.

\paragraph{Second-stage corrections and cost.}
Against DFT, the final interconnect, photonic, mobility and photovoltaic values of UniTB differ by medians of 0.36\%, 0.62\%, 1.12\% and 0.61\%, against 0.95\%, 0.77\%, 3.62\% and 0.62\% for UniTB-Flash; UniTB is the closer of the two for 14, 12, 13 and 11 of the 20 candidates, and for 12 of the 20 $n$-type and 9 of the 19 $p$-type thermoelectric candidates with defined DFT values. Among the first-stage top 200, UniTB also revises values such as the conduction-band mass of Rb$_{14}$ZnGe$_{16}$, which rises from 0.083 to 0.401 $m_{\rm e}$ and removes it from the final list. Over the five pools, the first stage took about 57 GPU-hours on NVIDIA H200 GPUs and the second stage about 18 GPU-hours, including Hamiltonian inference; the DFT checks of the 114 final structures took about 370 CPU core-hours.

\providecommand{\tminus}{\textminus}
\begin{center}\begin{minipage}{\linewidth}
\captionof{table}{\textbf{Structure counts of the library screen.} \textbf{a},~Removals by the pool rules are counted in the order listed. The thermoelectric pool requires a UniTB-Flash gap of 0.1--1.0 eV, and 415 of its 694 structures are also in a pool of the other four domains. \textbf{b},~First-stage valid counts structures with a valid UniTB-Flash value that pass the gap window or the metal condition and the short-contact check. Share is the UniTB stage as a fraction of the first-stage valid structures. Hits are structures of the final list whose DFT value passes the hit line (Supplementary Table~\ref{S-tab:hit-lines}). Thermoelectric $n$-type and $p$-type doping are ranked separately.}
\label{S-tab:library-funnel}
\centering\small
\renewcommand{\arraystretch}{1.08}
\setlength{\tabcolsep}{3pt}
\begin{tabular}{@{}lrrrr@{}}
\toprule
\textbf{a}\enspace Library and pool & \multicolumn{2}{c}{Four domains} & \multicolumn{2}{c}{Thermoelectrics}\\
\cmidrule(lr){2-3}\cmidrule(l){4-5}
 & Removed & Structures & Removed & Structures\\
\midrule
Library &  & 16,195 &  & 16,195\\
\quad Materials Project (MP) &  & 11,384 &  & \\
\quad AFLOW-ICSD &  & 2,103 &  & \\
\quad COD &  & 2,374 &  & \\
\quad JARVIS-ICSD &  & 334 &  & \\
\addlinespace[2pt]
Domain pools before the pool rules &  & 14,525 &  & 1,249\\
\addlinespace[2pt]
No MP match & 2,392 & 12,133 & 252 & 997\\
Not nonmagnetic & 784 & 11,349 & 70 & 927\\
$E_{\rm hull}>0.1$ eV/atom & 2,484 & 8,865 & 220 & 707\\
Radioactive elements & 21 & 8,844 & 6 & 701\\
Duplicate of an MP entry & 4 & 8,840 & 1 & 700\\
Training or test structure & 55 & 8,785 & 6 & 694\\
\midrule
\textbf{Pool after the rules} &  & \textbf{8,785} &  & \textbf{694}\\
Five pools together, distinct structures & \multicolumn{4}{r}{9,064}\\
\bottomrule
\end{tabular}
\par\vspace{14pt}
{\setlength{\tabcolsep}{5pt}%
\begin{tabular}{@{}lrrrrr@{}}
\toprule
\textbf{b}\enspace Stages per domain & Interconnects & Photonics & Photovoltaics & Mobility & \shortstack[r]{Thermoelectrics\\($n$ / $p$)}\\
\midrule
Pool before the pool rules & 3,604 & 10,921 & 4,258 & 5,609 & 1,249\\
Pool after the pool rules & 1,470 & 7,315 & 2,729 & 3,611 & 694\\
First-stage valid & 1,470 & 7,296 & 2,715 & 2,999 & 653 / 655\\
UniTB stage & 200 & 200 & 200 & 200 & 200 / 200\\
\quad share of first-stage valid & 13.6\% & 2.7\% & 7.4\% & 6.7\% & 30.6\% / 30.5\%\\
Final list & 20 & 20 & 20 & 20 & 20 / 20\\
With DFT & 20 & 20 & 20 & 20 & 20 / 19\\
Hits by DFT & 20 & 20 & 20 & 20 & 19 / 13\\
UniTB closer to DFT than UniTB-Flash & 14/20 & 12/20 & 11/20 & 13/20 & 12/20 / 9/19\\
\bottomrule
\end{tabular}}
\end{minipage}\end{center}

\begin{center}\begin{minipage}{\linewidth}
\captionof{table}{\textbf{Reference materials and hit lines.} \textbf{a},~Values come from the same property codes as for the screened structures. \textbf{b},~DFT PF/$\tau$ of the 13 thermoelectric reference materials at 10$^{20}$~cm$^{-3}$ and 300~K, on the dense $k$ meshes of the final lists. Bold with $\dagger$ marks the classic reference materials of each carrier type, whose defined values give its median line; undef., no defined value (indirect gap of at most 0.02 eV).}
\label{S-tab:hit-lines}
\centering\small
\renewcommand{\arraystretch}{1.08}
\setlength{\tabcolsep}{3pt}
\begin{tabular}{@{}llrrr@{}}
\toprule
\textbf{a}\enspace Reference material & Role & \shortstack[r]{UniTB-\\Flash} & UniTB & DFT\\
\midrule
\multicolumn{5}{@{}l@{}}{\textit{Interconnects, $\rho_0\lambda$ (10$^{-16}$\,$\Omega$\,m$^2$), lower is better}}\\
\quad W (mp-91) & hit line & 10.78 & 10.71 & 10.77\\
\quad Cu (mp-30) & second line & 6.589 & 6.597 & 6.595\\
\addlinespace[2pt]
\multicolumn{5}{@{}l@{}}{\textit{Photonics, refractive index $n$, higher is better}}\\
\quad Si (mp-149) & hit line & 3.992 & 3.928 & 3.920\\
\addlinespace[2pt]
\multicolumn{5}{@{}l@{}}{\textit{Photovoltaics, SLME (\%), higher is better}}\\
\quad CdTe (mp-406) & hit line & 2.768 & 3.101 & 3.356\\
\addlinespace[2pt]
\multicolumn{5}{@{}l@{}}{\textit{Mobility, conduction-band mass $m_{\rm c}$ ($m_{\rm e}$), lower is better}}\\
\quad Si (mp-149) & hit line & 0.2482 & 0.2503 & 0.2500\\
\bottomrule
\end{tabular}
\par\vspace{14pt}
\begin{tabular}{@{}llrr@{}}
\toprule
\textbf{b}\enspace Thermoelectric reference & MP id & \multicolumn{2}{c}{\shortstack{DFT PF/$\tau$\\(W\,m$^{-1}$\,K$^{-2}$\,s$^{-1}$)}}\\
\cmidrule(lr){3-4}
 &  & $n$ type & $p$ type\\
\midrule
Bi$_2$Te$_3$ & mp-34202 & undef.$^{\dagger}$ & undef.$^{\dagger}$\\
PbTe & mp-19717 & $\mathbf{4.52\times10^{10}}^{\dagger}$ & $\mathbf{5.24\times10^{11}}^{\dagger}$\\
PbSe & mp-2201 & $\mathbf{4.40\times10^{10}}^{\dagger}$ & $6.39\times10^{10}$\\
Mg$_3$Sb$_2$ & mp-2646 & undef.$^{\dagger}$ & undef.\\
Mg$_2$Si & mp-1367 & $\mathbf{2.72\times10^{11}}^{\dagger}$ & $2.28\times10^{11}$\\
ZrNiSn & mp-924129 & $\mathbf{1.29\times10^{11}}^{\dagger}$ & $1.61\times10^{11}$\\
CoSb$_3$ & mp-1317 & $\mathbf{1.39\times10^{11}}^{\dagger}$ & $1.22\times10^{11}$\\
Sb$_2$Te$_3$ & mp-1201 & undef. & undef.$^{\dagger}$\\
SnTe & mp-1883 & $1.74\times10^{11}$ & $\mathbf{1.79\times10^{11}}^{\dagger}$\\
GeTe & mp-938 & $1.09\times10^{11}$ & $\mathbf{1.92\times10^{11}}^{\dagger}$\\
SnSe & mp-691 & $6.87\times10^{10}$ & $\mathbf{9.98\times10^{10}}^{\dagger}$\\
NbFeSb & mp-9437 & $6.26\times10^{10}$ & $\mathbf{2.23\times10^{11}}^{\dagger}$\\
ZnSb & mp-753 & $1.30\times10^{11}$ & $\mathbf{3.04\times10^{10}}^{\dagger}$\\
\midrule
Median line &  & $1.291\times10^{11}$ & $1.859\times10^{11}$\\
\bottomrule
\end{tabular}
\end{minipage}\end{center}

\begin{center}\begin{minipage}{\linewidth}
\captionof{table}{\textbf{Final interconnect candidates.} Final 20, ranked by the UniTB value of $\rho_0\lambda$, in 10$^{-16}$\,$\Omega$\,m$^2$. $E_{\rm hull}$ refers to the matched Materials Project entry. DFT/hit line is the DFT value divided by that of W (mp-91, 10.77); below 1 is better than the line.}
\label{S-tab:final-metal}
\centering\small
\renewcommand{\arraystretch}{1.08}
\setlength{\tabcolsep}{3pt}
\begin{tabular}{@{}rllrrrrr@{}}
\toprule
Rank & Formula & ID & \shortstack[r]{$E_{\rm hull}$\\(eV/atom)} & \shortstack[r]{UniTB-\\Flash} & UniTB & DFT & \shortstack[r]{DFT/\\hit line}\\
\midrule
1 & Cu & mp-30 & 0.000 & 6.589 & 6.597 & 6.595 & 0.612\\
2 & CuZn & mp-987 & 0.000 & 7.068 & 7.083 & 7.079 & 0.657\\
3 & TiRh$_3$ & mp-1152 & 0.000 & 7.201 & 7.252 & 7.171 & 0.666\\
4 & MgIn & mp-2313 & 0.000 & 7.968 & 7.894 & 7.891 & 0.733\\
5 & CuAs & aflow-ee1a1210f72febbc & 0.053 & 7.855 & 7.977 & 7.988 & 0.742\\
6 & MgCu$_{4}$In & mp-30587 & 0.000 & 8.077 & 8.063 & 8.023 & 0.745\\
7 & Tl$_{2}$Ba$_{2}$Ca$_{3}$Cu$_{4}$O$_{12}$ & mp-556733 & 0.007 & 8.236 & 8.238 & 8.369 & 0.777\\
8 & NiIn & mp-20997 & 0.092 & 8.411 & 8.430 & 8.431 & 0.783\\
9 & NbS & mp-647 & 0.066 & 8.509 & 8.574 & 8.552 & 0.794\\
10 & LiIr$_3$ & mp-30738 & 0.000 & 8.379 & 8.731 & 8.482 & 0.788\\
11 & NbPt & mp-999376 & 0.000 & 8.911 & 8.975 & 9.012 & 0.837\\
12 & MgHg & mp-1408 & 0.000 & 9.010 & 8.983 & 8.988 & 0.835\\
13 & VCr$_{2}$C$_{2}$ & mp-4253 & 0.083 & 9.362 & 9.051 & 9.124 & 0.847\\
14 & Hf$_{2}$PdH$_{2}$ & mp-1069811 & 0.100 & 9.322 & 9.136 & 9.238 & 0.858\\
15 & Cu$_3$Pt & mp-12086 & 0.000 & 9.333 & 9.178 & 9.216 & 0.856\\
16 & MgAu & mp-437 & 0.000 & 9.154 & 9.304 & 9.285 & 0.862\\
17 & NbS & mp-2243 & 0.083 & 9.308 & 9.359 & 9.164 & 0.851\\
18 & Li$_{2}$PtIn & mp-21026 & 0.000 & 9.777 & 9.708 & 9.732 & 0.904\\
19 & Mo$_3$Ti & aflow-dbd2b2ac1b4a0bab & 0.000 & 9.763 & 9.828 & 9.461 & 0.879\\
20 & CdCu$_2$ & mp-568315 & 0.010 & 9.830 & 9.851 & 9.820 & 0.912\\
\bottomrule
\end{tabular}
\end{minipage}\end{center}

\begin{center}\begin{minipage}{\linewidth}
\captionof{table}{\textbf{Final photonic candidates.} Final 20, ranked by the UniTB refractive index $n$. $E_{\rm hull}$ refers to the matched Materials Project entry. DFT/hit line is the DFT value divided by that of Si (mp-149, 3.920); above 1 is better than the line.}
\label{S-tab:final-optics}
\centering\small
\renewcommand{\arraystretch}{1.08}
\setlength{\tabcolsep}{3pt}
\begin{tabular}{@{}rllrrrrr@{}}
\toprule
Rank & Formula & ID & \shortstack[r]{$E_{\rm hull}$\\(eV/atom)} & \shortstack[r]{UniTB-\\Flash} & UniTB & DFT & \shortstack[r]{DFT/\\hit line}\\
\midrule
1 & NaTe$_3$ & mp-28478 & 0.000 & 5.411 & 5.607 & 5.816 & 1.484\\
2 & Re$_4$Ge$_7$ & mp-1201064 & 0.040 & 5.553 & 5.581 & 5.595 & 1.427\\
3 & NaSn & mp-11051 & 0.000 & 5.142 & 5.404 & 5.352 & 1.366\\
4 & Co$_{2}$Sn$_{3}$Te$_{3}$ & mp-866481 & 0.000 & 5.384 & 5.312 & 5.374 & 1.371\\
5 & CoAsSe & mp-505511 & 0.000 & 5.182 & 5.096 & 5.146 & 1.313\\
6 & CoPSe & mp-10368 & 0.000 & 4.825 & 4.851 & 4.874 & 1.243\\
7 & RuAs$_{2}$ & mp-766 & 0.000 & 4.880 & 4.778 & 4.799 & 1.224\\
8 & SrGe$_2$ & mp-1244 & 0.015 & 4.668 & 4.749 & 4.851 & 1.238\\
9 & Ru$_{2}$Si$_{3}$ & mp-22192 & 0.000 & 4.769 & 4.747 & 4.764 & 1.215\\
10 & Ba$_{5}$Ga$_{6}$H$_{2}$ & mp-643683 & 0.000 & 4.780 & 4.746 & 4.757 & 1.214\\
11 & Ir$_{2}$Sn$_{3}$Te$_{3}$ & mp-5142 & 0.045 & 4.650 & 4.741 & 4.668 & 1.191\\
12 & Si$_{19}$Te$_8$ & mp-31135 & 0.015 & 5.094 & 4.722 & 5.002 & 1.276\\
13 & Ir$_{3}$Si$_{5}$ & mp-2084 & 0.000 & 4.692 & 4.681 & 4.715 & 1.203\\
14 & TaRuSb & mp-31454 & 0.000 & 4.624 & 4.639 & 4.662 & 1.189\\
15 & IrPTe & mp-1191625 & 0.037 & 4.598 & 4.548 & 4.549 & 1.161\\
16 & Rh$_2$O$_3$ & mp-613620 & 0.027 & 4.457 & 4.447 & 4.458 & 1.137\\
17 & Pb$_{4}$In$_{2}$Sb$_{4}$Se$_{13}$ & mp-1205220 & 0.013 & 4.281 & 4.372 & 4.349 & 1.110\\
18 & Rh$_{2}$Ge$_{3}$Se$_{3}$ & mp-976401 & 0.000 & 4.325 & 4.355 & 4.340 & 1.107\\
19 & Sr$_{11}$InSb$_9$ & mp-578752 & 0.000 & 4.334 & 4.298 & 4.358 & 1.112\\
20 & RbPb & mp-21525 & 0.000 & 4.186 & 4.254 & 4.295 & 1.096\\
\bottomrule
\end{tabular}
\end{minipage}\end{center}

\begin{center}\begin{minipage}{\linewidth}
\captionof{table}{\textbf{Final photovoltaic candidates.} Final 20, ranked by the UniTB SLME, in \%. $E_{\rm hull}$ refers to the matched Materials Project entry. DFT/hit line is the DFT value divided by that of CdTe (mp-406, 3.356); above 1 is better than the line.}
\label{S-tab:final-slme}
\centering\small
\renewcommand{\arraystretch}{1.08}
\setlength{\tabcolsep}{3pt}
\begin{tabular}{@{}rllrrrrr@{}}
\toprule
Rank & Formula & ID & \shortstack[r]{$E_{\rm hull}$\\(eV/atom)} & \shortstack[r]{UniTB-\\Flash} & UniTB & DFT & \shortstack[r]{DFT/\\hit line}\\
\midrule
1 & BaRh$_{2}$Ge$_{4}$S$_{6}$ & mp-1194551 & 0.020 & 33.09 & 33.17 & 33.22 & 9.899\\
2 & Ga$_2$Te$_2$Br$_7$ & mp-570186 & 0.000 & 32.85 & 33.06 & 32.70 & 9.746\\
3 & ZrSe$_{4}$Cl$_{6}$ & mp-30113 & 0.000 & 32.87 & 33.02 & 32.80 & 9.774\\
4 & Rb$_7$NaGe$_8$ & mp-14407 & 0.000 & 32.87 & 32.99 & 33.08 & 9.858\\
5 & C$_{2}$H$_{10}$N$_{2}$Se$_{4}$Zn & mp-1198883 & 0.058 & 31.81 & 32.91 & 32.89 & 9.800\\
6 & TeI$_4$ & mp-651155 & 0.002 & 32.82 & 32.87 & 32.68 & 9.740\\
7 & Bi$_2$Se$_4$Cl$_7$ & mp-505564 & 0.000 & 32.58 & 32.84 & 29.70 & 8.851\\
8 & Hg$_{6}$InSb$_{4}$Br$_{7}$ & mp-1203885 & 0.000 & 33.10 & 32.82 & 31.51 & 9.391\\
9 & Hg$_6$BiSb$_4$Br$_7$ & mp-568828 & 0.000 & 33.39 & 32.78 & 30.63 & 9.129\\
10 & K$_5$TlO$_4$ & mp-14439 & 0.000 & 31.98 & 32.77 & 31.06 & 9.257\\
11 & TeI$_4$ & mp-567269 & 0.002 & 32.68 & 32.76 & 32.88 & 9.799\\
12 & K$_{3}$CuNb$_{2}$Se$_{12}$ & mp-6168 & 0.000 & 32.78 & 32.74 & 32.47 & 9.675\\
13 & Cs$_{12}$Sn$_8$GeO$_4$ & mp-1198905 & 0.001 & 32.47 & 32.74 & 32.39 & 9.652\\
14 & Bi$_2$Te$_4$Cl$_7$ & mp-29991 & 0.005 & 32.83 & 32.71 & 32.74 & 9.758\\
15 & Cs$_3$SnTe$_3$ & mp-1193021 & 0.000 & 32.92 & 32.69 & 32.90 & 9.804\\
16 & Cd$_4$Sb$_2$I$_3$ & mp-29044 & 0.000 & 31.88 & 32.66 & 24.09 & 7.179\\
17 & Hg$_{5}$Sb$_{2}$I$_{6}$ & mp-28889 & 0.000 & 32.99 & 32.57 & 32.57 & 9.707\\
18 & Bi$_6$PtBr$_{10}$ & mp-583499 & 0.020 & 32.57 & 32.54 & 32.56 & 9.702\\
19 & AlSe$_{2}$Cl$_{4}$ & mp-574529 & 0.000 & 31.72 & 32.54 & 32.65 & 9.730\\
20 & TeI$_4$ & mp-570884 & 0.000 & 32.52 & 32.52 & 32.51 & 9.689\\
\bottomrule
\end{tabular}
\end{minipage}\end{center}

\begin{center}\begin{minipage}{\linewidth}
\captionof{table}{\textbf{Final high-mobility candidates.} Final 20, ranked by the UniTB conduction-band mass $m_{\rm c}$, in $m_{\rm e}$. $E_{\rm hull}$ refers to the matched Materials Project entry. DFT/hit line is the DFT value divided by that of Si (mp-149, 0.2500); below 1 is better than the line.}
\label{S-tab:final-emass}
\centering\small
\renewcommand{\arraystretch}{1.08}
\setlength{\tabcolsep}{3pt}
\begin{tabular}{@{}rllrrrrr@{}}
\toprule
Rank & Formula & ID & \shortstack[r]{$E_{\rm hull}$\\(eV/atom)} & \shortstack[r]{UniTB-\\Flash} & UniTB & DFT & \shortstack[r]{DFT/\\hit line}\\
\midrule
1 & Na$_{2}$KSb & mp-15724 & 0.000 & 0.09099 & 0.08530 & 0.08503 & 0.340\\
2 & In$_4$SnSe$_4$ & mp-628768 & 0.002 & 0.1202 & 0.09775 & 0.09928 & 0.397\\
3 & Na$_3$InP$_2$ & mp-21577 & 0.000 & 0.1033 & 0.1007 & 0.1050 & 0.420\\
4 & Na$_2$SnAs$_2$ & mp-574684 & 0.000 & 0.09301 & 0.1032 & 0.1032 & 0.413\\
5 & ZnIn$_{2}$Se$_{4}$ & mp-22607 & 0.000 & 0.1224 & 0.1173 & 0.1167 & 0.467\\
6 & GaHg$_2$SbCl$_4$ & mp-568031 & 0.001 & 0.1319 & 0.1198 & 0.1187 & 0.475\\
7 & Na$_2$LiGaAs$_2$ & mp-9722 & 0.000 & 0.1134 & 0.1204 & 0.1167 & 0.467\\
8 & NaSrAs & mp-9775 & 0.000 & 0.1164 & 0.1223 & 0.1183 & 0.473\\
9 & RbNa$_{8}$Ga$_{3}$As$_{6}$ & mp-1200468 & 0.000 & 0.1148 & 0.1229 & 0.1207 & 0.483\\
10 & K$_3$Ga$_3$As$_4$ & mp-567524 & 0.000 & 0.1218 & 0.1242 & 0.1223 & 0.489\\
11 & Rb$_2$In$_2$Sb$_3$ & mp-21760 & 0.000 & 0.1197 & 0.1251 & 0.1237 & 0.495\\
12 & K$_{11}$Cd$_2$Sb$_5$ & mp-1194660 & 0.000 & 0.1310 & 0.1306 & 0.1308 & 0.523\\
13 & In$_2$Se$_3$ & aflow-39d46486341c0aa9 & 0.000 & 0.1388 & 0.1314 & 0.1320 & 0.528\\
14 & K$_3$In$_2$As$_3$ & mp-583615 & 0.000 & 0.1309 & 0.1338 & 0.1358 & 0.543\\
15 & Na$_4$Cd$_3$Se$_5$ & mp-28851 & 0.000 & 0.1322 & 0.1339 & 0.1326 & 0.530\\
16 & NaIn$_3$Te$_5$ & cod-7721154 & 0.000 & 0.1606 & 0.1342 & 0.1358 & 0.543\\
17 & K$_2$In$_2$Sb$_3$ & mp-579637 & 0.000 & 0.1292 & 0.1357 & 0.1341 & 0.536\\
18 & Ag$_{2}$CdGeS$_{4}$ & mp-1197203 & 0.000 & 0.1438 & 0.1372 & 0.1371 & 0.548\\
19 & Ca$_2$CdAs$_2$ & mp-1191256 & 0.021 & 0.1400 & 0.1377 & 0.1357 & 0.543\\
20 & K$_2$LiGaAs$_2$ & mp-9703 & 0.000 & 0.1407 & 0.1381 & 0.1395 & 0.558\\
\bottomrule
\end{tabular}
\end{minipage}\end{center}

\begin{center}\begin{minipage}{\linewidth}
\captionof{table}{\textbf{Final thermoelectric candidates.} Final lists of 20 per carrier type, ranked by the UniTB value of PF/$\tau$. PF/$\tau$ is in 10$^{11}$~W\,m$^{-1}$\,K$^{-2}$\,s$^{-1}$ at 10$^{20}$~cm$^{-3}$ and 300~K, on the dense $k$ meshes of the final lists. The ranking used the second-stage values on the size-dependent meshes, which differ by more than 5\% for CsY$_2$CuTe$_4$ ($n$, 19), SnSb$_{2}$Te$_{4}$ ($p$, 4) and K$_6$CdPb$_8$ ($p$, 12). Ratio is the DFT value over the median line of the reference materials of the carrier type, $1.291\times10^{11}$ ($n$) and $1.859\times10^{11}$ ($p$). Undef. marks no defined DFT value (gap of at most 0.02 eV), counted as below the line. $E_{\rm hull}$ refers to the matched Materials Project entry.}
\label{S-tab:final-te}
\centering\footnotesize
\renewcommand{\arraystretch}{1.0}
\setlength{\tabcolsep}{3pt}
\begin{tabular}{@{}lrllrrrrrl@{}}
\toprule
Type & Rank & Formula & ID & $E_{\rm hull}$ & \shortstack[r]{UniTB-\\Flash} & UniTB & DFT & Ratio & \shortstack[l]{Above\\line}\\
\midrule
$n$ & 1 & LiTe$_{3}$ & mp-27466 & 0.000 & 10.7 & 11.3 & 11.7 & 9.07 & yes\\
$n$ & 2 & HfTe$_{5}$ & mp-1168 & 0.000 & 6.53 & 7.14 & 7.25 & 5.62 & yes\\
$n$ & 3 & NaTe$_{3}$ & mp-28478 & 0.000 & 4.67 & 4.64 & 4.53 & 3.51 & yes\\
$n$ & 4 & Ta$_{2}$Hg$_{2}$F$_{11}$ & mp-581413 & 0.000 & 3.69 & 3.83 & 3.75 & 2.90 & yes\\
$n$ & 5 & Al$_{6}$Ge$_{5}$ & mp-1190320 & 0.044 & 4.09 & 2.86 & 3.11 & 2.41 & yes\\
$n$ & 6 & Ta$_{4}$SiTe$_{4}$ & mp-28509 & 0.000 & 3.21 & 2.78 & 3.16 & 2.45 & yes\\
$n$ & 7 & TlPt$_{2}$Se$_{3}$ & mp-541487 & 0.000 & 2.63 & 2.62 & 2.63 & 2.04 & yes\\
$n$ & 8 & YWB$_{4}$ & mp-1190755 & 0.000 & 2.56 & 2.6 & 2.39 & 1.85 & yes\\
$n$ & 9 & YMoB$_{4}$ & mp-7691 & 0.000 & 2.63 & 2.5 & 2.39 & 1.85 & yes\\
$n$ & 10 & CaPdAs & mp-28763 & 0.000 & 2.31 & 2.37 & 2.33 & 1.81 & yes\\
$n$ & 11 & Nb$_{4}$SiTe$_{4}$ & mp-1232409 & 0.000 & 3.02 & 2.36 & 2.61 & 2.02 & yes\\
$n$ & 12 & Tl$_{2}$S & cod-9012277 & 0.001 & 2.33 & 2.34 & 2.35 & 1.82 & yes\\
$n$ & 13 & RuSi & mp-189 & 0.000 & 2.34 & 2.29 & 2.29 & 1.77 & yes\\
$n$ & 14 & CaI$_{10}\cdot$7H$_{2}$O & mp-721650 & 0.034 & 2.26 & 2.25 & 2.42 & 1.88 & yes\\
$n$ & 15 & KCuZrSe$_{3}$ & mp-9318 & 0.000 & 1.97 & 2.04 & 2.09 & 1.62 & yes\\
$n$ & 16 & Cs$_{2}$PdSb$_{2}$ & mp-1078965 & 0.000 & 1.06 & 1.93 & 1.97 & 1.52 & yes\\
$n$ & 17 & SrGe$_{2}$ & mp-1244 & 0.015 & 1.97 & 1.93 & 2.02 & 1.57 & yes\\
$n$ & 18 & Pt$_{3}$I$_{8}$ & mp-28320 & 0.000 & 1.79 & 1.89 & 2.21 & 1.71 & yes\\
$n$ & 19 & CsY$_{2}$CuTe$_{4}$ & mp-1199735 & 0.000 & 0.553 & 0.501 & 0.302 & 0.23 & no\\
$n$ & 20 & AgVSeO$_{5}$ & mp-557135 & 0.000 & 1.78 & 1.85 & 1.74 & 1.35 & yes\\
\addlinespace[2pt]
$p$ & 1 & Ta$_{2}$Hg$_{2}$F$_{11}$ & mp-581413 & 0.000 & 3.72 & 3.83 & 3.75 & 2.02 & yes\\
$p$ & 2 & AgInS$_{2}$ & mp-20162 & 0.090 & 2.35 & 3.77 & 3.73 & 2.01 & yes\\
$p$ & 3 & AgCl & mp-22922 & 0.000 & 3.74 & 3.75 & 3.73 & 2.01 & yes\\
$p$ & 4 & SnSb$_{2}$Te$_{4}$ & mp-27947 & 0.000 & 2.24 & 2.59 & 2.56 & 1.38 & yes\\
$p$ & 5 & HgAs$_{2}$O$_{6}$ & mp-3810 & 0.000 & 3 & 3.03 & 3.02 & 1.62 & yes\\
$p$ & 6 & I$_{2}$ & cod-1010091 & 0.000 & 2.51 & 2.62 & 2.55 & 1.37 & yes\\
$p$ & 7 & Ag$_{3}$AuSe$_{2}$ & mp-3172 & 0.000 & 2.45 & 2.35 & 2.48 & 1.34 & yes\\
$p$ & 8 & BaRh$_{2}$Si$_{9}$ & mp-1190822 & 0.000 & 2.13 & 2.23 & 2.06 & 1.11 & yes\\
$p$ & 9 & LiTe$_{3}$ & mp-27466 & 0.000 & 2.12 & 2.09 & 2.2 & 1.18 & yes\\
$p$ & 10 & Bi$_{14}$Te$_{13}$S$_{8}$ & jv-JVASP-29442 & 0.016 & 2.05 & 2.08 & 2.11 & 1.13 & yes\\
$p$ & 11 & Au$_{7}$P$_{10}$I & aflow-2b79959ceebc4df6 & 0.000 & 1.9 & 2.01 & 1.94 & 1.04 & yes\\
$p$ & 12 & K$_{6}$CdPb$_{8}$ & mp-680480 & 0.045 & 1.34 & 1.32 & undef. & -- & no\\
$p$ & 13 & PtSe$_{2}$ & mp-1115 & 0.000 & 1.94 & 1.92 & 1.91 & 1.03 & yes\\
$p$ & 14 & Na$_{2}$OsO$_{4}$ & mp-1190635 & 0.000 & 1.85 & 1.89 & 1.81 & 0.97 & no\\
$p$ & 15 & Ba$_{2}$Ir$_{4}$Ge$_{5}$ & mp-1190188 & 0.000 & 1.92 & 1.85 & 1.81 & 0.97 & no\\
$p$ & 16 & Ta$_{4}$SiTe$_{4}$ & mp-28509 & 0.000 & 2.59 & 1.85 & 2.47 & 1.33 & yes\\
$p$ & 17 & K$_{2}$ZrTe$_{3}$ & mp-29602 & 0.000 & 1.71 & 1.81 & 1.66 & 0.89 & no\\
$p$ & 18 & NaSn & mp-11051 & 0.000 & 1.25 & 1.8 & 1.75 & 0.94 & no\\
$p$ & 19 & TlCu$_{7}$S$_{4}$ & mp-21964 & 0.014 & 1.77 & 1.79 & 1.79 & 0.96 & no\\
$p$ & 20 & RhGeTe & mp-607818 & 0.048 & 1.52 & 1.76 & 1.73 & 0.93 & no\\
\bottomrule
\end{tabular}
\end{minipage}\end{center}

\paragraph{Inference timing.}
Main-text Fig.~\ref{fig:prior}c uses PbTe supercells of 16--3,456 atoms (Methods). At 3,456 atoms, full inference takes 5.11~s for UniTB and 1.65~s for UniTB-Flash, the prior assembly 4.08 and 0.66~s, and the network forward pass, charge equilibration included, 0.955 and 0.924~s. The network times of the two models differ by 0.6--3.8\% at every size.

\end{document}